\documentclass[12pt, a4paper]{scrbook}
\usepackage[utf8]{inputenc}
\usepackage[brazil]{babel}
\usepackage{fancyhdr}
\usepackage{indentfirst}
\usepackage{graphicx, fancyhdr}
\usepackage{amsmath,amssymb,amsfonts}
\usepackage{axodraw4j}

\usepackage{fancyhdr}                          
 \fancypagestyle{plain}{
   \fancyhead{}                               
 }

\begin{document}
\baselineskip 8.75 mm

\begin{titlepage}

\vspace*{-2cm}

\begin{center}
\vspace{-0.5cm}
{\Large \textsf{UNIVERSIDADE DE SÃO PAULO}}\\
\vspace{0.21cm}
{\large \textsf{INSTITUTO DE FÍSICA} \\}

\vspace{2.71cm}
 { \Huge \textbf{Quebra da Simetria de Lorentz na \\}}
\vspace{0.8cm}
 { \Huge \textbf{ Eletrodinâmica Quântica} \\}
\vspace{2.3cm} {\Large \textsf{Denny Mauricio de Oliveira}}
\end{center}
\vspace{1 cm}

\begin{flushright}
\parbox{7.9cm}{Dissertação de Mestrado apresentada  ao Ins\-tituto de Física para a obtenção do título de Mestre em Ciências.}
\end{flushright}

\vspace{5.1 cm}

\noindent \textbf{Banca Examinadora} \vspace{0.15 cm}

\noindent Prof. Dr. Ad\'ilson Jos\'e da Silva     (IFUSP) - Orientador \\
\noindent Prof. Dr. Alex Gomes Dias     (UFABC)     \\
\noindent Prof. Dr. Marcelo Otávio Caminha Gomes     (IFUSP)

\vspace{0.3 cm}

\begin{center}
{ São Paulo \\ 2010}
\end{center}

\end{titlepage}

\thispagestyle{empty}

\vspace*{-0.7cm}

 \hspace{-0.8cm}\includegraphics[scale=0.78]{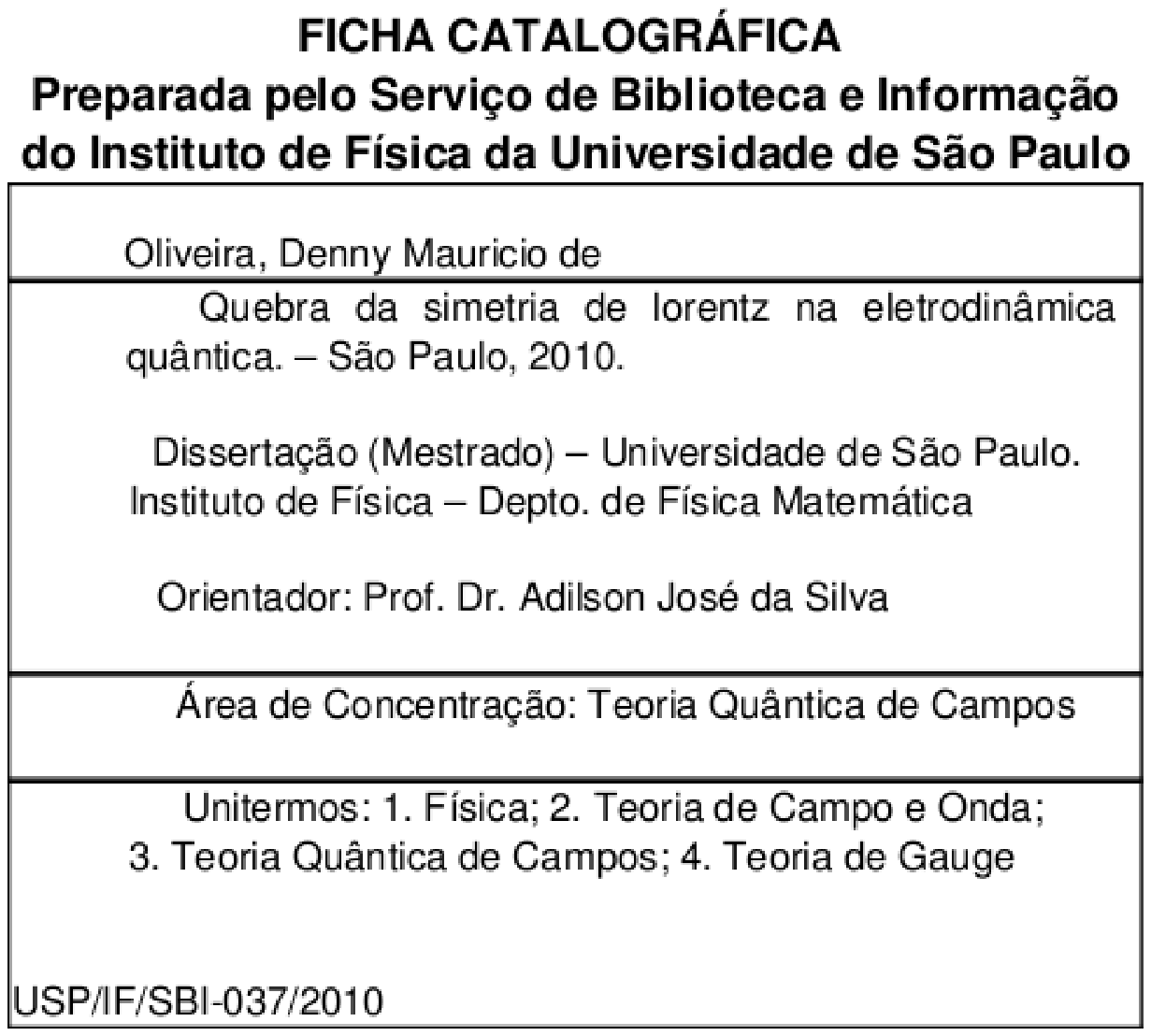}

\thispagestyle{empty}

\newpage
\thispagestyle{empty}

\vspace*{17.0 cm}
\begin{flushright}
 \parbox{5.4in}{
        \begin{center}
         \begin{flushleft}       \textit{\hspace{8eM}Dedico este trabalho à minha querida esposa Patrícia.}
\end{flushleft}
        \end{center}}
\vspace{1.0 cm}
\end{flushright}

\newpage
\thispagestyle{empty}
\frontmatter

\newpage
\thispagestyle{empty}

\vspace*{17.0 cm}
\begin{flushright}
 \parbox{5.4in}{
        \begin{center}
         \begin{flushleft}       \textit{\hspace{8eM}Agrade\c{c}o a Deus pela d\'adiva de continuar vivendo.}
\end{flushleft}
        \end{center}}
\vspace{1.0 cm}
\end{flushright}

\newpage
\thispagestyle{empty}
\frontmatter

\vspace*{7cm}

\hspace{3.6cm}\parbox[c]{8.2cm}{ {\it Debulhar o trigo\\
Recolher cada bago do trigo\\
Forjar no trigo o milagre do pão\\
E se fartar de pão
\vspace{1cm}

Decepar a cana\\
Recolher a garapa da cana\\
Roubar da cana a doçura do mel,\\
Se lambuzar de mel
\vspace{1cm}

Afagar a terra\\
Conhecer os desejos da terra\\
Cio da terra a propícia estação, e fecundar o chão
}}
\vspace{0.7cm}

\hspace{4cm}{\footnotesize Milton Nascimento e Chico Buarque de Hollanda, \textquotedblleft \textit{Cio da terra}``}

\thispagestyle{empty}

\newpage
\thispagestyle{empty}
\frontmatter

\bigskip
\bigskip
\begin{center}
{\Large{\bf Agradecimentos}}
\end{center}
\vspace{2cm}

\begin{quote}
Ao Professor Ad\'ilson Jos\'e da Silva, orientador, pela paciência e oportunidade concedida para a realiza\c{c}\~ao deste trabalho;
\medskip

Ao Professor Marcelo Otávio Caminha Gomes, pelas aulas de Teoria Quân\-tica de Campos, e aos professores do Instituto de Física, em especial aos do Departamento de F\'isica Matem\'atica, pelas li\c{c}\~oes ministradas;
\medskip

 Ao Professor Jos\'e Roberto S. do Nascimento, pela sugest\~ao do tema de estudo desta disserta\c{c}\~ao;
\medskip

Aos colegas do \textquotedblleft grupo de campos\textquotedblright do Departamento de Física Matemática: Fernando, Liner, Juliano, Walney, Carlos Stechhann, Roberto Maluf, Bruno, André, Eduardo, pelas discussões sobre Física, dicas, seminários e aulas assistidas;
\medskip

À minha querida esposa Patrícia, por seu carinho, amor e compreensão nos momentos mais difíceis da minha vida, e por me fazer acreditar que realmente era possível;
\medskip

Às minhas filhas, Isabela e Eduarda, recém-chegada a este mundo, eternas fontes de motivação e inspiração;
\medskip

A todos aqueles que, em um gesto de solidariedade, contribuiram com uma parte de si para salvar minha vida;
\medskip

Ao CNPq (Conselho Nacional de Desenvolvimento Científico e Tecnológico), pelo apoio concedido durante o período de realização deste trabalho.
\end{quote}

\newpage

\hspace{1cm}

\newpage

\centerline{\large \textbf{RESUMO}}

\vspace{3.0cm}

\begin{quote}
Nesta dissertação, estudamos implicações geradas pela quebra da simetria de Lorentz na Eletrodinâmica Quântica. Analisamos férmions interagindo com um campo eletromagnético nos contextos da mecânica quântica e ao efetuar correções radiativas. Na mecânica quântica, os termos de quebra da simetria de Lorentz foram tratados como perturbações à equação de Dirac, e seus valores esperados no vácuo foram obtidos. Nas correções radiativas, a quebra da simetria de Lorentz foi introduzida nessa interação para que o termo tipo Chern-Simons pudesse ser induzido em (3+1) dimensões. Também discutimos as consequências geradas por este termo sobre as velocidades de propagação de fótons clássicos.
\end{quote}

\newpage

\centerline{\large \textbf{ABSTRACT}} \vspace{3.0cm}

\begin{quote}
 In this dissertation, we study the implications generated by the Lorentz breaking symmetry in quantum electrodynamics. We
analyze fermions interacting with an electromagnetic field in the contexts of quantum mechanics and make radiative corrections. In
quantum mechanics, the terms of the Lorentz breaking symmetry were treated as perturbations to the Dirac equation, and their
expected values were obtained in a vacuum. In the radiative corrections, the Lorentz breaking symmetry was introduced in this
interaction for the Chern-Simons like term could be induced in (3 +1) dimensions. We also discussed the consequences generated by this term on the propagation speeds of classic photons.
\end{quote}
\newpage

\renewcommand{\contentsname}{\'Indice}
\tableofcontents
\mainmatter

\chapter{Introdu\c{c}\~ao}

H\'a pouco mais de cem anos, Albert Einstein \cite{Ein} estabeleceu uma teoria em que o papel do observador assume uma certa relev\^ancia na determina\c{c}\~ao dos fen\^omenos f\'isicos e na aplica\c{c}\~ao das pr\'oprias leis da F\'isica, quando as velocidades dos objetos em questão são próximas da velocidade da luz. Esta teoria \'e denominada Relatividade Restrita ou Especial. O ponto de partida para o estudo desta teoria corresponde aos postulados da const\^ancia da velocidade da luz e a validade das leis da Física que, quando aplicadas a referenciais inerciais distintos, devem descrever os fenômenos de maneiras semelhantes. Os fenômenos físicos são então descritos por equações que têm a mesma forma, ou seja, a Relatividade Restrita deve ser uma teoria \emph{covariante}, e a conex\~ao entre sistemas de  referências descritos por equações covariantes \'e feita atrav\'es de transforma\c{c}\~oes de Lorentz, como veremos a seguir.

\smallskip
Como a Relatividade Restrita é uma teoria elaborada em um espaço-tempo quadri\-dimensional, as transforma\c{c}\~oes de Lorentz podem ser definidas, basicamente, em dois tipos: as rota\c{c}\~oes e as impulsões (\emph{boosts}). As primeiras correspondem aos tr\^es tipos b\'asicos de rota\c{c}\~oes em torno dos tr\^es eixos espaciais, enquanto que a segunda \'e determinada por uma mudan\c{c}a de velocidades. H\'a ainda tr\^es tipos de impulsões, cada uma ocorrendo em um dos tr\^es eixos espaciais. Assim, um sistema \'e dito ter simetria de Lorentz se as leis da F\'isica n\~ao s\~ao alteradas por transforma\c{c}\~oes de Lorentz dos tipos descritos anteriormente.
\smallskip

A simetria de Lorentz e as propriedades da Mecânica Quântica correspondem a base para a formulação da Teoria Quântica de Campos, que descreve partículas como excitações locali\-zadas de um campo imerso no espaço-tempo. O desenvolvimento dessas ideias, no século passado, desencadeou a formulação do {\it Modelo Padrão} (MP), que des\-creve todas as interações conhecidas no universo (forças eletromagnética, nuclear forte e nuclear fraca), exceto a gravidade. Um resultado importante, propriedade do MP e obtido pela teoria quântica, é a {\it simetria} CPT. Nesta sigla, C quer dizer conjuga\c{c}\~ao de \emph{carga}, ou seja, esta transforma\c{c}\~ao converte a part\'icula em sua antipart\'icula; P corresponde \`a \emph{paridade}, isto \'e, deve ocorrer uma invers\~ao espacial quando esta transformação é realizada e, finalmente, T muda a dire\c{c}\~ao de fluxo do tempo. Devido a essas defini\c{c}\~oes, um sistema \'e dito ter simetria CPT se as tr\^es transforma\c{c}\~oes juntas n\~ao afetarem a f\'isica do sistema. Isso \'e denominado \emph{teorema} CPT, inicialmente proposto por Schwinger \cite{Sch}. Uma das consequ\^encias deste teorema é o fato de que partículas e suas antipartículas devem ter cargas elétricas de mesmo módulo e sinais opostos, tempos de vida exa\-tamente iguais, assim como massas e momentos magnéticos. As transformações discretas sofridas pelo campo de Dirac na figura de seus covariantes bilineares s\~ao estudadas no Capítulo 2 desta dissertação.
\smallskip

Os f\'isicos te\'oricos atuais esperam que, no limite de altas energias, uma teoria unificada possa descrever a Natureza de maneira simétrica. No entanto, a simetria de Lorentz é violada na tentativa de incorporar a Teoria da Relatividade Geral\footnote{Como é bem conhecido, Einstein não obteve sucesso ao tentar formular uma teoria que engloba o eletromagnetismo e a gravidade pois, nesta última, não existe força de repulsão entre as massas.} ao  MP. O MP com gravidade descreveria sistemas físicos que devem ocorrer em escalas de altas energias, que são da ordem de grandeza da escala Planck, cuja massa \'e $m_{Pl}=\sqrt{\dfrac{\hbar c}{G_N}}\approx10^{19}\,GeV$, sendo $G_N$ a constante universal de Newton. As distâncias típicas nessas circunstâncias de escala seriam da ordem de $10^{-33}\,cm$.
\smallskip

A quebra da simetria de Lorentz surge também em outra área da física teórica contemporânea de altas energias: a Teoria de Cordas. Nesta teoria, a partícula pontual passa a ser um objeto unidimensional. Desta forma, quando se move no espaço-tempo, ao invés de traçar uma linha (linha mundo), esta part\'icula descreve uma superfície denominada \emph{folha de mundo}. Assim, os modos normais de vibração desta superfície de mundo recuperariam as informações de descrição das partículas. Tal ideia de violação da invariância de Lorentz, no contexto da Teoria de Cordas, foi lançada pioneiramente, em 1989, por Kostelecký e Samuel \cite{Sam}. Neste trabalho, os autores argumentam que tal violação pode ser estendida ao MP. Então, na segunda metade dos anos 90, surge o {\it Modelo Padrão Extendido} (MPE).
\smallskip

O MPE é uma teoria que possui todas as propriedades do MP usual - tais como a estrutura de calibre $SU(3)\times SU(2)\times U(1)$ e renormalizabilidade -, e a estens\~ao que permite estabelecer viola\c{c}\~oes de simetrias de Lorentz e CPT. Esta teoria então fornece uma descri\c{c}\~ao quantitativa das viola\c{c}\~oes das simetrias de Lorentz e CPT, controladas por coeficientes cujos valores s\~ao determinados por experimentos espec\'ificos\footnote{Em relação a esses experimentos, na referência \cite{Gib}, é possível encontrar alguns detalhes, tais como a razão $\dfrac{m_K-m_{\bar K}}{m_K}\lesssim10^{-18}$ para um sistema de kaons neutros. A massa do kaon é 0,5 $GeV$.}. Como tais experimentos são sensíveis a testes de transformações CPT, a teoria será baseada em efeitos de baixa energia, sem gravidade. Uma caracter\'istica marcante desta teoria, como se sabe da literatura, \'e que a quebra da simetria CPT implica na quebra da simetria de Lorentz\footnote{Para efeito de linguagem, o termo {\it quebra da simetria de Lorentz} ser\'a utilizado de uma maneira geral, enquanto que o termo {\it quebra da simetria} CPT ser\'a utilizado de forma espec\'ifica.} \cite{Gre}. Esse fato permite que qualquer viola\c{c}\~ao observ\'avel da simetria CPT seja descrita pelo MPE.
\smallskip

Em 1997, \cite{Col} Colladay, Kostelecký e colaboradores propuseram, para o setor de férmions, uma base teórica para estudos de modelos que envolvem violação de simetria CPT. Nesse trabalho, os autores argumentam que resultados obtidos através de c\'alculos te\'oricos, utilizando-se correções em mecânica quântica relativística e teoria de perturbações (usualmente até primeira ordem de aproximação, no regime de baixas energias), devem servir como motiva\c{c}\~ao para a busca, via experimentos, de limites para os campos de quebra de Lorentz na teoria de férmions. Os autores também redefiniram vários elementos importantes para o estudo da extensão da teoria de Dirac, modificados pela nova teoria, como lagrangianas, equações de Dirac, energias, propagadores, espinores de Dirac, etc. Esses elementos serão apresentados no Capítulo 3. No Cap\'itulo 4, a teoria de viola\c{c}\~ao CPT ser\'a aplicada a mode\-los teóricos em mecânica quântica, tais como mudan\c{c}as de energia de el\'etrons e p\'ositrons relativ\'isticos e ao efeito Zeeman an\^omalo. 
\smallskip

Nesta dissertação, estudaremos os possíveis efeitos que a quebra da simetria de Lorentz podem causar em sistemas físicos da EDQ\footnote{Maiores detalhes de outros testes envolvendo quebra da simetria de Lorentz na EDQ são encontrados na referência \cite{Edq}.}. Assim, sistemas compreendidos no acoplamento de férmions com um campo de calibre serão analisados.

\smallskip
No início da década de 80, S. Deser, R. Jackiw e S. Templeton \cite{Des} escreveram a teoria eletromagnética de Maxwell em uma variante planar em (2+1)$D$, que preserva a invariância de Lorentz e transformações de calibre. Este modelo, denominado {\it teoria de Maxwell-Chern-Simons} (MCS), tem aplicabilidade a fen\^omenos planares de mat\'eria condensada, com grande destaque na literatura aos supercondutores e ao efeito Hall qu\^antico fracion\'ario \cite{Dun}. No Capítulo 5 discutimos algumas das propriedades e calculamos o termo induzido de CS proveniente do acoplamento de férmions com um campo de calibre no contexto do espaço-tempo planar.
\smallskip

Embora esta teoria s\'o exista em dimens\~oes \'impares, em 1990, Carroll, Field e Jackiw perceberam, em um trabalho pioneiro \cite{Car}, que \'e poss\'ivel formular uma teoria semelhante em (3+1)$D$, através da adição de
\begin{equation}
 S_{CS}^{(3+1)D}=\frac{1}{2}\int d^4x\,\varepsilon^{\mu\nu\rho\sigma}\eta_\mu A_\nu \partial_\rho A_\sigma
\end{equation}
à ação de Maxwell convencional. Este termo é conhecido na literatura como termo de Carrol-Field-Jackiw. Tal termo é CPT-ímpar\footnote{O MPE tamb\'em admite a adi\c{c}\~ao do termo $-\frac{1}{4}\eta_{\mu\nu\rho\sigma}\,F^{\mu\nu}F^{\rho\sigma}$ \`a lagrangiana de Maxwell. Apesar de violar a simetria de Lorentz, este termo é CPT-par.}. Apesar de transformações de calibre serem preservadas, a simetria de Lorentz é violada, porque \'e necessário acoplar ao termo tipo Chern-Simons (CS) um quadrivetor constante $\eta_\mu$, que gera uma anisotropia do espaço-tempo.
\smallskip

No Capítulo 6 calculamos as correções radiativas, na aproximação de um laço, provenientes do acoplamento axial de férmions com um campo de calibre na presença da quebra da simetria de Lorentz.  Tal acoplamento gera uma indução de um termo semelhante ao de CS \cite{Jac}, como na equação (1.1), na ação da EDQ. Essa indução é ambígua, uma vez que a proporcionalidade entre os campos de matéria e radiação depende exclusivamente do esquema de regularização adotado (tais esquemas podem ser: regularização dimensional, regularização de Pauli-Villars \cite{Alt}, método de {\it cut-off} e método do tempo próprio de Schwinger \cite{Lin}). Assim, calcularemos o termo induzido utilizando o método de regularização dimensional e discutiremos algumas consequências desse resultado sobre as velocidades de propagação de fótons clássicos. 

\smallskip
As lagrangianas (e consequentemente as grandezas delas derivadas) ser\~ao sempre representadas em unidades naturais $\hbar=c=1$ nesta disserta\c{c}\~ao. Quando necessário, a métrica utilizada será dada, em quatro dimensões, por $g_{\mu\nu}=diag(1,-1,-1,-1)$. As matrizes de Dirac, quando contraídas com um quadrivetor constante, obedecem as relações $\not\!c=\gamma^\mu c_\mu=\gamma_\mu c^\mu$, onde $\not\!c^2=c^2$.

\chapter{Simetrias discretas do campo de Dirac}

\section{O campo escalar como motivação}

\'E bem conhecido que a mec\^anica qu\^antica n\~ao relativ\'istica descreve satisfatoriamente os fen\^omenos que envolvem part\'iculas cujas velocidades s\~ao pequenas quando comparadas \`a velocidade da
luz. Este papel \'e bem realizado pela equa\c{c}\~ao de Schr\"odin\-ger:
\begin{equation}
i\frac{\partial\phi}{\partial t}=H\phi\,.
\end{equation}

Historicamente, sabe-se que Schr\"odinger n\~ao encontrou inicialmente esta equa\c{c}\~ao. Seu resultado foi o que hoje conhecemos por {\it equa\c{c}\~ao de Klein-Gordon} (KG)\footnote{Mais informações históricas podem ser obtidas em {\it Scientific American, Gênios da Ciência - Quânticos: Os Homens que Mudaram a Física, edição especial (2005)}.}. Schr\"o\-dinger n\~ao a publicou porque n\~ao conseguiu aplic\'a-la aos el\'etrons no átomo de hidrogênio, pois se trata de uma equa\c{c}\~ao relativ\'istica.  A equa\c{c}\~ao de KG \'e obtida atrav\'es da rela\c{c}\~ao entre massa, momento e energia relativ\'isticos:
\begin{equation}
p^2=p^\mu
p_\mu=E^2-|{\mbox{\boldmath{$p$}}}|^2=m^2\hspace{1eM}\rightarrow\hspace{1eM}E^2=|{\mbox{\boldmath{$p$}}}|^2+m^2\,,
\end{equation}
e tamb\'em atrav\'es das substitui\c{c}\~oes
$\displaystyle{E\rightarrow i\frac{\partial}{\partial t}}$ \,e\,
$\displaystyle{{\mbox{\boldmath{$p$}}}\rightarrow
-i{\mbox{\boldmath{$\nabla$}}}}$:
\begin{equation}
(\partial^\mu\partial_\mu+m^2)\phi\equiv(\square+m^2)\phi=0\,.
\end{equation}
onde $\square=\partial^2/\partial t^2-\nabla^2$ \'e o operador de
D'Alembert.
\smallskip

Em linguagem contempor\^anea, a equa\c{c}\~ao de Schr\"odinger (2.1) \'e interpretada como o limite n\~ao relativ\'istico da equa\c{c}\~ao de KG.
\smallskip

As solu\c{c}\~oes da equa\c{c}\~ao de KG s\~ao do tipo ondas planas
\begin{equation}
\phi({\mbox{\boldmath{$p$}}},t)\sim e^{-iEt+i{\bf p}\cdot{\bf x}}\,,
\end{equation}
onde, de acordo com (2.2), $\displaystyle{E=\pm\sqrt{|{\mbox{\boldmath{$p$}}}|^2+m^2}}$.
\smallskip

A equa\c{c}\~ao de KG apresenta dois problemas aparentes: o espectro de energia admite \emph{valores negativos} e a
probabilidade n\~ao \'e \emph{positiva definida}, pois a equa\c{c}\~ao de KG \'e de segunda ordem em
$\partial\phi/\partial t$. A mec\^anica qu\^antica n\~ao relativ\'istica fornece a probabilidade e a corrente de
probabilidade:
\begin{eqnarray}
\rho&=&\phi^\ast\phi\\
{\mbox{\boldmath{$J$}}}&=&-\frac{i}{2m}\,\left(\phi^\ast{\mbox{\boldmath{$\nabla$}}}\phi-\phi{\mbox{\boldmath{$\nabla$}}}\phi^\ast\right)\,,
\end{eqnarray}
que satisfazem a equa\c{c}\~ao da continuidade:
\begin{equation}
\frac{\partial\rho}{\partial
t}+{\mbox{\boldmath{$\nabla$}}}\cdot{\mbox{\boldmath{$J$}}}=0\,,
\end{equation}
que ainda pode ser escrita na forma quadrivetorial $\partial_\mu J^\mu=0$, com o quadrivetor corrente dado por
$J^\mu=(\rho,{\mbox{\boldmath{$J$}}})$.
\smallskip

No caso relativ\'istico, a equa\c{c}\~ao de KG exige que a corrente de probabilidade seja dada por\footnote{Esta nota\c{c}\~ao abreviada significa $A\stackrel{\leftrightarrow}{\partial^\mu} B=A\partial^\mu
B-(\partial^\mu A)B$.} (onde $\phi$ e $\phi^\ast$ satisfazem a equa\c{c}\~ao de KG):
\begin{equation}
J^\mu=i\phi^\ast\stackrel{\leftrightarrow}{\partial^\mu}\phi\,\,,\hspace{1eM}\mbox{ou}\hspace{1eM}J^\mu=\left(i\phi^\ast\stackrel{\leftrightarrow}{\partial_0}\phi,-i\phi^\ast\stackrel{\leftrightarrow}{{\mbox{\boldmath{$\nabla$}}}}\phi\right)\,.
\end{equation}

A solu\c{c}\~ao para os dois problemas aparentes \'e reinterpretar $\phi$, elevando seu \emph{status} de {\it fun\c{c}\~ao de onda de uma part\'icula} para um {\it operador que descreve campos}. Este operador de campo deve ser quantizado, utilizando-se regras espec\'ificas. A outra solu\c{c}\~ao \'e admitir que as energias positivas e negativas representam, respectivamente, as \emph{part\'iculas} e \emph{antipart\'iculas}. Esta \'e uma das principais caracter\'isticas da teoria de Dirac para o tratamento relativ\'istico do el\'etron, como veremos mais adiante. Definiremos agora alguns elementos de quantiza\c{c}\~ao can\^onica do campo bos\^onico de KG, que ser\~ao \'uteis quando as quantiza\c{c}ões do campo fermi\^onico de Dirac forem  realizadas.
\smallskip

O campo escalar de KG pode ser escrito como uma transformada de Fourier da seguinte maneira:
\begin{eqnarray}
\phi(x)=\int\frac{d^4p}{(2\pi)^4}\,e^{-ip\cdot x}\phi(p)\,.
\end{eqnarray}

Este campo é solução da equação de KG. Então, podemos decompô-lo em modos de Fourier,
\begin{eqnarray}
\phi(p)=f({\bf p})\delta(p_0-\omega_p)+g({\bf
p})\delta(p_0+\omega_p)\hspace{2eM},\hspace{2eM}p_0=\pm\omega_p\,,
\end{eqnarray}
onde $f({\bf p})$ e $g({\bf p})$ são operadores que serão determinados mais adiante. Assim,
\begin{eqnarray}
\phi(x)&=&\int\frac{d^3p}{(2\pi)^4}dp_0[f({\bf
p})\delta(p_0-\omega_p)+g({\bf
p})\delta(p_0+\omega_p)]e^{-ip_0x_0+i{\bf p}\cdot{\bf x}}\nonumber\\
&&\nonumber\\
&=&\int\frac{d^3p}{(2\pi)^4}[f({\bf p})e^{-i\omega_px_0+i{\bf p}\cdot{\bf
x}}+g(-{\bf p})e^{i\omega_px_0-i{\bf p}\cdot{\bf x}}]\nonumber\\
&&\nonumber\\
&=&\int\frac{d^3p}{(2\pi)^4}[f({\bf p})e^{-ip\cdot x}+g(-{\bf
p})e^{ip\cdot x}]\bigg|_{p_0=\omega_p}\,.
\end{eqnarray}

Uma escolha conveniente é
\begin{equation}
\frac{f({\bf p})}{(2\pi)^4}=\frac{1}{(2\pi)^{3/2}}\frac{a({\bf
p})}{\sqrt{2\omega_p}}\hspace{1.5eM}\mbox{e}\hspace{1.5eM}\frac{g({-\bf
p})}{(2\pi)^4}=\frac{1}{(2\pi)^{3/2}}\frac{b^\dagger({\bf
p})}{\sqrt{2\omega_p}}\,,
\end{equation}
e o campo (2.9) fica
\begin{equation}
\phi(x)=\frac{1}{(2\pi)^{3/2}}\int\frac{d^3p}{\sqrt{2\omega_p}}[a({\bf
p})e^{-ip\cdot x}+b^\dagger({\bf p})e^{ip\cdot x}]\,.
\end{equation}

Os operadores $a^\dagger({\bf p})$ e $a({\bf p})$ são operadores que criam e aniquilam partículas, res\-pectivamente. Já os operadores $b^\dagger({\bf p})$ e $b({\bf p})$ são os mesmos operadores para as antipartículas.

Se $a({\bf p})=b({\bf p})$, o campo é real, e sua lagrangiana\footnote{Como tratamos de uma teoria em que os campos são locais, é conveniente definir uma função ${\cal L}$, denominada densidade de lagrangiana, onde $L=\int d^3x{\cal L}(\phi,\partial_\mu\phi)$\,. Portanto, a terminologia {\it lagrangiana} se refere simplesmente à {\it densidade de lagrangiana}.} é dada por
\begin{equation}
{\cal L}=\frac{1}{2}(\partial_\mu\phi)^2-\frac{m^2}{2}\phi^2\,.
\end{equation}

Definindo o momento $\pi({\bf x},t)$ canonicamente conjugado ao campo $\phi({\bf x},t)$ como
\begin{equation}
\pi({\bf x},t)=\frac{\partial{\cal L}}{\partial(\partial_0\phi({\bf
x},t))}=\dot{\phi}({\bf x},t)\,,
\end{equation}
e adotando a regra de comutação em tempos iguais,
\begin{equation}
[\phi(x),\pi(x')]=i\delta^{(3)}({\bf x}-{\bf x}')\,,
\end{equation}
temos que
\begin{equation}
[a({\bf p}),a^\dagger({\bf p}')]=\delta^{(3)}({\bf p}-{\bf p}')\,.
\end{equation}

Considere o campo escalar real. Sabemos que o operador $a({\bf p})$ aniquila o vácuo, ou seja, $a({\bf p})|0\rangle=0$. Então, podemos fazer a seguinte substituição
\begin{equation}
\langle0|a({\bf p})a^\dagger({\bf
p})|0\rangle\,\,\to\,\,\langle0|[a({\bf p}),a^\dagger({\bf
p})]|0\rangle\,.
\end{equation}

Assim, obtemos a seguinte relação:
\begin{eqnarray}
\langle0|[\phi(x),\phi(y)]|0\rangle&=&\!\int\!\frac{d^3p\,d^3p'}{(2\pi)^3}\frac{1}{\sqrt{2\omega_p\,2\omega_{p'}}}\,\langle0|[a_p\,e^{-ip\cdot x}+a^\dagger_p\,e^{ip\cdot x},\,a_{p'}\,e^{ip'\cdot y}+a^\dagger_{p'}\,e^{-ip'\cdot y}]|0\rangle\nonumber\\
&=&\int\!\frac{d^3p}{(2\pi)^3}\left\{\frac{1}{2\omega_p}\,e^{-ip\cdot(x-y)}\Bigg|_{p^0=+\omega_p}-\frac{1}{2\omega_p}\,e^{-ip\cdot(x-y)}\Bigg|_{p^0=-\omega_p}\right\}\nonumber\\
&=&\int\!\frac{d^3p}{(2\pi)^3}\!\int\frac{dp^0}{2\pi i}\,\frac{-1}{p^2-m^2}\,e^{-ip\cdot(x-y)}\nonumber\\
&=&\int\!\frac{d^4p}{(2\pi)^4}\,\frac{i}{p^2-m^2}\,e^{-ip\cdot(x-y)}\hspace{2eM}\mbox{para}\hspace{1eM}x^0>y^0\,.
\end{eqnarray}

Definimos, então, o {\it propagador de Feynman} para o campo bosônico
\begin{equation}
\Delta_F(x-y)=\int\!\frac{d^4p}{(2\pi)^4}\,\frac{i}{p^2-m^2+i\epsilon}\,e^{-ip\cdot(x-y)}\,.
\end{equation}

Quando aplicamos a equação de KG a este propagador, obtemos a seguinte identidade:
\begin{equation}
(\square+m^2)\Delta_F(x-y)=-i\delta^{(4)}(x-y)\,.
\end{equation}

\section{A equa\c{c}\~ao de Dirac}

Motivado pelo desejo de encontrar uma equa\c{c}\~ao de primeira ordem relativisticamente covariante, Dirac escreveu uma  equa\c{c}\~ao linear na derivada temporal e no momento. Com $H_0=\alpha_1p_1+\alpha_2p_2+\alpha_3p_3+\beta m$ representando o \emph{hamiltoniano livre de Dirac}, a equa\c{c}\~ao (2.1) toma a forma:
\begin{equation}
i\frac{\partial\psi}{\partial
t}=\left[-i{\mbox{\boldmath{$\alpha$}}}\cdot{\mbox{\boldmath{$\nabla$}}}+\beta
m\right]\psi\,,
\end{equation}
onde ${\mbox{\boldmath{$\alpha$}}}$ e $\beta$ s\~ao matrizes $n\times n$, que devem ser determinadas, e $\psi$ uma matriz coluna de ordem $n$. Para descobrir esta rela\c{c}\~ao, aplicamos a
derivada temporal em ambos os lados de (2.22) e encontramos:
\begin{equation}
-\frac{\partial^2\psi}{\partial
t^2}=\left[-\alpha^i\alpha^j\nabla^i\nabla^j-im(\beta\alpha^i+\alpha^i\beta)\nabla^i+\beta^2M^2\right]\psi\,.
\end{equation}

Comparando com a equa\c{c}\~ao de KG, vemos claramente que ${\mbox{\boldmath{$\alpha$}}}$ e $\beta$ devem
satisfazer as condi\c{c}\~oes:
\begin{eqnarray}
\alpha^i\alpha^j+\alpha^j\alpha^i=2\delta^{ij}\,,&&\nonumber\\
\beta\alpha^i+\alpha^i\beta=0\,,&&\\
\beta^2=1\,.&&\nonumber
\end{eqnarray}

As rela\c{c}\~oes (2.24) nos indicam, de fato, que ${\mbox{\boldmath{$\alpha$}}}$ e $\beta$ \emph{n\~ao s\~ao}
n\'umeros ordin\'arios, ou seja, s\~ao \emph{matrizes}. No entanto, como sabemos, os operadores que representam fen\^omenos f\'isicos devem ser dados por matrizes quadradas\footnote{Esta consequ\^encia caracteriza uma \emph{transforma\c{c}\~ao linear} em seu caso particular, a \emph{opera\c{c}\~ao linear}.}. A combina\c{c}\~ao linear mais simples, a $2\times2$, n\~ao \'e poss\'ivel, pois as matrizes ${\mbox{\boldmath{$\alpha$}}}$ e $\beta$ 2$\times$2 {\it não existem}. A pr\'oxima op\c{c}\~ao mais simples \'e a de dimens\~ao $4\times4$ que, na \emph{representa\c{c}\~ao de Dirac} ou representa\c{c}\~ao padr\~ao\footnote{Veja Ap\^endice A.}, fornece:
\begin{equation}
{\mbox{\boldmath{$\alpha$}}}=\left(\begin{array}{cc}0&{\mbox{\boldmath{$
\sigma$}}}\\{\mbox{\boldmath{$
\sigma$}}}&0\end{array}\right)\hspace{2eM}\mbox{e}\hspace{2eM}\beta=\left(\begin{array}{cc}1&0\\0&-1\end{array}\right)\,.
\end{equation}

Esta escolha \'e feita com base em um teorema devido a Pauli. Isto implica que as matrizes ${\mbox{\boldmath{$\alpha$}}}$ e $\beta$ s\~ao hermitianas, e as propriedades (2.24) tamb\'em se aplicam a
${\mbox{\boldmath{$\alpha$}}}^\dagger$ e $\beta^\dagger$. Portanto, o hamiltoniano de Dirac em (2.22) \'e hermitiano. Para escrevermos a equa\c{c}\~ao de Dirac em nota\c{c}\~ao moderna, devemos considerar
${\mbox{\boldmath{$\alpha$}}}=\gamma^0{\mbox{\boldmath{$\gamma$}}}$ e $\beta=\gamma^0$, e então temos que:
\begin{equation}
(i\gamma^\mu\partial_\mu-m)\psi=0\,.
\end{equation}

De posse da equa\c{c}\~ao de Dirac acima, calcularemos agora suas autoenergias e suas solu\c{c}\~oes do tipo ondas-planas na forma $\psi=N\,u(p)\,e^{-ip\cdot x}$, cuja constante de normali\-za\c{c}\~ao ser\'a calculada mais adiante. Considere a equa\c{c}\~ao de autovalores $H_0\,u^{(s)}({\mbox{\boldmath{$p$}}})=E\,u^{(s)}({\mbox{\boldmath{$p$}}})$,
com o hamiltoniano de Dirac $H_0$ e as autoenergias (2.2), al\'em da representa\c{c}\~ao padr\~ao para ${\mbox{\boldmath{$\alpha$}}}$ e $\beta$:
\begin{equation}
\left(\begin{array}{cc}m&{\mbox{\boldmath{$\sigma$}}}\cdot{\mbox{\boldmath{$p$}}}\\{\mbox{\boldmath{$\sigma$}}}\cdot{\mbox{\boldmath{$p$}}}&-m\end{array}\right)\left(\begin{array}{c}\chi_s\\\eta_s\end{array}\right)=E\left(\begin{array}{c}\chi_s\\\eta_s\end{array}\right)\,,
\end{equation}
onde $s=1,2$ rotulam os spins para cima e para baixo, respectivamente,
$\chi_1=\left(\begin{array}{c}1\\0\end{array}\right)$,
$\chi_2=\left(\begin{array}{c}0\\1\end{array}\right)$ configuram uma
base ortonormal, e $\eta_s$ deve ser determinado por (2.27).
\smallskip

As solu\c{c}\~oes com energia positiva, para as part\'iculas com
$E=+|\sqrt{|{\mbox{\boldmath{$p$}}}|^2+m^2}|$, s\~ao dadas por:
\begin{equation}
u^{(s)}(p)=N\,\left(\begin{array}{c}\chi_s\\\frac{{\mbox{\boldmath{$\sigma$}}}\cdot{\mbox{\boldmath{$p$}}}}{{\displaystyle{E+m}}}\,\chi_s\end{array}\right)\,.
\end{equation}

Admitindo a existência de antipartículas na teoria de Dirac, as mesmas devem ter soluções com energias negativas, ou seja, $E=-|\sqrt{|{\mbox{\boldmath{$p$}}}|^2+m^2}|$. Assim, as antipart\'iculas devem ser pensadas como part\'iculas com energias $E\,\rightarrow\,-E$ e momentos ${\mbox{\boldmath{$p$}}}\,\rightarrow\,-{\mbox{\boldmath{$p$}}}$. 
Sendo a equa\c{c}\~ao de autovalores agora dada por
$H_0\,\upsilon^{(s)}({\mbox{\boldmath{$p$}}})=-E\,\upsilon^{(s)}({\mbox{\boldmath{$p$}}})$, obtemos a solu\c{c}\~ao\footnote{O espinor escolhido para os estados de energias negativas é $\chi_{-s}=-i\sigma^2\chi^\ast_{s}$ para haver concordância com a transformação por conjugação de carga que
será feita mais adiante.}
\begin{equation}
\upsilon^{(s)}(p)=N'\,\left(\begin{array}{c}\frac{{\mbox{\boldmath{$\sigma$}}}\cdot{\mbox{\boldmath{$p$}}}}{{\displaystyle{E+m}}}\,\chi_{-s}\\\chi_{-s}\end{array}\right)\,.
\end{equation}

A normaliza\c{c}\~ao do espinor (2.28), dentre outras poss\'iveis, com $\bar{u}=u^\dagger\gamma^0$, é:
\begin{eqnarray}
&&\bar{u}^{(r)}(p)u^{(s)}(p)=\delta^{rs}\,,\\
&&\bar{\upsilon}^{(r)}(p)\upsilon^{(s)}(p)=-\delta^{rs}\,.
\end{eqnarray}
de onde segue que $\displaystyle{N=N'=\sqrt{\frac{E+m}{2m}}}$.
\smallskip

\'E conveniente ainda escrevermos as seguintes rela\c{c}\~oes entre os espinores obtidos na teoria de Dirac:
\begin{eqnarray}
u^{(r)\dagger}(p)u^{(s)}(p)&=&\frac{E}{m}\,\delta^{rs}\,,\nonumber\\
&&\nonumber\\
\upsilon^{(r)\dagger}(p)\upsilon^{(s)}(p)&=&\frac{E}{m}\,\delta^{rs}\,,\\
&&\nonumber\\
u^{(r)\dagger}(p)\upsilon^{(s)}(p)&=&\upsilon^{(r)\dagger}(p)u^{(s)}(p)=0\nonumber\,.
\end{eqnarray}

Portanto, a equa\c{c}\~ao de Dirac no espa\c{c}o dos momentos e seus espinores normalizados para as part\'iculas e antipart\'iculas s\~ao, respectivamente, dados por:

\begin{eqnarray}
(\not\!p-m)u^{(s)}(p)&=&0\hspace{1.5eM}\rightarrow\hspace{1.5eM}u^{(s)}(p)=\sqrt{\frac{E+m}{2m}}\,\left(\begin{array}{c}\chi_s\\\frac{{\mbox{\boldmath{$\sigma$}}}\cdot{\mbox{\boldmath{$p$}}}}{{\displaystyle{E+m}}}\,\chi_s\end{array}\right)\,,\\
&&\nonumber\\
(\not\!p+m)\upsilon^{(s)}(p)&=&0\hspace{1.5eM}\rightarrow\hspace{1.5eM}\upsilon^{(s)}(p)=\sqrt{\frac{E+m}{2m}}\,\left(\begin{array}{c}\frac{{\mbox{\boldmath{$\sigma$}}}\cdot{\mbox{\boldmath{$p$}}}}{{\displaystyle{E+m}}}\,\chi_{-s}\\\chi_{-s}\end{array}\right)\,.
\end{eqnarray}

Como vimos nas se\c{c}\~oes anteriores, as solu\c{c}\~oes com energias negativas para as equa\c{c}\~oes de onda relativ\'isticas causaram um grande desconforto no in\'icio da cons\-tru\c{c}\~ao da teoria. No entanto, Dirac {\it reinterpretou} as soluções de ener\-gias negativas para o problema. Dirac prop\^os um estado de v\'acuo, denominado \emph{mar de Dirac}, com todos os estados de energias negativas preenchidos. Um el\'etron no mar de Dirac com energia nega\-tiva pode absorver radia\c{c}\~ao e ser transportado para um estado com energia positiva. O buraco (\emph{hole}) deixado pelo el\'etron \'e reinterpretado como uma aus\^encia de carga $+|e|$ e energia $-|E|$. Esta \textquotedblleft aus\^encia\textquotedblright \'e ent\~ao considerada uma evidência da exist\^encia de uma \emph{antipart\'icula} com carga $+|e|$ e energia $-|E|$. Ent\~ao, a teoria de Dirac prev\^e a exist\^encia do \emph{p\'ositron}, antipart\'icula do el\'etron, descoberto experimentalmente em 1932 por Carl Anderson.
\vspace{-0.9cm}

\begin{center}
\begin{picture}(115,160)
\LongArrow(30,0)(30,100)\Text(5,90)[]{$Energia$}
\LongArrow(90,45)(120,45)\Text(170,45)[]{Região Proibida}
\GBox(30,30)(100,60){0.8}\Text(18,62)[]{$Mc^2$}\Text(18,28)[]{-$Mc^2$}
\GCirc(55,17){4}{1.0}\Text(90,13)[]{Buraco}
\GCirc(55,79){4}{0.0}\Text(90,82)[]{Elétron}
\ArrowArc(56,47)(30,280,80)
\end{picture}\\ \vspace{0.5cm} {\sl Representação pictórica do mar de Dirac.}
\end{center}

\section{O propagador do férmion}

A equa\c{c}\~ao de Dirac admite uma solu\c{c}\~ao arbitr\'aria, que engloba as energias positivas e negativas, que \'e dada por:
\begin{equation}
\psi(x)=\frac{1}{(2\pi)^{3/2}}\sum\limits_s\int\frac{d^3p}{E/m}\left[c^{(s)}(p)u^{(s)}(p)e^{-ip\cdot
x}+d^{(s)\dagger}(p)\upsilon^{(s)}(p)e^{ip\cdot x}\right]\,,
\end{equation}
e seu operador adjunto
\begin{equation}
\bar{\psi}(x)=\frac{1}{(2\pi)^{3/2}}\sum\limits_s\int\frac{d^3p}{E/m}\left[c^{(s)\dagger}(p)\bar{u}^{(s)}(p)e^{ip\cdot
x}+d^{(s)}(p)\bar{\upsilon}^{(s)}(p)e^{-ip\cdot x}\right]\,,
\end{equation}
onde $c^{(s)}$ e $c^{(s)\dagger}$ são operadores de aniquilação e criação de partículas, enquanto que $d^{(s)}$ e $d^{(s)\dagger}$ são s mesmos operadores para as antipartículas. Como os férmions obedecem a estatística de Fermi-Dirac, estes operadores satisfazem a seguinte regra de anticomutação
\begin{equation}
\{c^{(s)}(p),c^{(r)\dagger}(p')\}=\{d^{(s)}(p),d^{(r)\dagger}(p')\}=\frac{E}{m}\delta^{rs}\,\delta({\bf
p}-{\bf p}')\,.
\end{equation}

Então,
\begin{eqnarray}
\langle0|\psi(x)\bar{\psi}(y)|0\rangle&=&\int\!\frac{d^3p}{(2\pi)^3}\,\frac{m}{E}\sum\limits_su^{(s)}(p)\bar{u}^{(s)}(p)\,e^{-ip\cdot(x-y)}\nonumber\\
&&\nonumber\\
&=&\int\!\frac{d^3p}{(2\pi)^3}(i\!\not\!\partial_x+m)\frac{1}{2E}\,e^{-ip\cdot(x-y)}\\
&&\nonumber\\
\langle0|\bar{\psi}(x)\psi(y)|0\rangle&=&\int\!\frac{d^3p}{(2\pi)^3}\,\frac{m}{E}\sum\limits_s\upsilon^{(s)}(p)\bar{\upsilon}^{(s)}(p)\,e^{-ip\cdot(y-x)}\nonumber\\
&&\nonumber\\
&=&\int\!\frac{d^3p}{(2\pi)^3}(-i\!\not\!\partial_x+m)\frac{1}{2E}\,e^{-ip\cdot(y-x)}\,.
\end{eqnarray}

Da mesma forma que foi feito com o campo de KG, obteremos, para o campo de Dirac, a função de Green retardada:
\begin{equation}
S_R(x-y)\equiv\theta(x^0-y^0)\langle0|\{\psi(x),\bar{\psi}(y)\}|0\rangle\,,
\end{equation}
que satisfaz
\begin{equation}
S(x-y)=(i\!\not\!\partial_x+m)\Delta(x-y)\,,
\end{equation}
com $\Delta(x-y)$ sendo o propagador para os bósons. Aplicando o operador $-i\!\not\!\partial_x+m$ na expressão acima em ambos os lados pela esquerda, segue que
\begin{eqnarray}
(-i\!\not\!\partial_x+m)S(x-y)&=&(\partial^\mu\partial_\mu+m^2)\Delta(x-y)\nonumber\\
&=&-i\delta^{(4)}(x-y)\,,
\end{eqnarray}
onde foi utilizada a express\~ao (2.21). 
\smallskip

Portanto, no espaço dos momentos,
\begin{eqnarray*}
i\delta^{(4)}(x-y)=\int\frac{d^4p}{(2\pi)^4}\,i\,e^{-ip\cdot(x-y)}=\int\frac{d^4p}{(2\pi)^4}(\not\!p-m)e^{-ip\cdot(x-y)}S(p)\,,
\end{eqnarray*}
ou seja,
\begin{equation}
S(p)=\frac{i}{\not\!p-m}\,.
\end{equation}

Utilizando os mesmos argumentos do campo de KG, o propagador de Feynman para o campo fermiônico é definido como se
segue:
\begin{eqnarray}
S(x-y)&\equiv&\langle0|T\psi(x)\bar{\psi}(y)|0\rangle\nonumber\\
&&\nonumber\\
&=&i\int\frac{d^4p}{(2\pi)^4}\,\frac{\not\!p+m}{p^2-m^2+i\epsilon}\,e^{-ip\cdot(x-y)}\,.
\end{eqnarray}

\section{Simetrias discretas}

A teoria de Dirac apresenta, além das transformações de Lorentz, outras duas importantes simetrias tipo espaço-tempo: a transformação por {\it paridade} e a {\it inversão temporal}. A primeira, de caráter espacial, troca o sinal da parte espacial de um quadrivetor qualquer: $x^\mu=(x^0,\mathbf{x})\,\to\,\tilde{x}_\mu=(x^0,-\mathbf{x})$ e o
mesmo vale para o momento $\tilde{p}_\mu=(p_0,-{\bf p})$. A segunda, estritamente temporal, inverte o fluxo do tempo no cone de luz: $-\tilde{x}_\mu=(-p^0,\mathbf{x})$ e $-p_\mu=(-x^0,\mathbf{p})$. No entanto, apesar dessas trocas, a norma no espaço-tempo de Minkowiski se conserva.
\smallskip

Como foi visto nas seções anteriores, a teoria quântica tem como uma de suas principais virtudes a capacidade de explicar a existência de partículas e antipart\'iculas na Natureza. Este fato acarreta em outra simetria discreta da teoria: a conversão da partícula em sua antipartícula e vice-versa. Esta simetria é denominada transformação por {\it conjugação de carga}. Acredita-se que a Natureza procura perservar estas três simetrias combinadas, denominada {\it simetria} CPT. No entanto, existem alguns processos que podem violar esta simetria. Antes de discutí-los, observaremos a atuação destas três simetrias sobre os
campos e partículas de Dirac através dos covariantes bilineares.

\subsection{Paridade}

Considerando os operadores de campo de Dirac (2.35) e (2.36), procuramos por um operador unitário que troca o sinal da parte espacial dos momentos nos operadores de criação e aniquilação de partículas e antipartículas, sem \emph{alterar seus estados de spin}:
\begin{eqnarray}
{\cal P}c^{(s)}(p){\cal P}^{-1}=\alpha_p\,c^{(s)}(\tilde{p})\hspace{1.5eM}&,&\hspace{1.5eM}{\cal P}d^{(s)}(p){\cal P}^{-1}=\beta_p\,d^{(s)}(\tilde{p})\,,\nonumber\\
&&\\
{\cal P}c^{(s)\dagger}(p){\cal
P}^{-1}=\alpha_p^\ast\,c^{(s)\dagger}(\tilde{p})\hspace{1.5eM}&,&\hspace{1.5eM}{\cal
P}d^{(s)\dagger}(p){\cal
P}^{-1}=\beta_p^\ast\,d^{(s)\dagger}(\tilde{p})\,,\nonumber
\end{eqnarray}
onde $\alpha_p$ e $\beta_p$ são as fases dessa transformação. Sendo o produto escalar conservado, ou seja, $p\cdot x=\tilde{p}\cdot\tilde{x}$, com $u^{(s)}(p)=\gamma^0u^{(s)}(\tilde{p})$ e $\upsilon^{(s)}(p)=-\gamma^0\upsilon^{(s)}(\tilde{p})$,  a transformação que procuramos é:

\begin{eqnarray}
{\cal P}\psi(x){\cal P}^{-1}&=&\frac{1}{(2\pi)^{3/2}}\!\int\!\!\frac{d^3p}{E/m}\sum_s\left[\alpha_p\,c^{(s)}(\tilde{p})u^{(s)}(p)e^{-i\tilde{p}\cdot\tilde{x}}+\beta^\ast_p\,d^{(s)\dagger}(\tilde{p})\upsilon^{(s)}(p)e^{i\tilde{p}\cdot\tilde{x}}\right]\nonumber\\
&&\nonumber\\
&=&\frac{1}{(2\pi)^{3/2}}\!\int\!\!\frac{d^3\tilde{p}}{E/m}\left[\alpha_p\,c^{(s)}(\tilde{p})\gamma^0u^{(s)}(\tilde{p})e^{-i\tilde{p}\cdot\tilde{x}}-\beta^\ast_p\,d^{(s)\dagger}(\tilde{p})\gamma^0\upsilon^{(s)}(\tilde{p})e^{i\tilde{p}\cdot\tilde{x}}\right]\,,\nonumber\\
\end{eqnarray}

Adotando-se $\alpha_p=-\beta^\ast_p$, temos que
\begin{equation}
{\cal P}\psi(x){\cal
P}^{-1}\equiv\psi^p(x)=\alpha_p\,\gamma^0\psi(\tilde{x})\,,
\end{equation}
e, analogamente para o campo adjunto,
\begin{equation}
{\cal P}\bar{\psi}(x){\cal
P}^{-1}\equiv\bar{\psi}^p(x)=\alpha^\ast_p\,\bar{\psi}(\tilde{x})\gamma^0\,.
\end{equation}

\subsection{Inversão temporal}

A inversão do fluxo do tempo inverte o spin das partículas e antipartículas. Assim, a ação do operador de inversão temporal ${\cal T}$, anti-unitário,  sobre os operadores de criação e aniquilação de partículas e antipartículas fornece
\begin{eqnarray}
{\cal T}c^{(s)}(p){\cal T}^{-1}=\alpha_t\,c^{(-s)}(\tilde{p})\hspace{1.5eM}&,&\hspace{1.5eM}{\cal T}d^{(s)}(p){\cal T}^{-1}=\beta_t\,d^{(-s)}(\tilde{p})\,,\nonumber\\
&&\\
{\cal T}c^{(s)\dagger}(p){\cal
T}^{-1}=\alpha_t^\ast\,c^{(-s)\dagger}(\tilde{p})\hspace{1.5eM}&,&\hspace{1.5eM}{\cal
T}d^{(s)\dagger}(p){\cal
T}^{-1}=\beta_t^\ast\,d^{(-s)\dagger}(\tilde{p})\,.\nonumber
\end{eqnarray}

Os espinores $\chi_s$, com $s=1,2$ foram definidos na seção 2.2.1.: $\chi_1=\left(\begin{array}{c}1\\0\end{array}\right)$,
$\chi_2=\left(\begin{array}{c}0\\1\end{array}\right)$. Assim,

\begin{eqnarray}
{\cal T}\psi(x){\cal T}^{-1}&=&\frac{1}{(2\pi)^{3/2}}\!\int\!\!\frac{d^3p}{E/m}\sum_{-s}\left[\alpha_t\,c^{(-s)}(\tilde{p})u^{(s)}(p)e^{-ip\cdot x}+\beta^\ast_t\,d^{(-s)\dagger}(\tilde{p})\upsilon^{(s)}(p)e^{ip\cdot x}\right]\nonumber\\
&&\nonumber\\
&=&\frac{1}{(2\pi)^{3/2}}\!\int\!\!\frac{d^3p}{E/m}\sum_{-s}\left[\alpha_t\,c^{(-s)}(\tilde{p})u^{(s)\ast}(p)e^{i\tilde{p}\cdot x}+\beta^\ast_t\,d^{(-s)\dagger}(\tilde{p})\upsilon^{(s)\ast}(p)e^{-i\tilde{p}\cdot x}\right]\,.\nonumber\\
\end{eqnarray}

Como o operador inversão temporal gira o spin, é necessária a operação que executa este papel:
\begin{equation}
\chi_{-s}=-i\sigma^2\,\chi_s^\ast\hspace{1.5eM}\Rightarrow\hspace{1.5eM}\chi^\ast_{s}=i\sigma^2\,\chi_{-s}\,.
\end{equation}

De acordo com a identidade $\sigma^2{\mbox{\boldmath{$\sigma$}}}^\ast=-{\mbox{\boldmath{$\sigma$}}}\sigma^2$,
s\~ao convenientes as relaç\~oes:
\begin{eqnarray}
u^{(s)\ast}(p)&=&\gamma^1\gamma^3\,u^{(-s)}(\tilde{p})\,,\nonumber\\
&&\\
\upsilon^{(s)\ast}(p)&=&\gamma^1\gamma^3\,\upsilon^{(-s)}(\tilde{p})\,.\nonumber
\end{eqnarray}

Assim, voltando à equação (2.50), utilizando (2.52), segue que
\begin{eqnarray}
{\cal T}\psi(x){\cal T}^{-1}&=&\frac{1}{(2\pi)^{3/2}}\!\int\!\!\frac{d^3p}{E/m}\,\alpha_t\gamma^1\gamma^3\sum_{-s}\left[c^{(-s)}(\tilde{p})u^{(-s)}(\tilde{p})e^{-ip\cdot\tilde{x}}+d^{(-s)\dagger}(\tilde{p})\upsilon^{(-s)}(\tilde{p})e^{ip\cdot\tilde{x}}\right]\,,\nonumber\\
\end{eqnarray}
onde escolhemos $\alpha_t=\beta^\ast_t$. Portanto,
\begin{equation}
{\cal T}\psi(x){\cal
T}^{-1}\equiv\psi^t(x)=\alpha_t\,\gamma^1\gamma^3\psi(-\tilde{x})\,,
\end{equation}
enquanto que, analogamente,
\begin{equation}
{\cal T}\bar{\psi}(x){\cal
T}^{-1}\equiv\bar{\psi}^t(x)=-\alpha^\ast_t\,\bar{\psi}(-\tilde{x})\gamma^1\gamma^3\,.
\end{equation}

\subsection{Conjugação de carga}

A opera\c{c}\~ao por conjuga\c{c}\~ao de carga age sobre os operadores de aniquila\c{c}\~ao e cria\c{c}\~ao de part\'iculas e antipart\'iculas, sem mudar seus estados de spin, da seguinte maneira:
\begin{eqnarray}
{\cal C}\,c^{(s)}(p)\,{\cal C}^{-1}=\alpha_c\,d^{(s)}(p)\hspace{1.5eM}&,&\hspace{1.5eM}{\cal C}\,d^{(s)}(p)\,{\cal C}^{-1}=\beta_c\,c^{(s)}(p)\,,\nonumber\\
&&\\
{\cal C}\,c^{(s)\dagger}(p)\,{\cal
C}^{-1}=\alpha_c^\ast\,d^{(s)\dagger}(p)\hspace{1.5eM}&,&\hspace{1.5eM}{\cal
C}\,d^{(s)\dagger}(p)\,{\cal
C}^{-1}=\beta_c^\ast\,c^{(s)\dagger}(p)\,.\nonumber
\end{eqnarray}

Então, o campo de Dirac transformado é:
\begin{eqnarray}
{\cal C}\,\psi(x)\,{\cal
C}^{-1}&=&\frac{1}{(2\pi)^{3/2}}\!\int\!\!\frac{d^3p}{E/m}\,\alpha_c\sum_{s}\left[d^{(s)}(p)u^{(s)}(p)e^{-ip\cdot
x}+c^{(s)\dagger}(p)\upsilon^{(s)}(p)e^{ip\cdot x}\right]\,.\nonumber\\
\end{eqnarray}

Como o operador por conjugação de carga não muda o estado de spin das partículas e antipartículas, é definido um operador $C=i\gamma^2\gamma^0$, que deve agir sobre um espinor que contém $\chi^\ast_{-s}$. Assim, de acordo com (2.51), temos que
\begin{eqnarray}
{\cal C}\,\bar{u}^{(s)T}(p)\,{\cal C}^{-1}&=&\upsilon^{(s)}(p)\,,\nonumber\\
&&\\
{\cal C}\,\bar{\upsilon}^{(s)T}(p)\,{\cal
C}^{-1}&=&u^{(s)}(p)\,\nonumber
\end{eqnarray}
onde
\begin{equation}
{\cal C}\,\psi(x)\,{\cal
C}^{-1}\equiv\psi^c(x)=\alpha_c\,C\,\bar{\psi}^T(x)\,,
\end{equation}
e
\begin{equation}
 {\cal C}\,\bar{\psi}(x)\,{\cal C}^{-1}\equiv\bar{\psi}^c(x)=\alpha_c^\ast\,\psi^T(x)C\,.
\end{equation}

\section{Covariantes bilineares}

Nesta se\c{c}\~ao, construiremos e classificaremos os covariantes bilineares, que s\~ao objetos que n\~ao carregam \'indices espinoriais e envolvem apenas dois campos espinoriais. Essas combina\c{c}\~oes t\^em propriedades de transforma\c{c}\~oes bem definidas no grupo de Lorentz. As matrizes $\gamma$ de Dirac formam uma base de 16 matrizes $4\times4$ linearmente independentes. A base \'e formada pelos seguintes elementos:

\begin{center}
\begin{tabular}{cc|c}
1&escalar&1\\
$\gamma^\mu$&vetor&4\\
$\sigma^{\mu\nu}=\frac{i}{2}[\gamma^\mu,\gamma^\nu]$&tensor&6\\
$\gamma^\mu\gamma^5$&pseudo-vetor&4\\
$\gamma^5$&pseudo-escalar&1\\
\hline &TOTAL:&16
\end{tabular}
\end{center}

A teoria de Dirac \'e constru\'ida com base em densidades de lagrangianas. Assim, \'e poss\'ivel construir e classificar os seguintes covariantes bilineares, na forma $\bar{\psi}\Gamma\psi$, com $\Gamma$ representando as 16 matrizes acima, de acordo com sua natureza:

\begin{center}
\begin{tabular}{rcll}
$E(x)$&=&$\bar{\psi}(x)\psi(x)$&\hspace{1cm}(escalar)\\
&&&\\
$V^\mu(x)$&=&$\bar{\psi}(x)\gamma^\mu\psi(x)$&\hspace{1cm}(vetor)\\
&&&\\
$P(x)$&=&$i\bar{\psi}(x)\gamma_5\psi(x)$&\hspace{1cm}(pseudoescalar)\\
&&&\\
$A^\mu(x)$&=&$\bar{\psi}(x)\gamma_5\gamma^\mu\psi(x)$&\hspace{1cm}(vetor axial)\\
&&&\\
$T^{\mu\nu}(x)$&=&$\bar{\psi}(x)\sigma^{\mu\nu}\psi(x)$&\hspace{1cm}(tensor)
\end{tabular}
\end{center}
\medskip

Na pr\'oxima subse\c{c}\~ao h\'a uma tabela que mostra como estes covariantes bilinerares se comportam frente as transforma\c{c}\~oes por paridade, invers\~ao temporal e conjuga\c{c}\~ao de carga, bem
como as transforma\c{c}\~oes combinadas por simetrias CPT.

\section{A transforma\c{c}\~ao CPT}

De acordo com (2.47), (2.54) e (2.59), o campo de Dirac se comporta
da seguinte maneira frente as transforma\c{c}\~oes T, PT e CPT:
\begin{eqnarray}
T:\hspace{0.5eM}\psi^t(x)&=&{\cal T}\psi(x){\cal T}^{-1}=\alpha_t\,\gamma^1\gamma^3\psi(-\tilde{x})\\
&&\nonumber\\
PT:\hspace{0.5eM}\psi^{pt}(x)&=&{\cal P}\psi^t(x){\cal P}^{-1}=\alpha_{pt}\,\gamma^0\gamma^1\gamma^3\psi(-x)\\
&&\nonumber\\
CPT:\hspace{0.5eM}\psi^{cpt}(x)&=&{\cal C}\psi^{pt}(x){\cal
C}^{-1}=\alpha_{cpt}\,\gamma^5\gamma^0\bar{\psi}^T(-x)
\end{eqnarray}
e, de forma an\'aloga para o campo de Dirac adjunto,
\begin{eqnarray}
T:\hspace{0.5eM}\bar{\psi}^t(x)&=&{\cal T}\bar{\psi}(x){\cal T}^{-1}=-\alpha^\ast_t\,\bar{\psi}(-\tilde{x})\gamma^1\gamma^3\\
&&\nonumber\\
PT:\hspace{0.5eM}\bar{\psi}^{pt}(x)&=&{\cal P}\bar{\psi}^t(x){\cal P}^{-1}=\alpha^\ast_{pt}\,\bar{\psi}(-x)\gamma^3\gamma^1\gamma^0\\
&&\nonumber\\
CPT:\hspace{0.5eM}\bar{\psi}^{cpt}(x)&=&{\cal
C}\bar{\psi}^{pt}(x){\cal
C}^{-1}=\alpha^\ast_{cpt}\,\psi^T(-x)\gamma^5\gamma^0
\end{eqnarray}

A tabela a seguir sumariza os resultados das transforma\c{c}\~oes C, P, T e CPT sofridas pelos covariantes bilineares.
\vspace{1cm}

\begin{center}
\begin{tabular}{||c||ccc||c||}
\hline\hline
&$C$&$P$&$T$&$CPT$\\
\hline\hline
$E(x)$&$E(x)$&$E(\tilde{x})$&$E(-\tilde{x})$&$E(-x)$\\
\hline
$V^\mu(x)$&$-V^\mu(x)$&$V_\mu(\tilde{x})$&$V_\mu(-\tilde{x})$&$-V^\mu(-x)$\\
\hline
$P(x)$&$P(x)$&$-P(\tilde{x})$&$-P(-\tilde{x})$&$P(-x)$\\
\hline
$A^\mu(x)$&$A^\mu(x)$&$-A_\mu(\tilde{x})$&$A_\mu(-\tilde{x})$&$-A^\mu(-x)$\\
\hline
$T^{\mu\nu}(x)$&$-T^{\mu\nu}(x)$&$T_{\mu\nu}(\tilde{x})$&$-T_{\mu\nu}(-\tilde{x})$&$T^{\mu\nu}(-x)$\\
\hline\hline
\end{tabular}
\end{center}

\newpage

O {\it Teorema} CPT diz que, em uma teoria de campos relativ\'istica, deve existir a invari\^ancia de transforma\c{c}\~ao por paridade (invers\~ao espacial) e invers\~ao temporal ({\it time reversal}) seguida por uma transforma\c{c}\~ao por conjuga\c{c}\~ao de carga. O teorema CPT assume a veracidade das leis quânticas e invariância de Lorentz. Especificamente, o teorema CPT afirma que fenômenos descritos por qualquer teoria quântica de campo, local e invariante de Lorentz com um hamiltoniano hermitiano, devem ter esta simetria preservada\footnote{Este teorema \'e cuidadosamente demonstrado em \cite{Str} por Streater e Wightman.}. Como pode ser facilmente verificado nesta tabela, os covariantes bilineares de Dirac que n\~ao preservam a simetria CPT s\~ao o vetor e o vetor axial. No entanto, vemos que $\bar{\psi}\gamma^\mu\psi$ n\~ao preserva a transforma\c{c}\~ao por conjuga\c{c}\~ao de carga e $\bar{\psi}\gamma_5\gamma^\mu\psi$ n\~ao \'e invariante frente transforma\c{c}\~oes por paridade. No próximo capítulo, apresentaremos um modelo de quebra de simetria de Lorentz para férmions. No Capitulo 4, investigaremos possíveis efeitos provocados pela quebra da simetria de Lorentz em sistemas utilizando mec\^anica qu\^antica relativ\'istica.

\chapter{Apresenta\c{c}\~ao do modelo de violação de Lorentz para férmions}

\section{Introdução}

No Cap\'itulo 2 estudamos as transforma\c{c}\~oes de simetrias sofridas pelos covariantes bilineares do campo fermi\^onico de Dirac. Tais transforma\c{c}\~oes s\~ao a paridade, a invers\~ao temporal e a conjuga\c{c}\~ao de carga (veja tabela na seção 2.2.4). Vimos que os bilineares de Dirac que sofrem viola\c{c}ões das simetrias CPT, no setor fermi\^onico ou da mat\'eria, s\~ao o \emph{vetor} e o \emph{pseudovetor} ou \emph{vetor axial}.
\smallskip

O objetivo deste capítulo é apresentar um modelo efetivo para a quebra de simetria CPT. O ponto de partida, no contexto da mecânica quântica relativística em quatro dimensões, é acoplar campos de fundo\footnote{Entende-se por campo de fundo um campo cujas fontes não são acessíveis.} aos covariantes bilineares.
Se a lagrangiana da EDQ convencional é dada por
\begin{equation}
{\cal L}_{EDQ}=\bar{\psi}(i\!{\not\!\partial}-e\!{\not\!\!A}-m)\psi-\frac{1}{4}F_{\mu\nu}F^{\mu\nu}\,,
\end{equation}
devemos adicionar a ela as seguintes lagrangianas
\begin{multline}
 {\cal
L}_{Fermion}=-\bar{\psi}\!{\not\!a\psi}-\bar{\psi}\!{\not\!b\gamma_5\psi}-\frac{1}{2}H_{\mu\nu}\bar{\psi}\sigma^{\mu\nu}\psi+ic_{\mu\nu}\bar{\psi}\gamma^\mu
D^\nu\psi+id_{\mu\nu}\bar{\psi}\gamma^\mu\gamma_5D^\nu\psi\,,\\
\\
{\cal
L}_{Foton}=\frac{1}{2}\varepsilon^{\mu\nu\rho\sigma}\eta_\mu A_\nu\partial_\rho A_\sigma+\frac{1}{4}\eta^{\mu\nu\rho\sigma}F_{\mu\nu}F_{\rho\sigma}\,, \hspace{3cm}                      
\end{multline}
onde $iD_\mu=i\partial_\mu+eA_\mu$ \'e a derivada covariante. 
\smallskip

A estrutura da equação de Dirac é modificada pela introdução desses termos na lagrangiana. As correções surgem nas matrizes de Dirac e na massa do férmion:
\begin{eqnarray}
(i\Gamma^\mu D_\mu-M)\psi=0\,,
\end{eqnarray}
onde
$\Gamma_\nu=\gamma_\nu+c_{\mu\nu}\gamma^\mu+d_{\mu\nu}\gamma^\mu\gamma_5$
\,e\, $M=m+\not\!a+\not\!b\gamma_5+\frac{1}{2}H_{\mu\nu}\sigma^{\mu\nu}$. $c_{\mu\nu}$, $d_{\mu\nu}$, $a_\mu$, $b_\mu$ e $H_{\mu\nu}$ são campos de fundo especificados num referencial escolhido.
\smallskip

Cada um dos termos adicionais na lagrangiana contém um parâmetro constante cuja ordem de grandeza é muito pequena \footnote{A ordem de grandeza dessas constantes é muito pequena quando comparadas à massa dos férmions, uma vez que violações da simetria de Lorentz, se existirem, são efeitos muito pequenos, e ainda não observados experimentalmente.}. São eles que controlam as medidas de violações de simetrias CPT e Lorentz nos experimentos. Todos os cinco mon\^omios envolvidos, presentes na lagrangiana adicional (3.2), violam a simetria de Lorentz. Como os mon\^omios que cont\^em $a_\mu$ e $b_\mu$ violam a simetria CPT, somente eles estão presentes nesta dissertação. Já os termos $H_{\mu\nu}$, $c_{\mu\nu}$ e $d_{\mu\nu}$, que preservam CPT, não são aqui estudados. 

\section{A equa\c{c}\~ao de Dirac e o propagador do férmion modificados}

A lagrangiana de fémions de massa $m$ com os termos de quebra de simetria $CPT$ é dada por:
\begin{equation}
{\cal
L}=\bar{\psi}(\not\!p-\not\!a -\not\!b\gamma_5-m)\psi\,,
\end{equation}
onde $a^\mu=(a_0,{\mbox{\boldmath{$a$}}})$ e $b^\mu=(b_0,{\bf b})$ são quadrivetores constantes.
\smallskip

Procuramos por soluções do tipo ondas-planas, em que $\psi^{(\alpha)}=N_u^{(\alpha)}\,u^{(\alpha)}(p)e^{-ip\cdot x}$
($\alpha=1,2)$, é um espinor de Dirac modificado com quatro componentes, onde $N_u^{(\alpha)}$ é uma constante de normalização a ser determinada.
\smallskip

Da equação (3.4), obtemos a equação de Dirac modificada
\begin{equation}
(\not\!p-\not\!a -\not\!b\gamma_5-m)\psi=0\,.
\end{equation}

Aplicando o {\it ansatz} para $\psi$ e multiplicando a equação acima pela esquerda por $(\not\!p-\not\!a -\not\!b\gamma_5+m)$, obtemos a seguinte relação:
\begin{equation}
\left\{(p-a)^2-b^2-m^2-[\not\! p-\not\!a,\not
b]\gamma_5\right\}u(p)=0\,.
\end{equation}

A expressão acima ainda não é diagonal, uma vez que na mesma aparecem termos com matrizes não diagonais. Para obter a equação algébrica para este modelo, \'e necess\'ario multiplicar a equa\c{c}\~ao acima por $(p-a)^2-b^2-m^2+[\not\! p-\not\!a,\not b]\gamma_5$ pela esquerda:
\begin{equation}
\{[(p-a)^2-b^2-m^2]^2-\left([\not\! p-\not\!a,\not b]\gamma_5\right)^2\}u(p)=0\,.
\end{equation}

Vamos trabalhar o comutador acima com $\not\!k=\not\!p-\not\!a$ momentaneamente:
\begin{eqnarray}
\left([\not\! p-\not\!a,\not b]\gamma_5\right)^2&=&(\not\!k\not\!b\,-\not\!b\not\!k)\gamma_5(\not\!k\not\!b\,-\not\!b\not\!k)\gamma_5\nonumber\\
&=&\not\!k\not\!b\not\!k\not\!b\,-\not\!k\not\!b\not\!b\not\!k\,-\not\!b\not\!k\not\!k\not\!b\,+\not\!b\not\!k\not\!b\not\!k\nonumber\\
&=&\not\!k\not\!b\left[-\not\!b\not\!k+2(k\cdot b)\right]-2k^2b^2+\not\!b\not\!k\left[-\not\!k\not\!b+2(k\cdot b)\right]\nonumber\\
&=&-4k^2b^2+2(k\cdot b)\not\!k\not\!b+2(k\cdot b)\not\!b\not\!k\nonumber\\
&=&-4k^2b^2+2(k\cdot b)\left[-\not\!b\not\!k+2(k\cdot b)\right]\,+2(k\cdot b)\not\!b\not\!k\nonumber\\
&=&-4k^2b^2+4(k\cdot b)^2\,,
\end{eqnarray}
onde foi utilizada a identidade $\not\!c\not\!d=-\not\!d\not\!c+2(c\cdot d)$.
\smallskip

Portanto, a rela\c{c}\~ao de dispers\~ao para o modelo \'e
\begin{equation}
[(p-a)^2-b^2-m^2]^2-4[b\cdot(p-a)]^2+4b^2(p-a)^2=0\,.
\end{equation}

Esta relação de dispersão é quártica na variável $p^0({\bf p})$. Ela possui duas raízes positivas $E_u^{(\alpha)}$ e duas negativas $E_v^{(\alpha)}$, onde $\alpha=1,2$.
\smallskip

A equação (3.9) é facilmente resolvida para os casos em que $b^\mu$ é estritamente temporal ou espacial. Para o caso $b^\mu=(b_0,{\bf 0})$, a relação de dispersão fornece
\begin{eqnarray}
E_u^{(\alpha)}&=&\sqrt{(|{\bf p}-{\mbox{\boldmath{$a$}}}|+(-1)^{\alpha}b_0)^2+m^2}+a_0\,,\nonumber\\
&&\\
E_\upsilon^{(\alpha)}&=&\sqrt{(|{\bf
p}+{\mbox{\boldmath{$a$}}}|-(-1)^{\alpha}b_0)^2+m^2}-a_0\,,\nonumber
\end{eqnarray}
em que $E_{u,\upsilon}^{(\alpha)}$ indicam as energias para as partículas e suas antipartículas, respectivamente. Note que o termo $a_\mu$ provoca uma redefini\c{c}\~ao dos zeros da energia e do momento para as part\'iculas, pois $E\,\,\to\,\,E-a_0$ \,e\, ${\mbox{\boldmath{$p$}}}\,\,\to\,\,{\mbox{\boldmath{$p$}}}-{\mbox{\boldmath{$a$}}}$, ou seja, há uma opera\c{c}\~ao por conjuga\c{c}\~ao de carga: $a_\mu\bar{\psi}\gamma^\mu\psi\to-a_\mu\bar{\psi}\gamma^\mu\psi$. 
\smallskip

Para o caso $b^\mu=(0,{\bf b})$, as solu\c{c}\~oes s\~ao
\begin{eqnarray}
E_u^{(\alpha)}&=&\sqrt{({\mbox{\boldmath{$p$}}}-{\mbox{\boldmath{$a$}}})^2+m^2+{\mbox{\boldmath{$b$}}}^2+(-1)^{\alpha}2\sqrt{[{\mbox{\boldmath{$b$}}}\cdot({\mbox{\boldmath{$p$}}}-{\mbox{\boldmath{$a$}}})]^2+m^2}}\,+a_0\,,\nonumber\\
&&\\
E_\upsilon&^{(\alpha)}=&\sqrt{({\mbox{\boldmath{$p$}}}+{\mbox{\boldmath{$a$}}})^2+m^2+{\mbox{\boldmath{$b$}}}^2-(-1)^{\alpha}2\sqrt{[{\mbox{\boldmath{$b$}}}\cdot({\mbox{\boldmath{$p$}}}+{\mbox{\boldmath{$a$}}})]^2+m^2}}\,-a_0\,.\nonumber
\end{eqnarray}

Quando multiplicamos a equação (3.5) pela esquerda por $\gamma^0$, esta equação de movimento pode ser escrita na forma hamiltoniana, onde $i\dfrac{\partial\psi}{\partial t}=H\psi$. Assim, temos que\footnote{$\displaystyle {\bf \Sigma}={\mbox{\boldmath{$
\alpha$}}}\gamma^5=\left(\begin{array}{cc}{\mbox{\boldmath{$
\sigma$}}}&0\\0&{\mbox{\boldmath{$ \sigma$}}}\end{array}\right)\,.$}
\begin{eqnarray}
H={\mbox{\boldmath{$\alpha$}}}\cdot({\bf
p}-{\mbox{\boldmath{$a$}}})+m\gamma^0+a_0+\gamma_5b_0+{\bf
\Sigma}\cdot{\bf b}\,.
\end{eqnarray}

Vamos construir os espinores para o caso $b^\mu$ puramente temporal. O hamiltoniano é então dado por:
\begin{equation}
H={\mbox{\boldmath{$\alpha$}}}\cdot(\mathbf{p}-{\mbox{\boldmath{$a$}}})+m\gamma^0+a_0+b_0\gamma_5\,.
\end{equation}

Na representação padrão das matrizes $\gamma$ de Dirac, como \'e usual, obtemos os seguintes espinores
\begin{equation}
u^{(\alpha)}(p)=N_u^{(\alpha)}\,\left(\begin{array}{c}\chi^{(\alpha)}\\\xi_u^{(\alpha)}\,\chi^{(\alpha)}\end{array}\right)
\end{equation}
para os estados de energia positiva e, para os estados de energia negativa,
\begin{equation}
\upsilon^{(\alpha)}(p)=N_\upsilon^{(\alpha)}\,\left(\begin{array}{c}\xi_\upsilon^{(\alpha)}\,\eta^{(\alpha)}\\\eta^{(\alpha)}\end{array}\right)\,,
\end{equation}
onde
\begin{equation}
\xi_u^{(\alpha)}=\frac{{\mbox{\boldmath{$\sigma$}}}\cdot({\mbox{\boldmath{$p$}}}-{\mbox{\boldmath{$a$}}})-b_0}{E_u^{(\alpha)}-a_0+m}\,.
\end{equation}

A solução para $\xi_\upsilon^{(\alpha)}$ é obtida pela troca, na expressão acima, $a_\mu\to-a_\mu$ e $b_\mu\to-b_\mu$, além da troca
$E_u^{(\alpha)}\to E_\upsilon^{(\alpha)}$.
\smallskip

O espinor (3.15) pode ser normalizado, como foi feito no Cap\'itulo 2, se escolhermos a mesma condi\c{c}\~ao de normaliza\c{c}\~ao do caso da teoria convencional:
\begin{equation}
\bar{u}^{(\alpha)}(p)u^{(\alpha')}(p)=\delta^{\alpha\alpha'}\,.
\end{equation}

Utilizando a definição $\bar{u}=u^\dagger\gamma^0$ e a autoenergia positiva (3.10) para a partícula, encontramos a constante $N_u^{(\alpha)}$ de normalização:

\begin{equation}
N_u^{(\alpha)}=\sqrt{\frac{E_u^{(\alpha)}-a_0+m}{2M}}\,.
\end{equation}

Se o espinor de duas componentes $\chi^{(\alpha)}$ for escolhido ser autovetor do operadoror ${\mbox{\boldmath{$\sigma$}}}\cdot\dfrac{({\mbox{\boldmath{$p$}}}-{\mbox{\boldmath{$a$}}})}{|{\mbox{\boldmath{$p$}}}-{\mbox{\boldmath{$a$}}}|}$ com autovalor $-(-1)^\alpha$, o espinor de Dirac modificado e normalizado fica

\begin{equation}
u^{(\alpha)}(p)=\sqrt{\frac{E_u^{(\alpha)}-a_0+m}{2m}}\left(\begin{array}{c}\chi^{(\alpha)}\\\displaystyle{\frac{-(-1)^{(\alpha)}|{\mbox{\boldmath{$p$}}}-{\mbox{\boldmath{$a$}}}|-b_0}{E_u^{(\alpha)}-a_0+m}}\,\chi^{(\alpha)}\end{array}\right)\,.
\end{equation}

Procedendo de modo análogo, obtemos o espinor para os antiférmions:

\begin{equation}
\upsilon^{(\alpha)}(p)=\sqrt{\frac{E_\upsilon^{(\alpha)}+a_0+m}{2m}}\left(\begin{array}{c}\displaystyle{\frac{-(-1)^{(\alpha)}|{\mbox{\boldmath{$p$}}}+{\mbox{\boldmath{$a$}}}|+b_0}{E_\upsilon^{(\alpha)}+a_0+m}}\,\chi^{(\alpha)}\\\chi^{(\alpha)}\end{array}\right)\,.
\end{equation}

Ainda podemos obter outro elemento de grande importância na teoria da EDQ extendida: o {\it propagador fermiônico modificado}. O propagador de Feynman escolhido deve satisfazer
\begin{equation}
(i\!\not\!\partial-\not\!a-\not\!b\gamma_5-m)S_{a,b}(x-y)=i\delta^{(4)}(x-y)\,,
\end{equation}
ou, escrita no espaço de Fourier,
\begin{equation}
\int\frac{d^4p}{(2\pi)^4}(\not\!p-\not\!a
-\not\!b\gamma_5-m)e^{-ip\cdot(x-y)}S_{a,b}(p)=i\delta^{(4)}(x-y)
\end{equation}
fornece, através da representação de Fourier da Delta de Dirac,
\begin{equation}
S_{a,b}(p)=\frac{i}{\not\!p-\not\!a -\not\!b\gamma_5-m}\,.
\end{equation}

Este propagador é escrito em sua forma invertida:
\begin{eqnarray}
 S_{a,b}(p)&=&\frac{i}{\not\!p-\not\!a -\not\!b\gamma_5-m}=\frac{i(\not\!p-\not\!a -\not\!b\gamma_5+m)}{(\not\!p-\not\!a -\not\!b\gamma_5-m)(\not\!p-\not\!a -\not\!b\gamma_5+m)}\nonumber\\
&&\nonumber\\
&=&  \frac{i(\not\!p-\not\!a -\not\!b\gamma_5+m)}  { \left\{(p-a)^2-b^2-m^2-[\not\! p-\not\!a,\not
b]\gamma_5\right\}}\,,
\end{eqnarray}
onde foi utilizada (3.6). Utilizando também (3.9), segue que

\begin{equation}
 S_{a,b}(p)=\frac{i(\not\!p-\not\!a -\not\!b\gamma_5+m)\{ (p-a)^2-b^2-m^2+[\not\! p-\not\!a,\not
b]\gamma_5 \}}{[(p-a)^2-b^2-m^2]^2-4[(p-a)\cdot b]^2+4(p-a)^2b^2}\,.
\end{equation}

Portanto, vemos que a quebra da simetria de Lorentz modifica as relações de dispersão, as autoenergias e os espinores da teoria de Dirac convencional, além de gerar uma perturbação no hamiltoniano de Dirac livre. No caso do propagador, vemos que a modificação deixa o propagador com uma forma mais complicada que o usual (2.44).

\chapter{Implicações em Mecânica Quântica}

\section{O problema de Landau e a quebra de simetria de Lorentz}

Neste capítulo, investigaremos possíveis efeitos que a quebra da simetria de Lorentz pode causar em sistemas quânticos de elétrons interagindo com um campo eletromagné\-tico. Iniciaremos nosso estudo com um sistema composto por elétrons interagindo com um campo magnético intenso e constante. Tal fenômeno é conhecido como {\it problema de Landau}.
\smallskip

Se consideramos os elétrons relativísticos, o hamiltoniano não perturbado deste problema é dado por 
\begin{eqnarray}
H_0={\mbox{\boldmath{$\alpha$}}}\cdot({\bf p}-e{\bf A})+m\gamma^0\,.
\end{eqnarray}

O problema terá como perturbação o hamiltoniano de interação
\begin{eqnarray}
H_{int}=a_0-{\mbox{\boldmath{$\alpha$}}}\cdot{\mbox{\boldmath{$a$}}}+b_0\gamma_5+{\mbox{\boldmath{$\Sigma$}}}\cdot{\mbox{\boldmath{$b$}}}\,.
\end{eqnarray}

As autoenergias e os autoestados exatos do problema de Landau são calculados no Apêndice B a partir do hamiltoniano não perturbado (4.1) e servirão como base para os cálculos perturbativos.
\smallskip

Os cálculos das corre\c{c}\~oes das energias do el\'etron em ordem mais baixa de aproxima\c{c}\~ao, com as autofunções (B.10-12), s\~ao dadas por

\begin{eqnarray}
\Delta E^-_{n,s}&=&\langle\psi^-_{n,s}|H^-_{int}|\psi^-_{n,s}\rangle=\int\psi^{-\dagger}_{n,s}(x)(H^-_{int})\psi^-_{n,s}(x)dx\nonumber\\
&&\nonumber\\
&=&{\cal A}_0(x)+{\cal A}_z(x)+{\cal B}_0(x)+{\cal
B}_z(x)\,,\nonumber
\end{eqnarray}
onde
\begin{eqnarray*}
{\cal A}_0(x)&=&\int\psi^{-\dagger}_{n,s}(x)(a_0)\psi^-_{n,s}(x)dx=a_0\int\psi^{-\dagger}_{n,s}(x)\psi^-_{n,s}(x)dx\nonumber\\
&&\nonumber\\
&=&a_0\nonumber\,,
\end{eqnarray*}

\begin{eqnarray*}
{\cal A}_z(x)&=&\langle\psi^-_{n,s}|(-{\mbox{\boldmath{$\alpha$}}}\cdot{\mbox{\boldmath{$a$}}})|\psi^-_{n,s}\rangle\nonumber\\
&&\nonumber\\
&=&\int\psi^{-\dagger}_{n,s}(x)(-a_z\alpha_z)\psi^-_{n,s}(x)dx\nonumber\\
&&\nonumber\\
&=&-\frac{a_z}{2^nn!\sqrt{\pi}}\frac{E_{n,s}^-+m}{2E_{n,s}^-}\left(\begin{array}{cc}1,&\frac{\sigma_zp_z}{E_{n,s}^-+m}\end{array}\right)\left(\begin{array}{cc}0&\sigma_z\\\sigma_z&0\end{array}\right)\left(\begin{array}{c}1\\\frac{\sigma_zp_z}{E_{n,s}^-+m}\end{array}\right)\int\limits_{-\infty}^{+\infty}d\xi\,e^{-\xi^2}[H_n(\xi)]^2\nonumber\\
&&\nonumber\\
&=&-a_z\frac{E_{n,s}^-+m}{2E_{n,s}^-}\left(\frac{p_z}{E_{n,s}^-+m}+\frac{p_z}{E_{n,s}^-+m}\right)\nonumber\\
&&\nonumber\\
&=&-a_z\frac{p_z}{E_{n,s}^-}\,,\nonumber
\end{eqnarray*}

\begin{eqnarray*}
{\cal B}_0(x)&=&\int\psi^{-\dagger}_{n,s}(x)(b_0\gamma_5)\psi^-_{n,s}(x)dx\nonumber\\
&&\nonumber\\
&=&\frac{b_0}{2^nn!\sqrt{\pi}}\frac{E_{n,s}^-+m}{2E_{n,s}^-}\left(\begin{array}{cc}1,&\frac{\sigma_zp_z}{E_{n,s}^-+m}\end{array}\right)\left(\begin{array}{cc}0&-1\\-1&0\end{array}\right)\left(\begin{array}{c}1\\\frac{\sigma_zp_z}{E_{n,s}^-+m}\end{array}\right)\int\limits_{-\infty}^{+\infty}d\xi\,e^{-\xi^2}[H_n(\xi)]^2\nonumber\\
&&\nonumber\\
&=&-b_0\frac{E_{n,s}^-+m}{2E_{n,s}^-}\left(\frac{sp_z}{E_{n,s}^-+m}+\frac{sp_z}{E_{n,s}^-+m}\right)\nonumber\\
&&\nonumber\\
&=&-s\frac{p_z}{E_{n,s}^-}b_0\,\nonumber
\end{eqnarray*}

\begin{eqnarray*}
{\cal B}_z(x)&=&\int\psi^{-\dagger}_{n,s}(x)(b_z\Sigma_z)\psi^-_{n,s}(x)dx\nonumber\\
&&\nonumber\\
&=&\frac{b_z}{2^nn!\sqrt{\pi}}\frac{E_{n,s}^-+m}{2E_{n,s}^-}\left(\begin{array}{cc}1,&\frac{\sigma_zp_z}{E_{n,s}^-+m}\end{array}\right)\left(\begin{array}{cc}\sigma_z&0\\0&\sigma_z\end{array}\right)\left(\begin{array}{c}1\\\frac{\sigma_zp_z}{E_{n,s}^-+m}\end{array}\right)\int\limits_{-\infty}^{+\infty}d\xi\,e^{-\xi^2}[H_n(\xi)]^2\nonumber\\
&&\nonumber\\
&=&b_z\frac{E_{n,s}^-+m}{2E_{n,s}^-}\left[s+\frac{sp_z^2}{(E_{n,s}^-+m)^2}\right]\nonumber\\
&&\nonumber\\
&=&sb_z\left[1-\frac{|e|B_0(2n+1-s)}{2E_{n,s}^-(E_{n,s}^-+m)}\right]\,.\nonumber
\end{eqnarray*}

Assim, somando todas as contribui\c{c}\~oes, obtemos \cite{Rus}
\begin{eqnarray}
\Delta
E^-_{n,s}=a_0-a_z\frac{p_z}{E^-_{n,s}}-sb_0\frac{p_z}{E^-_{n,s}}+sb_z\left[1-\frac{|eB_0|(2n+1-s)}{2E^-_{n,s}(E^-_{n,s}+m)}\right]\,.
\end{eqnarray}

Uma análise deste resultado nos fornece a ideia de que os campos $a_\mu$ e $b_\mu$ s\~ao grandezas observ\'aveis. No entanto, este resultado deve ser analisado com mais detalhes no que se refere justamente \`a ordem de grandeza de seus componentes. Seguindo argumentos de Kostelecký e colaboradores \cite{Rus}, os termos proporcionais ao campo magn\'etico $B_0$ devem ser negligenciados pois, para $B_0\simeq5\,T$, a raz\~ao $|eB_0|/m^2\simeq10^{-9}$ \'e muito pequena. No confinamento axial neste experimento, o momento axial \'e o momento de Landau $p_z$, que corresponde a um momento efetivo no eixo $z$. Como a frequ\^encia axial \'e bem menor que a frequ\^encia ciclotron, o termo $p_z/E^-_{n,s}$ deve ser desconsiderado. Neste contexto, a corre\c{c}\~ao para as energias de Landau em termos dominantes \'e simplesmente
\begin{equation}
\Delta E^-_{n,s}\approx a_0+s\,b_z\,.
\end{equation}

Uma proposta para tentar detectar a quebra da simetria de Lorentz consiste na comparação das mudan\c{c}as ({\it shifts}) de n\'iveis de energia\footnote{De acordo com a refer\^encia \cite{Ahi}, tais transi\c{c}\~oes energ\'eticas, com e sem invers\~ao de spin, s\~ao denomi\-nadas frequ\^encias de {\it anomalia} e {\it ciclotron}.}, {\it com} e {\it sem} invers\~ao de spin, de el\'etrons e p\'ositrons t\'ipicos de Landau.  Tais frequ\^encias são obtidas em experimentos conhecidos por \emph{armadilhas de Penning}\footnote{Esta armadilha foi imaginada por F.M. Penning e rendeu a Hans Georg Dehmelt o Pr\^emio Nobel de F\'isica em 1989 por sua utiliza\c{c}\~ao pr\'atica.}. Tais armadilhas s\~ao dispositivos que armazenam part\'iculas carregadas utilizando um campo magn\'etico constante e um campo el\'etrico est\'atico espacialmente n\~ao homog\^eneo.

Considerando $\hbar=1$ e as autoenergias (B.6), as frequ\^encias de transi\c{c}\~oes ener\-g\'eticas para o el\'etron sem e com invers\~ao de spin, n\~ao perturbadas, s\~ao dadas por 
\begin{equation}
\omega^-=E^{-}_{1,-1}-E^{-}_{0,-1}\hspace{1eM},\hspace{2eM}\bar{\omega}^-=E^{-}_{0,+1}-E^{-}_{1,-1}\,.
\end{equation}

O teorema $CPT$ afirma que ambas as frequ\^encias do el\'etron acima devem ser iguais \`as mesmas do p\'ositron. No entanto, as frequ\^encias {\it corrigidas} $\omega(\bar{\omega})^{\mp(CPT)}$, de acordo com a teoria de viola\c{c}\~ao de Lorentz, para os el\'etrons e p\'ositrons, s\~ao dadas por
\begin{equation}
\omega^{-(CPT)}\approx\omega^{+(CPT)}\approx\omega\hspace{1eM},\hspace{2eM}\bar{\omega}^{\mp(CPT)}\approx\bar{\omega}\pm
2b_z\,.
\end{equation}

Note que, no resultado obtido acima, as frequ\^encias corrigidas {\it n\~ao dependem} de $a_\mu$. Isso ocorre porque, como j\'a foi discutido, este campo apenas redefine os zeros da energia e momentos. O termo dominante na teoria de viola\c{c}\~ao CPT \'e dependente do campo $b_\mu$ e \'e proveniente da diferen\c{c}a entre
as frequ\^encias sem e com mudança de spin:
\begin{equation}
\Delta\omega\equiv\omega^{-(CPT)}-\omega^{+(CPT)}\approx0\hspace{1eM},\hspace{2eM}\Delta\bar{\omega}\equiv\bar{\omega}^{-(CPT)}-\bar{\omega}^{+(CPT)}\approx+4b_z\,.
\end{equation}

Portanto, o experimento de armadilhas de Penning discutido aqui \'e sens\'ivel apenas \`a parte espacial do vetor ${\mathbf b}$ na dire\c{c}\~ao do campo magn\'etico. 

\section{Expansão não relativística do hamiltoniano modificado}

Considere o hamiltoniano (3.12) de uma partícula sujeita a um campo eletromagnético (agora com o campo coulombiano) e os termos de quebra das simetrias CPT:
\begin{eqnarray}
H=m\gamma^0+{\mbox{\boldmath{$\alpha$}}}\cdot({\bf
p}-{\mbox{\boldmath{$a$}}}-e{\bf A})+eA_0+a_0+{\bf \Sigma}\cdot{\bf
b}+\gamma_5b_0.
\end{eqnarray}

Faremos a expansão utilizando o método Foldy-Wouthuysen\footnote{Vide Apêndice C.} (FW) \cite{Fol}. Este método propõe reescrever o hamiltoniano acima separado em duas partes. Como as matrizes $\gamma^5$ e ${\bf \Sigma}$ são, respectivamente, \textit{não diagonal} e \textit{diagonal} na representação padrão adotada, temos o hamiltoniano acima escrito em termos de operadores pares e ímpares:
\begin{eqnarray}
H=m\gamma^0+{\cal P}+{\cal I},
\end{eqnarray}
onde
\begin{eqnarray}
{\cal P}&=&eA_0+a_0+{\bf \Sigma}\cdot{\bf b}\hspace{8eM}\mbox{(operador par),}\\
{\cal I}&=&{\mbox{\boldmath{$ \alpha$}}}\cdot({\bf
p}-{\mbox{\boldmath{$a$}}}-e{\bf
A})+\gamma_5b_0\hspace{4eM}\mbox{(operador ímpar).}
\end{eqnarray}

Considere o \textit{ansatz} (vide Apêndice C):
\begin{equation}
S=-\frac{i}{2m}\gamma^0{\cal I}\,.
\end{equation}

Vamos expandir o hamiltoniano na representação FW escrevendo-o em séries de potências de $1/m$. Faremos a expansão através da fórmula de \textit{Baker-Campbell-Hausdorff}:
\begin{equation}
e^{iS}He^{-iS}=H+i[S,H]+\frac{i^2}{2!}[S,[S,H]]+\frac{i^3}{3!}[S,[S,[S,H]]]+...
\end{equation}

De outra forma,
\begin{equation}
H'=H+i[S,H]-\frac{1}{2}[S,[S,H]]-\frac{i}{6}[S,[S,[S,H]]]+\frac{1}{24}[S,[S,[S,[S,H]]]]...
\end{equation}

Utilizando as relações de (anti)comutações
\begin{equation}
\{\gamma^0,{\cal
I}\}=0\hspace{2em}\mbox{e}\hspace{2em}[\gamma^0,{\cal P}]=0\,,
\end{equation}
\noindent podemos calcular algumas relações de comutações envolvendo $S$ e $H$:
\begin{eqnarray}
i[S,H]&=&-{\cal I}+\frac{\gamma^0{\cal
I}^2}{m}+\frac{\gamma^0}{2m}[{\cal I},{\cal P}]\,,\\
\frac{i^2}{2}[S,[S,H]]&=&-\frac{1}{2m}\gamma^0{\cal
I}^2-\frac{1}{2m^2}{\cal I}^3-\frac{1}{8m^2}[{\cal
I},[{\cal I},{\cal P}]]\,,\\
\frac{i^3}{3!}[S,[S,[S,H]]]&=&\frac{1}{6m^2}{\cal
I}^3-\frac{1}{6m^3}\gamma^0{\cal
I}^4\,,\nonumber\\
&-&\frac{1}{48m^3}\gamma^0[{\cal I},[{\cal I},[{\cal
I},{\cal P}]]]\\
\frac{i^4}{4!}[S,[S,[S,[S,H]]]]&\simeq&\frac{1}{24m^3}\gamma^0{\cal
I}^4\,.
\end{eqnarray}

O último termo foi obtido por indução. Adicionando todos os termos de (4.16-19), obtemos:
\begin{eqnarray}
H&=&\gamma^0\left(M+\frac{1}{2m}\,{\cal I}^2-\frac{1}{8m^2}\,{\cal I}^4\right)+{\cal P}-\frac{1}{8m^2}[{\cal I},[{\cal I},{\cal P}]]\nonumber\\
&+&\frac{1}{2m}\,\gamma^0[{\cal I},{\cal P}]-\frac{1}{3m^2}\,{\cal
I}^3-\frac{1}{48m^3}\,\gamma^0[{\cal I},[{\cal I},[{\cal I},{\cal
P}]]]\,,\nonumber
\end{eqnarray}

O hamiltoniano é então reescrito retendo-se termos proporcionais até $1/n^3$:
\begin{eqnarray}
H'''=\gamma^0\left(m+\frac{1}{2m}\,{\cal I}^2-\frac{1}{8m^3}\,{\cal
I}^4\right)+{\cal P}-\frac{1}{8m^2}\,[{\cal I},[{\cal I},{\cal
P}]]\,.
\end{eqnarray}

Voltemos ao operador ímpar representado em (4.11). Utilizando a fórmula
\begin{equation}
({\mbox{\boldmath{$ \alpha$}}}\cdot\textbf{A})({\mbox{\boldmath{$
\alpha$}}}\cdot\textbf{B})=\textbf{A}\cdot\textbf{B}+i{\bf
\Sigma}\cdot(\textbf{A}\times\textbf{B})\,,
\end{equation}
vem que
\begin{eqnarray*}
{\cal I}^2&=&[{\mbox{\boldmath{$ \alpha$}}}\cdot(\textbf{p}-{\mbox{\boldmath{$a$}}}-e\textbf{A})+\gamma^5b_0]^2\\
&&\\
&=&(\textbf{p}-{\mbox{\boldmath{$a$}}}-e\textbf{A})^2-ie{\bf
\Sigma}\cdot(\textbf{p}\times\textbf{A}+\textbf{A}\times\textbf{p})+\{{\mbox{\boldmath{$
\alpha$}}},\gamma_5\}\cdot(\textbf{p}-{\mbox{\boldmath{$a$}}}-e\textbf{A})b_0+b_0^2\,.
\end{eqnarray*}

O anticomutador acima é dado por
\begin{eqnarray*}
\{{\mbox{\boldmath{$ \alpha$}}},\gamma_5\}=-2{\bf \Sigma}\,.
\end{eqnarray*}

Então, com $\textbf{p}=-i\nabla$ e $\textbf{P}=\textbf{p}-{\mbox{\boldmath{$a$}}}-e\textbf{A}$, temos que
\begin{equation}
\frac{1}{2m}\,{\cal I}^2=\frac{1}{2m}[\textbf{P}-{\bf
\Sigma}b_0]^2-\frac{e}{2m}\,{\bf \Sigma}\cdot\textbf{B}\,.
\end{equation}

Agora, calculemos o comutador
\begin{equation}
[{\cal I},{\cal P}]=ie\,{\mbox{\boldmath{$
\alpha$}}}\cdot\textbf{E}-2({\mbox{\boldmath{$ \alpha$}}}\cdot{\bf
\Sigma})(\textbf{P}\cdot\textbf{b})\,.
\end{equation}

Da mesma forma, obtemos o outro comutador:
\begin{eqnarray*}
[{\cal I},[{\cal I},{\cal P}]]&=&e\,\nabla\cdot\textbf{E}+i{\bf
\Sigma}\cdot(\nabla\times\textbf{E})+2{\bf
\Sigma}\cdot(\textbf{E}\times\textbf{p})-4({\bf
\Sigma}\cdot\textbf{P})(\textbf{P}\cdot\textbf{b})\,.
\end{eqnarray*}

Assim,
\begin{eqnarray}
\frac{1}{8m^2}\,[{\cal I}[{\cal I},{\cal P}]]&=&\frac{1}{8m^2}\left[e\,\nabla\cdot\textbf{E}+i{\bf \Sigma}\cdot(\nabla\times\textbf{E})+2{\bf \Sigma}\cdot(\textbf{E}\times\textbf{p})\right]\nonumber\\
&&\nonumber\\
&-&\frac{1}{2m^2}\,({\bf
\Sigma}\cdot\textbf{P})(\textbf{P}\cdot\textbf{b})\,.
\end{eqnarray}

Então, com todas as contribuições juntas, temos que:
\begin{eqnarray}
H'''&=&\gamma^0\left[m+\frac{1}{2m}(\textbf{P}-{\bf \Sigma}b_0)^2-\frac{1}{8m^3}\textbf{P}^4\right]\nonumber\\
&&\nonumber\\
&+&eA_0+a_0+{\bf \Sigma}\cdot\textbf{b}-\frac{e}{2m}{\bf \Sigma}\cdot\textbf{B}\nonumber\\
&&\nonumber\\
&-&\frac{e}{4m^2}\,{\bf \Sigma}\cdot(\textbf{E}\times\textbf{p})+\frac{i}{8m^2}\,{\bf \Sigma}\cdot(\nabla\times\textbf{E}) -\frac{e}{8m^2}\,\nabla\cdot\textbf{E}\nonumber\\
&&\nonumber\\
&-&\frac{1}{2m^2}\,({\bf
\Sigma}\cdot\textbf{P})(\textbf{P}\cdot\textbf{b})\,.
\end{eqnarray}

Assim, a expressão acima corresponde ao limite não relativístico do hamiltoniano (4.8) expresso na representação FW. Este resultado \'e idêntico ao obtido em \cite{Kha}, porém foi obtido por nós através de um método diferente (FW). O termo mais interessante é o de interação spin-órbita que envolve o campo de fundo $b_\mu$.  

\section{O efeito Zeeman anômalo modificado}

O efeito Zeeman corresponde \`a mudan\c{c}a das linhas espectrais de um certo elemento devido a a\c{c}\~ao de um campo magn\'etico externo. Para o caso de \'atomos cujo spin eletr\^onico total \'e nulo, as linhas de emiss\~ao se decomp\~oem em multipletos (dubletos, tripletos, etc.), cujas caracter\'isticas dependem essencialmente do elemento e do campo externo. Tal fen\^omeno \'e denominado {\it efeito Zeeman normal}, descoberto pelo f\'isico holand\^es Pieter Zeeman em 1896. Para o caso em que o spin eletr\^onico total n\~ao \'e nulo, tem-se o {\it efeito Zeeman an\^omalo}: as linhas de emiss\~ao n\~ao se decomp\~oem em dubletos ou tripletos, mas em multipletos de estrutura mais complicada. Isso ocorre porque o spin se acopla ao campo magn\'etico externo.

Nesta se\c{c}\~ao, o efeito Zeeman an\^omalo ser\'a estudado na presença da quebra de simetria de Lorentz, que possivelmente fornecerá modificações no espectro de emiss\~ao do hidrog\^enio. Investigaremos isso aplicando teoria de perturbações geradas, separadamente, pelos campos de fundo $a_\mu$ e $b_\mu$.

\subsection{O acoplamento vetorial como perturbação}

O hamiltoniano com o acoplamento vetorial no limite não relativístico do modelo de quebra de simetria de Lorentz é reescrito através de (4.25) somente com os termos $a^\mu$:
\begin{eqnarray}
H=\frac{1}{2m}(\mathbf{p}-e\mathbf{A}-{\mbox{\boldmath{$a$}}})^2-\frac{e}{2M}{\mbox{\boldmath{$
\sigma$}}}\cdot{\bf B}+eA_0+a_0\,.
\end{eqnarray}

Este hamiltoniano é escrito em termos do hamiltoniano de Pauli (n\~ao perturbado) como se segue:
\begin{eqnarray}
H=\left[\frac{1}{2m}(\mathbf{p}-e\mathbf{A})^2-\frac{e}{2m}{\mbox{\boldmath{$
\sigma$}}}\cdot\mathbf{B}+eA_0\right]+\left[-\frac{1}{m}(\mathbf{p}-e\mathbf{A})\cdot{\mbox{\boldmath{$a$}}}+a_0+\frac{1}{2m}{\mbox{\boldmath{$a$}}}^2\right]\,.
\end{eqnarray}

O primeiro termo do hamiltoniano acima \'e o bem conhecido hamiltoniano de Pauli. O segundo termo \'e proveniente do
acoplamento vetorial atrav\'es do campo $a^\mu$, e é o hamiltoniano de intera\c{c}\~ao. Assim,
\begin{eqnarray}
H_{int(a)}=\frac{i}{m}{\mbox{\boldmath{$a$}}}\cdot\nabla+\frac{e}{m}\mathbf{A}\cdot{\mbox{\boldmath{$a$}}}+a_0+\frac{{\mbox{\boldmath{$a$}}}^2}{2m}\,,
\end{eqnarray}
onde foi utilizada a rela\c{c}\~ao
$\mathbf{p}\cdot{\mbox{\boldmath{$a$}}}=-i\nabla\cdot{\mbox{\boldmath{$a$}}}-i{\mbox{\boldmath{$a$}}}\cdot\nabla=-i{\mbox{\boldmath{$a$}}}\cdot\nabla$. Note que, neste caso, a quebra de conjuga\c{c}\~ao de carga n\~ao
\'e mais manifesta, pois existe uma \'unica express\~ao que representa o hamiltoniano para as part\'iculas e antipart\'iculas.
\smallskip

Os \'ultimos dois termos em (4.28) s\~ao apenas duas constantes que n\~ao representam mudan\c{c}as f\'isicas nos n\'iveis de energia, pois n\~ao se manifestam nas transi\c{c}\~oes energ\'eticas. Para aplicar a teoria de perturbações, escrevemos a fun\c{c}\~ao de onda $\psi$ do \'atomo de hidrog\^enio em coordenadas esf\'ericas para uma \'unica part\'icula:
\begin{eqnarray*}
\psi_{n\ell m}(r,\theta,\phi)=R_{n\ell}(r)\Theta_{\ell m}(\theta)\Phi_m(\phi)\,.
\end{eqnarray*}

Utilizando  o operador gradiente escrito em coordenadas esf\'ericas
\begin{eqnarray*}
{\mbox{\boldmath{$\nabla$}}}=\textbf{\^e}_r\,\frac{\partial}{\partial r}+\textbf{\^e}_\theta\,\frac{1}{r}\,\frac{\partial}{\partial\theta}+\textbf{\^e}_\phi\,\frac{1}{r\sin\theta}\,\frac{\partial}{\partial\phi}\,,
\end{eqnarray*}
segue que:
\begin{eqnarray}
\Delta E_{(a),1}&=&\frac{i}{M}\langle n\ell m|{\mbox{\boldmath{$a$}}}\cdot\nabla|n\ell m\rangle\nonumber\\
&&\\
&=&\frac{i}{M}\int\left\{R_{n\ell}^\ast(r)\frac{\partial R_{n\ell}(r)}{\partial r}|\Theta_{\ell m}(\theta)|^2|\Phi_m(\phi)|^2\,{\mbox{\boldmath{$a$}}}\cdot\textbf{\^e}_r\right.\nonumber\\
&&\left. \right. \\
&+&\left.\frac{|R_{n\ell}(r)|^2}{r}\Theta_{\ell m}^\ast(\theta)\frac{\partial\Theta_{\ell m}(\theta)}{\partial\theta}|\Phi_m(\phi)|^2{\mbox{\boldmath{$a$}}}\cdot\textbf{\^e}_\theta\right.\nonumber\\
&&\left. \right.\\
&+&\left. im\frac{|R_{n\ell}(r)|^2|\Theta_{\ell
m}(\theta)|^2}{r\sin\theta}|\Phi_m(\phi)|^2{\mbox{\boldmath{$a$}}}\cdot\textbf{\^e}_\phi\right\}d^3r\nonumber\,.
\end{eqnarray}

Para facilitar os c\'alculos, escolhemos o campo ${\mbox{\boldmath{$a$}}}$ ao longo do eixo $z$, onde
${\mbox{\boldmath{$a$}}}\cdot\textbf{\^e}_r=a_z\cos\theta$, ${\mbox{\boldmath{$a$}}}\cdot\textbf{\^e}_\theta=-a_z\sin\theta$ e
${\mbox{\boldmath{$a$}}}\cdot\textbf{\^e}_\phi=0$. O primeiro termo \'e ent\~ao escrito explicitamente:
\begin{eqnarray}
\Delta E_{(a),1}=\frac{ia_z}{m}\int\left[R_{n\ell}^\ast(r)\frac{\partial
R_{n\ell}(r)}{\partial r}r^2dr\right]|\Theta_{\ell m}(\theta)|^2\sin\theta\cos\theta d\theta\,.
\end{eqnarray}

Esta corre\c{c}\~ao \'e nula, pois
\begin{eqnarray*}
\int\limits_0^\pi|\Theta_{\ell m}(\theta)|^2\sin\theta\cos\theta d\theta=0\,,
\end{eqnarray*}
para todas as fun\c{c}\~oes associadas de Legendre.
\smallskip

Agora, o segundo termo \'e
\begin{eqnarray*}
\Delta E_{(a),2}=-\frac{ia_z}{m}\int\frac{|R_{n\ell}(r)|^2}{r}\Theta_{\ell
m}^\ast(\theta)\frac{\partial\Theta_{\ell m}(\theta)}{\partial\theta}\sin^2\theta d\theta r^2dr\,.
\end{eqnarray*}

Observando a integra\c{c}\~ao angular, verifica-se que
\begin{eqnarray*}
\int\limits_0^\pi\Theta_{\ell m}^\ast(\theta)\frac{\partial\Theta_{\ell m}(\theta)}{\partial\theta}\sin^2\theta d\theta=\int\limits_{-1}^{+1}\Theta_{\ell m}(x)\frac{\partial\Theta_{\ell m}(x)}{\partial x}(x^2-1)dx=0\,,
\end{eqnarray*}
resultado proveniente da f\'ormula de recorr\^encia
\begin{eqnarray*}
(x^2-1)\frac{d}{dx}\Theta_{\ell m}(x)=\ell x\Theta_{\ell m}(x)-(\ell+m)\Theta_{\ell-1,m}(x)
\end{eqnarray*}
e da rela\c{c}\~ao de ortogonalidade dos polin\^omios de Legendre:
\begin{eqnarray*}
\int\limits_{-1}^{+1}\Theta_{km}(x)\Theta_{\ell m}(x)dx=0\,\,,\hspace{2eM}\mbox{para todo $k\neq\ell$.}
\end{eqnarray*}

Portanto, 
\begin{equation}
\Delta E_{(a),1}=0\,,
\end{equation}

Agora, analisaremos o termo que depende do potencial vetor:
\begin{eqnarray*}
\Delta E_{(a),2}=\frac{e}{m}\int\Psi^\ast(\mathbf{A}\cdot{\mbox{\boldmath{$a$}}})\Psi d^3r\,.
\end{eqnarray*}

Para um campo magn\'etico intenso representado por $\mathbf{B}=B_0\,\textbf{\^e}_z$, o potencial vetor associado \'e
$\mathbf{A}=-B_0\left(\dfrac{y}{2},-\dfrac{x}{2},0\right)$. Isto implica que:
\begin{equation}
\Delta E_{(a),2}=-\frac{eB_0}{2m}\int\Psi^\ast(ya_x-xa_y)\Psi
d^3r\,.
\end{equation}

Ap\'os c\'alculos expl\'icitos, obtem-se que
\begin{equation}
\Delta E_{(a),2}=0\,.
\end{equation}

Portanto, conclu\'imos que a presen\c{c}a do campo de fundo $a_\mu$ n\~ao afeta o espectro de energia do hidrog\^enio, pois o hamiltoniano (4.28) n\~ao produz nenhuma corre\c{c}\~ao aos deslocamento nos n\'iveis de energia do efeito Zeeman anômalo.

\subsection{O acoplamento axial como perturbação}

Neste caso, o hamiltoniano é dado por
\begin{eqnarray}
H&=&\left[\frac{1}{2m}(\mathbf{p}-e\mathbf{A})^2-\frac{e}{2m}{\mbox{\boldmath{$\sigma$}}}\cdot\mathbf{B}+eA_0\right]+\left[{\mbox{\boldmath{$\sigma$}}}\cdot\mathbf{b}-\frac{b_0}{m}{\mbox{\boldmath{$\sigma$}}}\cdot(\mathbf{p}-e\mathbf{A})+\frac{b_0^2}{2m}\right]\nonumber\\
&&\nonumber\\
H&\equiv&H_{Pauli}+H_{int(b)}\,.
\end{eqnarray}

O terceiro termo de $H_{int(b)}$ será negligenciado porque é uma constante e não causa efeito algun. Assim,
\begin{eqnarray}
H_{int(b)}={\mbox{\boldmath{$\sigma$}}}\cdot\mathbf{b}-\frac{b_0}{m}{\mbox{\boldmath{$\sigma$}}}\cdot(\mathbf{p}-e\mathbf{A})\,.
\end{eqnarray}

Para obtermos as possíveis mudanças no espectro de energia do hidrogênio, é necessário utilizarmos teoria de perturbações independente do tempo até primeira ordem, cujo hamiltoniano de interação será dado por (4.37). Assim,
\begin{equation}
\Delta E_{(b),1}=\langle n\ell jm_jm_s|{\mbox{\boldmath{$\sigma$}}}\cdot\mathbf{b}|n\ell
jm_jm_s\rangle\,,
\end{equation}
onde $n$, $j$, $m_j$ são os números quânticos associados ao hidrogênio no caso em que ocorre a soma entre os momentos angulares $\mathbf{L}$ e $\mathbf{S}$ (veja o Apêndice D). Considerando-se a situação em que $j=\ell+1/2$ e $m_j=m+1/2$, com suas autofunções sendo representadas por (C.20) e (C.21), esta contribuição de energia pode ser calculada explicitamente:

\begin{eqnarray}
\Delta E_{(b),1}&=&\langle n\ell
jm_jm_s|{\mbox{\boldmath{$\sigma$}}}\cdot\mathbf{b}|n\ell
jm_jm_s\rangle=\langle jm_j|b_z\sigma_z|jm_j\rangle\nonumber\\
&&\nonumber\\
&=&\int\left(\begin{array}{cc}\sqrt{\frac{\ell+m+1}{2\ell+1}}\,Y_\ell^{m\,\ast}&\sqrt{\frac{\ell-m}{2\ell+1}}\,Y_\ell^{m+1\,\ast}\end{array}\right)\left(\begin{array}{cc}b_z&0\\0&-b_z\end{array}\right)\left(\begin{array}{c}\sqrt{\frac{\ell+m+1}{2\ell+1}}\,Y_\ell^{m}\\\sqrt{\frac{\ell-m}{2\ell+1}}\,Y_\ell^{m+1}\end{array}\right)d\Omega\nonumber\\
&&\nonumber\\
&=&b_z\int\left(\begin{array}{cc}\sqrt{\frac{\ell+m+1}{2\ell+1}}\,Y_\ell^{m\,\ast}&\sqrt{\frac{\ell-m}{2\ell+1}}\,Y_\ell^{m+1\,\ast}\end{array}\right)\left(\begin{array}{c}\sqrt{\frac{\ell+m+1}{2\ell+1}}\,Y_\ell^{m}\\-\sqrt{\frac{\ell-m}{2\ell+1}}\,Y_\ell^{m+1}\end{array}\right)d\Omega\nonumber\\
&&\nonumber\\
&=&\frac{\ell+m+1}{2\ell+1}b_z\int|Y_\ell^m|^2d\Omega-\frac{\ell-m}{2\ell+1}b_z\int|Y_\ell^{m+1}|^2d\Omega=\frac{2m+1}{2\ell+1}\,b_z\nonumber\\
&&\nonumber\\
&=&\frac{2m_jb_z}{2\ell+1}\,,
\end{eqnarray}

\noindent onde os $Y_\ell^m$ são os harmônicos esféricos normalizados. Para o caso em que $j=\ell-1/2$, obtemos $-\dfrac{2m_jb_z}{2\ell+1}$. Então, a contribuição total para o termo ${\mbox{\boldmath{$\sigma$}}}\cdot\mathbf{b}$, que representa o acoplamento axial sem campo eletromagnético, é
\begin{equation}
\Delta E_{\sigma\cdot b}=\pm2\frac{m_jb_z}{2\ell+1}\,.
\end{equation}

Este resultado coincide com a referência \cite{Kha}, mas é o dobro do resultado obtido em \cite{Man}.
\smallskip

 A energia é corrigida em primeira ordem de aproximação pelo fator $\pm m_j$. De fato, cada linha do espectro é separada em $2j+1$ linhas, cuja separação linear é dada por ${\displaystyle{\frac{b_z}{2\ell+1}}}$. Como a mudança de energia depende do módulo de ${\bf b}$, este resultado teórico para a correção de energia pode ser utilizado para se obter um limite superior para o parâmetro $b^\mu$ através de experimentos específicos.

A contribuição em primeira ordem do segundo termo do hamiltoniano (4.37) é escrito a seguir:
\begin{equation}
\Delta E_{(b),2}=-\frac{b_0}{m}\langle n\ell
jm_jm_s|{\mbox{\boldmath{$\sigma$}}}\cdot\nabla|n\ell
jm_jm_s\rangle\,.
\end{equation}

Aqui, a função de onda $\psi_{n\ell m}$ passa a depender de uma função $\chi_{m_s}$ que depende do spin do elétron. Neste caso, a função de onda total será $\Psi_{n\ell jm_j}=\psi_{n\ell m}\chi_{m_s}$. No entanto, devemos considerar que o operador nabla deve atuar em $R_{n\ell}(r)$, $\theta_{\ell m}(\theta)$ e $\Phi_m(\phi)$, enquanto que ${\mbox{\boldmath{$\sigma$}}}$ deve atuar na função de spin. Assim,
\begin{eqnarray}
&&\Delta E_{(b),2}=-\frac{b_0}{m}\int\left\{R_{n\ell}(r)^\ast\frac{\partial R_{n\ell}(r)}{\partial r}|\theta_{\ell
m}(\theta)|^2|\Phi_m(\phi)|^2\langle jm_j|{\mbox{\boldmath{$\sigma$}}}\cdot\textbf{ê}_r|jm_j\rangle\right.\nonumber\\
&&\nonumber\\
&&\left.+\frac{|R_{n\ell}(r)|^2}{r}|\Phi_m(\phi)|^2\Theta_{\ell m}(\theta)^\ast\frac{\partial\Theta_{\ell
m}(\theta)}{\partial\theta}\langle jm_j|{\mbox{\boldmath{$\sigma$}}}\cdot\textbf{ê}_\theta|jm_j\rangle\right.\nonumber\\
&&\nonumber\\
&&\left.+\frac{|R_{n\ell}(r)|^2|\Theta_{\ell m}(\theta)|^2}{r\sin\theta}|\Phi_m(\phi)|^2\langle jm_j|{\mbox{\boldmath{$\sigma$}}}\cdot\textbf{ê}_\phi|jm_j\rangle\right\}d^3r\,.
\end{eqnarray}

Os produtos escalares em coordenadas carte\-sianas que aparecem nos valores esperados entre os \emph{kets} e os \emph{bras} s\~ao dados por:
\begin{eqnarray*}
{\mbox{\boldmath{$\sigma$}}}\cdot\textbf{ê}_r&=&\sin\theta\cos\phi\,\sigma_x+\sin\theta\sin\phi\,\sigma_y+\cos\theta\,\sigma_z\,,\\
{\mbox{\boldmath{$\sigma$}}}\cdot\textbf{ê}_\theta&=&\cos\theta\cos\phi\,\sigma_x+\cos\theta\sin\phi\,\sigma_y-\sin\theta\,\sigma_z\,,\\
{\mbox{\boldmath{$\sigma$}}}\cdot\textbf{ê}_\phi&=&-\sin\theta\,\sigma_x+\cos\phi\,\sigma_y\,.
\end{eqnarray*}

Como vemos em (4.42), somente os termos proporcionais a $\sigma_z$ v\~ao contribuir com valores esperados n\~ao nulos. Assim,
\begin{eqnarray*}
\Delta E_{(b),2}&=&\pm\frac{ib_0m_j}{(2\ell+1)m}\int\left[R_{n\ell}(r)^\ast\frac{\partial R_{n\ell}(r)}{\partial r}|\Theta_{\ell m}(\theta)|^2\cos\theta\right.\\
&&\left. \right.\\
&-&\left.\frac{|R_{n\ell}(r)|^2}{r}\Theta_{\ell m}(\theta)^\ast\frac{\partial\Theta_{\ell m}(\theta)}{\partial \theta}\sin\theta\right]d^3r\\
&&\\
&=&\pm\frac{ib_0m_j}{(2\ell+1)m}\int\left[R(r)^\ast\frac{\partial R(r)}{\partial r}r^2dr\right]\int|\Theta(\theta)|^2\sin\theta\cos\theta d\theta\\
&&\\
&\pm&\frac{ib_0m_j}{(2\ell+1)m}\int\left[-\frac{|R_{n\ell}(r)|^2}{r}r^2dr\right]\int\Theta_{\ell
m}(\theta)^\ast\frac{\partial\Theta_{\ell m}(\theta)}{\partial
\theta}\sin^2\theta d\theta\,.
\end{eqnarray*}

As express\~oes obtidas acima s\~ao exatamente as mesmas obtidas para o c\'alculo da corre\c{c}\~ao de energia para o caso do acoplamento vetorial. Assim,
\begin{equation}
 \Delta E_{(b),2}=0\,.
\end{equation}

Observando novamente o hamiltoniano de intera\c{c}\~ao (4.37), o termo $eb_0\,{\mbox{\boldmath{$\sigma$}}}\cdot\mathbf{A}/m$ n\~ao deve oferecer nenhuma corre\c{c}\~ao ao efeito Zeeman para o caso em que o hidrog\^enio \'e livre, uma vez que $\mathbf{A}=0$. Considerando o caso em que o el\'etron \'e sujeito a um forte campo magn\'etico, este termo n\~ao pode ser negligenciado. Para um campo magn\'etico intenso representado por $\mathbf{B}=B_0\,\textbf{\^e}_z$, o potencial vetor associado \'e $\mathbf{A}=-B_0\left(\dfrac{y}{2},-\dfrac{x}{2},0\right)$. Isto implica que a corre\c{c}\~ao de energia, em primeira ordem de teoria de perturba\c{c}\~oes, \'e dado por:

\begin{multline}
 \Delta E_{\sigma\cdot A}=\frac{eb_0}{m}\langle n\ell jm_jm_s|{\mbox{\boldmath{$\sigma$}}}\cdot\mathbf{A}|n\ell jm_jm_s\rangle\\
\\
=-\frac{eb_0B_0}{2m}\langle n\ell jm_jm_s|y\sigma_x-x\sigma_y|n\ell jm_jm_s\rangle\\
\\
=-\frac{eb_0B_0}{2m}\langle jm_j|\left(\begin{array}{cc}1&0 \end{array}  \right)\left(\begin{array}{cc}0&y\\y&0 \end{array} \right) \left(\begin{array}{c}1\\0\end{array}\right) - \left(\begin{array}{cc}1&0 \end{array}  \right)\left(\begin{array}{cc}0&-ix\\ix&0 \end{array} \right) \left(\begin{array}{c}1\\0\end{array}\right)  |jm_j\rangle\\    
\\
\hspace{-9cm}=0\,.
\end{multline}

Portanto, o\'unico efeito perturbativo foi obtido em (4.40), cujo resultado \'e proporcional a $|\mathbf{b}|$, que foi gerado pelo termo de interação spin-órbita ${\mbox{\boldmath{$\sigma$}}}\cdot\mathbf{b}$. Podemos concluir também que um campo magn\'etico externo, mesmo intenso, n\~ao produz nenhum efeito de mudan\c{c}a de energia no espectro do hidrog\^enio.

\chapter{Férmions em um campo de calibre em (2+1) Dimens\~oes}

\section{Introdução}

Como é bem conhecido, a teoria de calibre de Maxwell usual é descrita a partir de um campo de calibre fundamental $A_\mu$, através da lagrangiana
\begin{equation}
{\cal L}_M=-\frac{1}{4}F_{\mu\nu}F^{\mu\nu}-A_\mu J^\mu\,,
\end{equation}
onde $F_{\mu\nu}=\partial_\mu A_\nu-\partial_\nu A_\mu$ é o tensor antisimétrico do campo eletromagnético e $J^\mu$ é a corrente de matéria. Esta lagrangiana é invariante frente a transformações de calibre $A_\mu\,\to\,A_\mu+\partial_\mu\Lambda$; e, consequentemente, as equações de Euler-Lagrange de movimento
\begin{equation}
 \partial_\mu F^{\mu\nu}=J^\nu
\end{equation}
também são invariantes frente a mesma transformação. A corrente de matéria se conserva, pois $\partial_\nu J^\nu=0$ devido a antisimetria de $F^{\mu\nu}$.  
\smallskip

Na teoria eletromagnética de Maxwell convencional, o tensor $F_{\mu\nu}$ é uma matriz quadrada antisimétrica de ordem $4\times4$. O número de campos nesta teoria é dado por $\dfrac{1}{2}D(D-1)$ que, em quatro dimensões, corresponde a três campos elétricos e três campos magnéticos. No entanto, esta teoria pode ser definida em {\it qualquer} dimensão, se os índices do campo de calibre $A_\mu$ variam de $\mu=0,1,2,...,D-1$, onde $D$ é a dimensão escolhida.
\smallskip

Em particular, em sistemas planares (duas dimensões espaciais e uma temporal), o campo magnético é dado por $B=\varepsilon^{ij}\partial_i A_j$, ou seja, este campo é um {\it escalar}. Isso acontece porque nesta teoria o potencial vetor é bidimensional, e o rotacional de um vetor em duas dimensões resulta em um escalar. Já o campo elétrico é um vetor espacial de duas componentes. Portanto, o número de campos em (2+1) dimensões é três, de acordo com a relação acima. 
\smallskip

Uma teoria que apresenta características distintas em relação à teoria de Maxwell simplesmente reduzida dimensionalmente é a {\it teoria de Chern-Simons} (CS), cuja lagrangiana é dada por
\begin{equation}
{\cal
L}_{CS}=\frac{\theta}{2}\varepsilon^{\mu\nu\rho}A_\mu\partial_\nu A_\rho-A_\mu J^\mu\,,
\end{equation}
onde $\theta$ \'e o par\^ametro de CS, cujo significado f\'isico ser\'a discutido mais adiante. O s\'imbolo $\varepsilon^{\mu\nu\rho}$ \'e o conhecido s\'imbolo de Levi-Civita (ou tensor completamente antisimétrico) em (2+1)$D$, cujas propriedades $\varepsilon^{012}=\varepsilon_{012}=1$ s\~ao comumente estabelecidas. Com a transforma\c{c}\~ao de calibre $A_\mu\,\to\,A_\mu+\partial_\mu\Lambda$, a lagrangiana (5.3) varia apenas por uma diverg\^encia:
\begin{equation}
{\cal L}_{CS}\,\,\to\,\,{\cal
L}_{CS}+\partial_{\rho}\left(\frac{\theta}{2}\varepsilon^{\mu\nu\rho}\partial_\nu A_\rho\Lambda\right)\,,
\end{equation}
mas a ação $\displaystyle{S=\int d^3x\,{\cal L}_{CS}}$ permanece invariante, pois os termos de superfície são desprezados.
\smallskip

As equações clássicas de Euler-Lagrange fornecem a seguinte equação de movimento
\begin{equation}
J^\mu=\frac{\theta}{2}\varepsilon^{\mu\nu\rho}F_{\nu\rho}\,,
\end{equation}
que tamb\'em goza da liberdade de calibre representada acima. Também é possível observar que a corrente se conserva devido a identidade de Bianchi: $\varepsilon^{\mu\nu\rho}\partial_\mu F_{\nu\rho}=0$. 
\smallskip

A teoria de CS pura apresenta características interessantes. As equações de Euler-Lagrange para esta teoria, em termos das densidades de carga e corrente, são dadas por
\begin{eqnarray}
\rho&=&\theta B\,,\nonumber\\
J^i&=& \theta\varepsilon^{ij}E_j\,.
\end{eqnarray}

A primeira indica que a densidade de carga \'e {\it localmente} proporcional ao campo magn\'etico, cuja constante de proporcionalidade \'e o par\^ametro de CS. Assim, o efeito produzido por este termo na teoria de CS pura \'e anexar fluxo magn\'etico \`a carga el\'etrica (tais part\'iculas s\~ao denominadas {\it \^anions}). Já a segunda relação diz que o campo elétrico é proporcional à corrente, cuja constante de proporcionalidade também é $\theta$. Estes efeitos estão em contraste com a teoria eletromagnética de Maxwell usual. 

\section{O propagador de Maxwell-Chern-Simons}

A lagrangiana desta teoria é descrita pelo acoplamento dos campos de Maxwell e CS\footnote{A leitura desta seção não é necessá para a compreensão do trabalho. Trata-se apenas de uma ilustração do modelo.}:
\begin{equation}
 {\cal L}_{MCS}=-\frac{1}{4}F_{\mu\nu}F^{\mu\nu}+\frac{\theta}{2}\varepsilon^{\mu\nu\rho}A_\mu\partial_\nu A_\rho-\dfrac{1}{2}\lambda\partial_\mu A^\mu\partial^\nu A_\nu\,,
\end{equation}
onde o último termo representa um termo de fixação de calibre para que o propagador possa ser determinado univocamente.
\smallskip

Assim, a a\c{c}\~ao desta teoria, em (2+1) dimens\~oes, agora \'e escrita na forma 
\begin{eqnarray}
S&=&\int d^3x\,\frac{1}{2}\left[\partial_\mu A^\mu\partial^\nu A_\nu-\partial_\mu A_\nu\partial^\nu A^\mu+\theta\varepsilon^{\mu\nu\rho}A_\mu \partial_\nu A_\rho-\lambda\partial_\mu A^\mu\partial^\nu A_\nu\right]\nonumber\\
&&\nonumber\\
&=&\int d^3x\,\frac{1}{2}\left[A^\mu g_{\mu\nu}\square A^\nu-A^\mu\partial_\nu\partial_\mu A^\nu+\theta\varepsilon_{\mu\nu\rho} A^\mu\partial^\rho A^\nu-\lambda A^\mu\partial^\mu\partial_\nu A^\nu\right]\nonumber\\
&&\nonumber\\
&=&\int d^3x\,\frac{1}{2}A^\mu\left[\square
g_{\mu\nu}-\partial_\nu\partial_\mu+\theta\varepsilon_{\mu\nu\rho}\partial^\rho+\lambda\partial_\mu\partial_\nu\right]
A^\nu\,,
\end{eqnarray}
onde os termos divergentes s\~ao negligenciados de acordo com o teorema de Gauss.
\smallskip

Sendo o termo entre colchetes o núcleo da ação e o propagador de Feynman uma fun\c{c}\~ao de Green, o mesmo pode ser calculado atrav\'es da identidade:
\begin{eqnarray}
\left(-\square
g_{\mu\nu}+\partial_\nu\partial_\mu-\theta\varepsilon_{\mu\nu\rho}\partial^\rho-\lambda\partial_\mu\partial_\nu\right)\Delta_F^{\nu\sigma}(x-y)=i\delta^\sigma_\mu\,\delta^3(x-y)\,.
\end{eqnarray}

Se aplicamos a transformada de Fourier sobre o propagador
\begin{eqnarray*}
 \Delta_F^{\mu\nu}(x-y)=\int\frac{d^4k}{(2\pi)^4}\Delta_F^{\mu\nu}(k)e^{ik\cdot(x-y)}\,,
\end{eqnarray*}
observamos que
\begin{eqnarray*}
 \partial_\mu\Delta_F^{\mu\nu}(x-y)=\int\frac{d^4k}{(2\pi)^4}\Delta_F^{\mu\nu}(k)(ik_\mu)e^{ik\cdot(x-y)}\,,
\end{eqnarray*}
e obtemos a expressão
\begin{equation}
\left(-k^2g_{\mu\nu}+k_\mu k_\nu-i\theta\varepsilon_{\mu\nu\rho}k^\rho-\lambda k_\mu k_\nu
\right)\Delta_F^{\nu\sigma}(k)=i\delta^\sigma_\mu\,.
\end{equation}

Considerando, para o c\'alculo de (5.10), o {\it ansatz} geral
\begin{equation}
\Delta_F^{\nu\sigma}(k)={\cal A}g^{\nu\sigma}+{\cal B}k^\nu k^\sigma+{\cal C}\varepsilon^{\nu\sigma\tau}k_\tau\,,
\end{equation}
segue que

\begin{eqnarray*}
&-&{\cal A}k^2\delta^\sigma_\mu-{\cal B}k^2k_\mu k^\sigma-{\cal C}k^2g_{\mu\nu}\varepsilon^{\nu\sigma\tau}k_\tau+{\cal A}k_\mu k^\sigma+{\cal B}k^2k_\mu k^\sigma+{\cal C}\varepsilon^{\nu\sigma\tau}k_\mu k_\nu k_\tau-i{\cal A}\theta\varepsilon_{\mu\nu\rho}g^{\nu\sigma}k^\rho\\
&&\\
&-&i{\cal B}\theta\varepsilon_{\mu\nu\rho}k^\rho k^\nu
k^\sigma-i{\cal
C}\theta\varepsilon_{\mu\nu\rho}\varepsilon^{\nu\sigma\tau}k^\rho k_\tau-{\cal A}\lambda k_\mu k^\sigma-{\cal B}\lambda k^2k_\mu
k^\sigma-{\cal C}\lambda\varepsilon^{\nu\sigma\tau}k_\mu k_\nu k_\tau=i\delta^\sigma_\mu\,.
\end{eqnarray*}

Observe o segundo termo da segunda linha na express\~ao acima. Utilizando a identidade
\begin{equation}
\varepsilon_{\mu\nu\rho}\varepsilon^{\nu\sigma\tau}=-\delta^\sigma_\mu
\delta^\tau_\rho+\delta^\sigma_\rho \delta^\tau_\mu \,,
\end{equation}
este termo fica
\begin{eqnarray*}
-i{\cal C}\theta\varepsilon_{\mu\nu\rho}\varepsilon^{\nu\sigma\tau}k^\rho k_\tau&=&-i{\cal C}\theta(-\delta^\sigma_\mu\delta^\tau_\rho+\delta^\sigma_\rho\delta^\tau_\mu)k^\rho k_\tau\\
&=&i{\cal C}\theta(k^2\delta^\sigma_\mu-k_\mu k^\sigma)\,.
\end{eqnarray*}

Assim, podemos identificar um sistema de tr\^es equa\c{c}\~oes com tr\^es inc\'ognitas ${\cal A}$, ${\cal B}$ e ${\cal C}$ de acordo com a forma do tensor:
\begin{eqnarray*}
&&[-{\cal A}k^2+i{\cal C}\theta k^2]\delta^\sigma_\mu=i\delta^\sigma_\mu\\
&&\nonumber\\
&&[{\cal A}(1-\lambda)-{\cal B}\lambda k^2-i{\cal C}\theta]k_\mu k^\sigma=0\\
&&\nonumber\\
&&(-i{\cal A}\theta-{\cal C}k^2)g_{\mu\nu}\varepsilon^{\nu\sigma\tau}k_\tau=0\,,
\end{eqnarray*}
que, resolvido, fornece
\begin{eqnarray*}
{\cal A}&=&-\dfrac{i}{k^2-\theta^2}\\
&&\nonumber\\
{\cal B}&=&\dfrac{i}{k^2(k^2-\theta^2)}-\dfrac{1}{\lambda}\dfrac{i}{k^4}\\
&&\nonumber\\
{\cal C}&=&-\dfrac{\theta}{k^4-k^2\theta^2}\,.
\end{eqnarray*}

Portanto, de acordo com (5.11) e os resultados acima, o propagador do f\'oton para a teoria de MCS em (2+1) dimens\~oes \'e escrito da seguinte maneira:
\begin{equation}
\Delta_F^{\mu\nu}(k)=-\frac{ig^{\mu\nu}}{k^2-\theta^2}+\frac{ik^\mu
k^\nu}{k^2(k^2-\theta^2)}-\frac{\theta\varepsilon^{\mu\nu\rho}k_\rho}{k^2(k^2-\theta^2)}-\dfrac{1}{\lambda}\dfrac{ik^\mu
k^\nu}{k^4}\,.
\end{equation}

Nessa express\~ao para o propagador, de modo semelhante ao que ocorre em campos massivos, \'e evidente a exist\^encia de um polo em $\theta=\sqrt{k^2}$, que representa uma massa. Note tamb\'em que, para o caso de $\lambda\to\infty$, temos o calibre de Landau, que fornece a condi\c{c}\~ao de transversalidade similar a do propagador do fóton usual. 

\section{Indução de Chern-Simons em (2+1)$D$}

A EDQ possibilita o estudo de sistemas que envolvem interações entre férmions e um campo de calibre externo. No entanto, é interessante observar que, mesmo partindo de uma teoria de interação de férmions de massa $m$ com um campo vetorial $A_\mu$ sem a presença de um termo de CS na dinâmica deste campo, um termo deste tipo é induzido por correções radiativas. Calcularemos este termo induzido, utilizando o formalismo de integrais de trajetórias para o campo de Dirac e expandindo o resultado até termos de segunda ordem. Assim, a lagrangiana que indica tal acoplamento é dada por:

\begin{equation}
 {\cal L}=\bar{\psi}(i\!{\not\!\partial}-e\not\!\!A-m)\psi\,.
\end{equation}

A ação efetiva para este modelo dependente do campo $A_\mu$, na aproximação de um laço \cite{Tqc}, é definida como se segue
\begin{equation}
 e^{iS_{ef}(A)}=N\int D\bar{\psi}D\psi\,exp\left[i\int d^3x\,\bar{\psi}(i\!{\not\!\partial}-e\not\!\!A-m)\psi\right]\,,
\end{equation}
onde $N$ é uma constante de normalização.
\smallskip

Integrando sobre os campos de férmions\footnote{Quando calculamos a integral gaussiana em variáveis ordinárias, o determinante surge no denominador. No entanto, em gaussianas de variáveis anticomutantes (variáveis de Grassmann), o determinante aparece no numerador. Esta demonstração está no Apêndice E.}, obtemos
\begin{equation}
 e^{iS_{ef}(A)}=N\,\mbox{det}\,(i\!{\not\!\partial}-e\not\!\!A-m)\,,
\end{equation}
ou seja\footnote{O Apêndice F contém a demonstração desta identidade.},
\begin{eqnarray}
 S_{ef}(A)&=&-i\ln\mbox{det}\,[i\!{\not\!\partial}-e\not\!\!A-m]\nonumber\\
&=&-iTr\,\ln[i\!{\not\!\partial}-e\not\!\!A-m]\,,
\end{eqnarray}
onde o termo constante foi negligenciado.

A expressão acima é expandida\footnote{$\displaystyle \ln(1-x)=-\sum\limits_{n=1}^\infty\frac{x^n}{n}\,.$} para se obter
\begin{eqnarray}
 S_{ef}[A,m]&=&-iTr\ln\left[(i\!{\not\!\partial}-m)\left(\frac{i\!{\not\!\partial}-e\not\!\!A-m}{i\!{\not\!\partial}-m} \right) \right]\nonumber\\
&&\nonumber\\
&=&-iTr\ln(i\!{\not\!\partial}-m)-iTr\ln\left[1- \frac{1}{i\!{\not\!\partial}-m}e\not\!\!A \right]\nonumber\\
&&\nonumber\\
&=&-iTr\ln(i\!{\not\!\partial}-m)+iTr\sum\limits_{n=1}^\infty\frac{1}{n}\left[ \frac{1}{i\!{\not\!\partial}-m}e\not\!\!A \right]^n\,.
\end{eqnarray}

O termo da expansão acima que dará origem ao termo tipo CS é o de segunda ordem, que fornece a contribuição bilinear em $A_\mu$. Então, temos:
\begin{equation}
 S_{ef}^{(2)}=\frac{ie^2}{2}Tr\left[\frac{1}{i\!{\not\!\partial}-m}\not\!\!A \frac{1}{i\!{\not\!\partial}-m}\not\!\!A \right]\,.
\end{equation}

Para o cálculo do traço da ação acima, considere ${\cal O}$ um operador que depende das matrizes de Dirac e dos índices internos do grupo de Lie. Então, seu traço total $Tr$ é definido por:
\begin{equation}
 Tr{\cal O}\,\dot{=}\,tr\,tr_D\int d^3x\langle x|{\cal O}|x'\rangle\bigg|_{x=x'}\,,
\end{equation}
onde o símbolo $tr_D$ indica que o traço será calculado sobre as matrizes $\gamma$ de Dirac.
\smallskip

Inserindo as relações de completeza ou fechamento nos espaços das posições e momentos
\begin{equation}
 \int d^3x|x\rangle\langle x|=1\quad,\quad  \int\frac{d^3p}{(2\pi)^3}|p\rangle\langle p|=1\,,
\end{equation}
onde $\langle x|p\rangle=\langle p|x\rangle^\ast=e^{ipx}$, vem que

\begin{eqnarray*}
 S_{ef}^{(2)}&=& \frac{ie^2}{2}tr\,tr_D \!\int\! d^3x \!\int\! d^3y \!\int\!\frac{d^3p}{(2\pi)^3} \!\int\!\frac{d^3q}{(2\pi)^3} \langle x|\frac{1}{i\!{\not\!\partial}-m}|p\rangle\langle p|\not\!\!A|y\rangle   \langle y|\frac{1}{i\!{\not\!\partial}-m}|q\rangle\langle q|\not\!\!A|x\rangle \\
&&\\
&=&\frac{ie^2}{2}tr\,tr_D \!\int\! d^3x \!\int\! d^3y \!\int\!\frac{d^3p}{(2\pi)^3} \!\int\!\frac{d^3q}{(2\pi)^3}\frac{1}{\not\!p-m}\not\!\!A(y)  \frac{1}{\not\!q-m}\not\!\!A(x)e^{ipx-ipy+iqy-iqx}\\
&&\\
&=& \frac{ie^2}{2}tr\,tr_D \!\int\! d^3x \!\int\! d^3y \!\int\!\frac{d^3p}{(2\pi)^3} \!\int\!\frac{d^3q}{(2\pi)^3}\frac{(\not\!p+m)\not\!\!A(y)(\not\!q+m)\not\!\!A(x)}{(p^ 2-m^2)(q^2-m^2)}e^{i(p-q)(x-y)}\\
&&\\
&=&-\frac{ie^2}{2}tr\,tr_D\int\frac{d^3k}{(2\pi)^3} \int\frac{d^3p}{(2\pi)^3}\frac{(\not\!p+m)\not\!\!A(-k)(\not\!p+\not\!k+m)\not\!\!A(k)}{[(p+k)^2-m^2](p^2-m^2)}\,,
\end{eqnarray*}
onde foi feita a mudança de variável $p-q\,\to\,-k$.
\smallskip

Vamos escrever explicitamente os termos que aparecem no numerador da integral acima:
\begin{eqnarray*}
&& (\not\!p+m)\not\!\!A(\not\!p+\not\!k+m)\not\!\!A=(\not\!p\not\!\!A)   (\not\!p\not\!\!A+\not\!k\not\!\!A+m\not\!\!A)\\
&=&\not\!p\not\!\!A\not\!p\not\!\!A+\not\!p\not\!\!A\not\!k\not\!\!A+m\not\!p\not\!\!A\not\!\!A+m\not\!\!A\not\!p\not\!\!A+m\not\!\!A\not\!k\not\!\!A+m^2\not\!\!A\not\!\!A
\end{eqnarray*}

O único termo que contribuirá com nossos cálculos é $m\!{\not\!\!A\!{\not\!k\!{\not\!\!A}}}$, uma vez que $m\!{\not\!\!A\!{\not\!p\!{\not\!\!A}}}$ anulará a integral em $p$. Assim,
\begin{eqnarray*}
 S_{CS}=-\frac{ie^2}{2}mtr_D\,tr\int\frac{d^3k}{(2\pi)^3}\int\frac{d^3p}{(2\pi)^3}\frac{\not\!\!A\not\!k\not\!\!A}{[(p+k)^2-m^2](p^2-m^2)}\,.
\end{eqnarray*}

Utilizando a parametrização de Feynman
\begin{equation}
\frac{1}{ab}=\int_0^1dz\frac{1}{[az+b(1-z)]^2}\,,
\end{equation}
onde $a=(k+p)^2-m^2$ e $b=p^2-m^2$, podemos escrever
\begin{equation}
az+b(1-z)=p^2+2(k\cdot p)z+k^2z-m^2=(p+kz)^2+k^2z(1-z)-m^2\,.
\end{equation}

Com a mudan\c{c}a de vari\'avel $p\,\to\,p-kz$, e considerando $\mu^2=m^2-k^2z(1-z)$ e a utilização da integral no momento $p$ (vide Apêndice G), temos que

\begin{eqnarray}
 S_{CS}&=&=-\frac{ie^2}{2}mtr_D\,tr\int\frac{d^3k}{(2\pi)^3}\not\!\!A(-k)\not\!k\not\!\!A(k)\int\limits_0^1dz\int\frac{d^3p}{(2\pi)^3}\frac{1}{(p^2-\mu^2)^2}\nonumber\\
&&\nonumber\\
&=&-\frac{ie^2}{2}mtr_D\,tr\int d^3x\int d^3y\int\frac{d^3k}{(2\pi)^3}\not\!\!A(y)(-i\!{\not\!\partial_x})\not\!\!A(x)e^{-ik(x-y)}\int\limits_0^1dz\frac{i}{8\pi|\mu|}\nonumber\\
&&\nonumber\\
&=&\frac{ie^2}{16\pi}m\,tr\,tr_D\int d^3x\int d^3y\not\!\!A(y)\!{\not\!\partial_x}\not\!\!A(x)g(x-y)\,, 
\end{eqnarray}
onde\footnote{Integral \cite{Gra} sobre o parâmetro $z$: \vspace{0.35cm} \\ $ \displaystyle \int\frac{dz}{|\mu|}=-\frac{1}{|k|}\arcsin\left[ \frac{(2z-1)|k|}{\sqrt{4m^2-k^2}} \right]\hspace{0.7cm},\hspace{0.7cm}\mu^2=m^2-k^2z(1-z)\,.$}

\begin{eqnarray*}
 g(x-y)=2\int\frac{d^3k}{(2\pi)^3}\frac{1}{|k|}\arcsin\left(\frac{|k|}{\sqrt{4m^2-k^2}}  \right)e^{-ik(x-y)}\,.
\end{eqnarray*}

Para extrairmos o termo local de CS induzido na ação acima,, expandimos o integrando de $g(x-y)$ em torno de $k\,\to\,
0$. Assim,

\begin{equation}
 g(x-y)=\frac{1}{|m|}\delta^{(3)}(x-y)\,.
\end{equation}

Substituindo este resultado em (5.24), obtemos

\begin{eqnarray*}
 S_{CS}&=&\frac{ie^2}{16\pi}\frac{m}{|m|}\,tr\,tr_D\int d^3x\int d^3y\not\!\!A(y)\!{\not\!\partial_x}\not\!\!A(x)\delta(x-y)\\
&&\\
&=&\frac{ie^2}{16\pi}\frac{m}{|m|}\,tr\,tr_D\int d^3x\,tr_D[\gamma^\mu\gamma^\nu\gamma^\rho]A_\mu\partial_\nu A_\rho\,.
\end{eqnarray*}

Portanto, aplicando o traço às matrizes de Dirac (veja (A.5)), obtemos a ação induzida

\begin{equation}
 S_{CS}^{(2+1)D}=-\frac{e^2}{8\pi}\frac{m}{|m|}tr\int d^3x\,\varepsilon^{\mu\nu\rho}A_\mu\partial_\nu A_\rho\,.
\end{equation}
\vspace{0.01cm}

Este resultado é a contribuição bilinear (abeliana) no campo de calibre encontrada em \cite{Dun}. Ele indica que a interação de férmions com um campo de calibre formulado em (2+1) dimensões gera um termo semelhante ao de Chern-Simons. Como foi discutido na seção 5.2, é possível afirmar então que os quanta deste campo de calibre podem ser massivos.

\vspace{0.15cm}
\begin{center}
\vspace{1.2cm}
\hspace{0.75cm}\begin{picture}(250,80)
\Photon(0,50)(80,50){5}{8}\Text(40,67)[]{$k$}\Text(88,50)[]{$\mu$}
\ArrowArcn(110,50)(30,180,0)\Text(110,92)[]{$p+k$}
\ArrowArcn(110,50)(30,360,180)\Text(110,8)[]{$p$}
\Photon(140,50)(220,50){5}{8}\Text(180,67)[]{$k$}\Text(132,50)[]{$\nu$}
\end{picture}\\ {\sl \footnotesize \hspace{2eM}Diagrama de Feynman de um laço de férmions utilizado para o cálculo da ação (5.26).}
\end{center}

\chapter{Correções Radiativas com Quebra da Simetria de Lorentz em (3+1)$D$}

\section{Expansão do propagador do férmion e a quebra de simetria de Lorentz}

Estudamos nos Capítulos 3 e 4 a influência dos termos de quebra de simetria de Lorentz sobre a equação de Dirac e todas as grandezas dela derivadas. Aplicamos a ideia sugerida por Kostelecký e Colladay, que corresponde em tratar os termos de quebra de simetria de Lorentz como perturbações na equação de Dirac. Aplicando teoria de perturbações, obtemos como resultados correções de energias em certos sistemas quânticos na presença de termos de quebra da simetria de Lorentz.
\smallskip

Além dessa aplicação à mecânica quântica, a indução do termo de CS (Chern-Simons) em (3+1) dimensões é um resultado interessante oferecido pela teoria de quebra de simetria de Lorentz. Neste Capítulo, procuraremos por essa indução calculando correções radiativas em um sistema de férmions acoplado a um campo de calibre no espaço quadrimensional. Iniciaremos essa busca reescrevendo o propagador de Feynman na presença da quebra da simetria de Lorentz.
\smallskip

No Cap\'itulo 2, encontramos o propagador fermi\^onico da EDQ usual $S(p)$ (veja (2.43)). No cap\'itulo seguinte, formulamos um propagador modificado pela quebra da simetria de Lorentz para o f\'ermion, representado nesta dissertação por (3.23). Neste capítulo, estudaremos correções radiativas geradas pela corrente quiral $\bar{\psi}\gamma^\mu\gamma_5\psi$ acoplada a um quadrivetor $b_\mu$, que acarreta numa violação da invariância de Lorentz, tal como foi apresentado no Capítulo 3. Assim, tomando $\not\!\!a=0$ no propagador (3.23), a quebra da simetria de Lorentz será representada pelo termo $\not\!b\!{\gamma_5}$:
\begin{equation}
S_b(p)=\frac{i}{\not\!p-m-\not\!b\gamma_5}\,.
\end{equation}

O propagador acima invertido é dado por
\begin{equation}
 S_{b}(p)=\frac{i(\not\!p-\not\!b\gamma_5+m)\{ p^2-b^2-m^2+[\not\! p,\not
b]\gamma_5 \}}{(p^2-b^2-m^2)^2-4(p\cdot b)^2+4p^2b^2}\,.
\end{equation}

Este propagador tem uma estrutura complicada e tornaria os cálculos perturbativos mais tediosos. Entretanto, o quadrivetor $b^\mu$, que sinaliza uma possível quebra da simetria de Lorentz na teoria, é muito pequeno e a correção por ele produzida no propagador pode ser tratada em forma perturbativa. Apliquemos assim a seguinte expansão em uma soma de termos infinitos de uma progress\~ao geom\'etrica cuja raz\~ao \'e $\not\!b\!{\gamma_5}$  
\begin{eqnarray}
S_b(p)&=&\frac{i}{\not\!p-m}\frac{1}{1-i\not\!b\gamma_5S(p)}\nonumber\\
&&\nonumber\\
&=&\frac{i}{\not\!p-m}\left[1+(-i\not\!b\gamma_5)\frac{i}{\not\!p-m}+(-i\not\!b\gamma_5)\frac{i}{\not\!p-m}(-i\not\!b\gamma_5)\frac{i}{\not\!p-m}+\cdots\right]\nonumber\\
&&\nonumber\\
&=&\frac{i}{\not\!p-m}+\frac{i}{\not\!p-m}(-i\not\!b\gamma_5)\frac{i}{\not\!p-m}+\frac{i}{\not\!p-m}(-i\not\!b\gamma_5)\frac{i}{\not\!p-m}(-i\not\!b\gamma_5)\frac{i}{\not\!p-m}+\cdots\nonumber\\
&&
\end{eqnarray}

Assim, com $\times$ representando cada inserção $-i\!{\not\!b\!{\gamma_5}}$ no propagador, seu gráfico é dado por
\medskip

\begin{picture}(150,50)
 
\ArrowLine(49,19)(51,19)
\Line(0,18)(100,18)\Text(120,20)[]{=}
\Line(0,20)(100,20)

\ArrowLine(140,20)(240,20)\Text(260,20)[]{+}

\ArrowLine(280,20)(330,20)\Text(330,20)[]{$\times$} 
\ArrowLine(330,20)(380,20)\Text(407,20)[]{+}

\ArrowLine(140,-20)(174,-20)\Text(174,-20)[]{$\times$} 
\ArrowLine(174,-20)(208,-20)\Text(208,-20)[]{$\times$}
\ArrowLine(208,-20)(242,-20)\Text(273,-20)[]{+\hspace{1eM}$\cdots$}
\end{picture}
\vspace{1.6cm}

\noindent onde o lado esquerdo representa $S_b(p)$ e os termos do lado direito ao propagador de Feynman expandido. Na próxima seção, vamos calcular as correções radiativas utilizando este propagador expandido.

\section{Indução de Chern-Simons pela quebra da simetria de Lorentz}

As correções radiativas à ação da EDQ usual serão calculadas considerando-se um sistema de férmions acoplados a um campo $A_\mu$ formulado no espaço-tempo em (3+1) dimensões, empregando-se a perturbação $\not\!b\gamma_5$.  Assim, a lagrangiana deste modelo é dada por
\begin{equation}
 {\cal L}=\bar{\psi}(i\!{\not\!\partial}-\not\!b\gamma_5-e\!{\not\!\!A} -m)\psi\,.
\end{equation}

Para calcular o termo induzido, seguiremos os mesmos passos utilizados no Capítulo 5. A ação efetiva para este modelo, dependente do termo de quebra de simetria de Lorentz $-\bar{\psi}\not\!b\gamma_5\psi$, na aproximação de um laço é definida como se segue
\begin{equation}
 e^{iS_{ef}[b,m]}=N\int D\bar{\psi}D\psi\,exp\left[i\int d^4x\,\bar{\psi}(i\!{\not\!\partial}-e\not\!\!A-\not\!b\gamma_5-m)\psi\right]\,,
\end{equation}
onde $N$ é uma constante de normalização.
\smallskip

Com a utilização das variáveis de Grassmann, integrando sobre os campos de férmions, obtém-se
\begin{equation}
 e^{iS_{ef}[b,m]}=N\,det(i\!{\not\!\partial}-e\not\!\!A-\not\!b\gamma_5-m)\,,
\end{equation}
ou seja,
\begin{equation}
 S_{ef}[b,m]=-iTr\ln[i\!{\not\!\partial}-e\not\!\!A-\not\!b\gamma_5-m]\,.
\end{equation}

Sendo $A$ e $B$ duas matrizes que não comutam, obtemos a seguinte identidade:
\begin{eqnarray*}
 \ln(B-A)&=&\ln B\left(1-\frac{1}{B}A  \right)\\
&&\\
&=&\ln B+\ln\left(1-\frac{1}{B}A\right)\\
&&\\
&=&\ln B-\frac{1}{B}A-\frac{1}{2}\frac{1}{B}A\frac{1}{B}A-\frac{1}{3}\frac{1}{B}A\frac{1}{B}A\frac{1}{B}A+\cdots\\
&&\\
&=&\ln B-\sum\limits_n \frac{1}{n}\left[\frac{1}{B}A\right]^n\,.
\end{eqnarray*}

Sendo $A=e\not\!\!A$ e $B=i\!{\not\!\partial}-\not\!b\gamma_5-m$ na expressão (6.7), temos que
\begin{eqnarray}
 S_{ef}[b,m]=-iTr\ln[i\!{\not\!\partial}-\not\!b\gamma_5-m]+iTr\sum\limits_{n=1}^{\infty}\frac{1}{n}\left[ \frac{1}{i\!{\not\!\partial}-\not\!b\gamma_5-m}e\not\!\!A  \right]^n\,.
\end{eqnarray}

O primeiro termo da expansão acima corresponde a um termo constante adicionado à ação e, por isso, não tem importância, uma vez que não depende do campo de calibre $\not\!\!A$. As contribuições virão de termos para $n\geqslant1$. Assim, para $n=1$ na expansão acima, temos que

\begin{equation}
  S_{ef}^{(1)}=ieTr\left[\frac{1}{i\!{\not\!\partial}-\not\!b\gamma_5-m}\not\!\!A  \right]
\end{equation}
ou
\begin{equation}
 S_{ef}^{(1)}=ie\,Tr\int d^4x\int\frac{d^4p}{(2\pi)^4}\frac{1}{\not\!p-\not\!b\gamma_5-m}\not\!\!A\,.
\end{equation}

As contribuições de (6.10) dão origem a termos do tipo {\it tadpoles}, que são lineares em $A_\mu$ e são divergentes no ultravioleta. Como esses termos não contribuem com a indução de CS, os mesmos serão desconsiderados nos cálculos a seguir. Entretanto, ilustramos seus gráficos em primeira ordem no campo de calibre.

\hspace{-1.1cm} \begin{picture}(500,250)(0,0)
\vspace{-3cm}
  \Photon(0,200)(80,200){3}{4} \BCirc(100,200){20} \BCirc(100,200){18.3} \ArrowLine(119,201)(119,199) \Text(140,200)[]{=}

\Photon(160,200)(240,200){3}{4} \BCirc(260,200){20} \ArrowLine(280,201)(280,199) \Text(300,200)[]{+}

\Photon(320,200)(400,200){3}{4} \ArrowArcn(420,200)(20,180,0) \ArrowArcn(420,200)(20,360,180)\Text(441,200)[]{$\times$}
\Text(460,200)[]{+}

\Photon(60,100)(140,100){3}{4}\ArrowArcn(160,100)(20,180,90) \ArrowArcn(160,100)(20,90,270) \ArrowArcn(160,100)(20,270,180)
\Text(161,120)[]{$\times$}  \Text(161,80)[]{$\times$} \Text(216,100)[]{+}

\Photon(250,100)(330,100){3}{4} \ArrowArcn(350,100)(20,180,90)   \ArrowArcn(350,100)(20,90,0)   \ArrowArcn(350,100)(20,0,270)  \ArrowArcn(350,100)(20,270,180)  
 
\Text(350,120)[]{$\times$} \Text(371,100)[]{$\times$}  \Text(350,80)[]{$\times$}   \Text(420,100)[]{+\hspace{3eM}$\cdots$} 
\end{picture}\\
\vspace{-3cm}

{\footnotesize Contribuições em primeira ordem de aproximação no campo de calibre denominadas \textit{tadpoles}. O gráfico do lado esquerdo refere-se a (6.10).}

 As contribuições em segunda ordem, ou seja, para $n=2$ fornecem
\begin{eqnarray}
S_{ef}^{(2)}&=&\frac{ie^2}{2}Tr\left[\frac{1}{i\!{\not\!\partial}-\not\!b\gamma_5-m}\not\!\!A  \frac{1}{i\!{\not\!\partial}-\not\!b\gamma_5-m}\not\!\!A \right]\nonumber\\
&&\nonumber\\
&=&-\frac{ie^2}{2}Tr[S_b(p)\not\!\!AS_b(p)\not\!\!A]\,.
\end{eqnarray}

O cálculo do traço da ação acima é semelhante ao cálculo de um operador ${\cal O}$ que depende das matrizes de Dirac e dos índices internos do grupo de Lie. Então, seu traço total $Tr$ é idêntico ao calculado no Capítulo 5 mas, em $(3+1)$ dimensões, é definido por:

\begin{equation}
 Tr{\cal O}\,\dot{=}\,tr\,tr_D\int d^4x\langle x|{\cal O}|x'\rangle\bigg|_{x=x'}\,,
\end{equation}
onde o símbolo $tr_D$ indica que o traço será calculado sobre as matrizes $\gamma$ de Dirac na representação padrão.
\smallskip

Inserindo conjuntos completos de operadores normalizados nos espaços das posições e momentos
\begin{equation}
 \int d^4x|x\rangle\langle x|=1\quad,\quad  \int\frac{d^4p}{(2\pi)^4}|p\rangle\langle p|=1\,,
\end{equation}
onde, novamente, $\langle x|p\rangle=\langle p|x\rangle^\ast=e^{ipx}$, vem que
\begin{eqnarray*}
 S_{ef}^{(2)}&=& \frac{ie^2}{2}tr\,tr_D \!\int\! d^4x \!\int\! d^4y \!\int\!\frac{d^4p}{(2\pi)^4} \!\int\!\frac{d^4q}{(2\pi)^4}\\
&&\\
&&\times\langle x|\frac{1}{i\!{\not\!\partial}-\not\!b\gamma_5-m}|p\rangle\langle p|\not\!\!A|y\rangle   \langle y|\frac{1}{i\!{\not\!\partial}-\not\!b\gamma_5-m}|q\rangle\langle q|\not\!\!A|x\rangle \\
&&\\
&=&\frac{ie^2}{2}tr\,tr_D \!\int\! d^4x \!\int\! d^4y \!\int\!\frac{d^4p}{(2\pi)^4} \!\int\!\frac{d^4q}{(2\pi)^4}\\
&&\\
&&\times\frac{1}{\not\!p-\not\!b\gamma_5-m}\not\!\!A(y)\frac{1}{\not\!q-\not\!b\gamma_5-m}\not\!\!A(x)\langle x|p\rangle \langle p|y\rangle\langle y|q\rangle\langle q|x\rangle\\
&&\\
&=& \frac{ie^2}{2}tr\,tr_D \!\int\! d^4x \!\int\!\frac{d^4p}{(2\pi)^4}\frac{1}{\not\!p-\not\!b\gamma_5-m}\!\int\! d^4y\not\!\!A(y) \!\int\!\frac{d^4k}{(2\pi)^4}e^{ik(x-y)}\frac{1}{\not\!p-i\!{\not\!\partial}-\not\!b\gamma_5-m}\not\!\!A(x)\\
&&\\
&=& \frac{ie^2}{2}tr\,tr_D \!\int\! d^4x \!\int\!\frac{d^4p}{(2\pi)^4}\frac{1}{\not\!p-\not\!b\gamma_5-m}\!\int\! d^4y\not\!\!A(y)\delta(x-y)\frac{1}{\not\!p-i\!{\not\!\partial}-\not\!b\gamma_5-m}\not\!\!A(x)\\
&&\\
&=&\frac{ie^2}{2}tr\,tr_D \!\int\! d^4x \!\int\!\frac{d^4p}{(2\pi)^4}\frac{1}{\not\!p-\not\!b\gamma_5-m}\not\!\!A(x)\frac{1}{\not\!p-i\!{\not\!\partial}-\not\!b\gamma_5-m}\not\!\!A(x)\,,
\end{eqnarray*}
onde, com a utilização da mudança de variável $p-q\,\to\,k$ e do propagador (6.1), obtemos
\begin{equation}
 S_{ef}^{(2)}=\frac{ie^2}{2}tr\,tr_D \!\int\! d^4x \!\int\!\frac{d^4p}{(2\pi)^4}S_b(p)\not\!\!AS_b(p-i\!{\not\partial})\not\!\!A\,.
\end{equation}

\vspace*{2.3cm}

\hspace{-1.5cm} \begin{picture}(500,250)(0,0)

\Photon(0,300)(80,300){3}{5} \BCirc(105,300){25} \BCirc(105,300){23} \ArrowLine(104,324)(106,324) \ArrowLine(106,276.2)(104,276.2)
\Photon(130,300)(210,300){3}{5} \Text(225,300)[]{=}

\Photon(240,300)(320,300){3}{5}\ArrowArcn(345,300)(25,180,0) \ArrowArcn(345,300)(25,360,180)\Photon(370,300)(450,300){3}{5}
\Text(470,300)[]{+}

\Photon(0,200)(80,200){3}{5} \ArrowArcn(105,200)(25,180,90) \ArrowArcn(105,200)(25,90,0) \ArrowArcn(105,200)(25,0,180)
\Photon(130,200)(210,200){3}{5} \Text(105,226)[]{$\times$} \Text(225,200)[]{+}

\Photon(240,200)(320,200){3}{5} \ArrowArcn(345,200)(25,180,0) \ArrowArcn(345,200)(25,0,270) \ArrowArcn(345,200)(25,270,180)
\Text(345,175)[]{$\times$} \Photon(370,200)(450,200){3}{5} \Text(470,200)[]{+}

\Photon(0,100)(80,100){3}{5} \ArrowArcn(105,100)(25,180,90) \ArrowArcn(105,100)(25,90,0) \ArrowArcn(105,100)(25,0,270) \ArrowArcn(105,100)(25,270,180) \Text(105,125)[]{$\times$} \Photon(130,100)(210,100){3}{5} \Text(105,75)[]{$\times$}   \Text(225,100)[]{+}

\Photon(240,100)(320,100){3}{5} \ArrowArcn(345,100)(25,180,120) \ArrowArcn(345,100)(25,120,60) \ArrowArcn(345,100)(25,60,0) \ArrowArcn(345,100)(25,0,180)  \Photon(370,100)(450,100){3}{5}  \Text(470,100)[]{+}  \Text(335,122)[]{$\times$} \Text(357.8,122)[]{$\times$}         

\Photon(0,0)(80,0){3}{5} \ArrowArcn(105,0)(25,180,120) \ArrowArcn(105,0)(25,120,60) \ArrowArcn(105,0)(25,60,0) \ArrowArcn(105,0)(25,0,180)  \Photon(130,0)(210,0){3}{5}  \Text(245,0)[]{+\hspace{2eM}$\cdots$}  \Text(95,-22)[]{$\times$} \Text(117.8,-22)[]{$\times$}

\end{picture}

\vspace{1.75cm}

\begin{center}
 \footnotesize Diagramas de Feynman na aproximação de um laço, em termos bilineares no campo de calibre, para o cálculo da ação (6.14) em termos do propagador fermiônico expandido. Cada termo $\times$ indica uma inserção $\not\!b\gamma_5$ no propagador do férmion.
\end{center}

As correções radiativas serão obtidas introduzindo-se os propagadores do férmion expandido (6.2). Assim, temos que
\begin{eqnarray}
 S_b(p)&=&\frac{i}{\not\!p-m}+\frac{i}{\not\!p-m}\not\!b\gamma_5\frac{1}{\not\!p-m}+\frac{1}{\not\!p-m}\not\!b\gamma_5\frac{1}{\not\!p-m}\not\!b\gamma_5\nonumber\\
&&\nonumber\\
&=&\frac{i(\not\!p+m)}{p^2-m^2}+\frac{i}{(p^2-m^2)^2}(\not\!p+m) \not\!b\gamma_5  (\not\!p+m)\nonumber\\
&&\nonumber\\
 &+&\frac{1}{(p^2-m^2)^3}(\not\!p+m) \not\!b\gamma_5  (\not\!p+m) \not\!b\gamma_5 (\not\!p+m) +\cdots
\end{eqnarray}

Utilizando as notações
\begin{eqnarray}
P&=&\not\!p+m\\
B_5&=&\not\!b\gamma_5\\
{\cal D}^2&=&p^2-m^2
\end{eqnarray}
o propagador é escrito em forma compacta até termos lineares em $B_5$:
\begin{equation}
 S_b(p)=\frac{i}{{\cal D}^2}P+\frac{i}{{\cal D}^4}PB_5P+\cdots
\end{equation}

Agora, expandindo o propagador $S_b(p-i\partial)$ em termos de $\tilde{B}_5=\not\!b\gamma_5+i\!{\not\!\partial}$, expandimos até termos lineares em $\tilde{B}_5$
\begin{equation}
 S_b(p-i\partial)=\frac{i}{{\cal D}^2}P+\frac{i}{{\cal D}^4}P\tilde{B}_5P+\cdots
\end{equation}

Desenvolvendo o integrando de (6.14) em termos de (6.16) e (6.17), temos que, em notação abreviada,
\begin{eqnarray}
 &&S_b(p)\not\!\!AS_b(p-i\partial)\not\!\!A=\left[ \frac{i}{{\cal D}^2}P+\frac{i}{{\cal D}^4}PB_5P+\cdots \right]\not\!\!A\left[   \frac{i}{{\cal D}^2}P+\frac{i}{{\cal D}^4}P\tilde{B}_5P+\cdots\right]\not\!\!A\nonumber\\
&&\nonumber\\
&&=-\frac{1}{{\cal D}^4}P\not\!\!AP\not\!\!A-\frac{1}{{\cal D}^6}P\not\!\!AP\tilde{B}_5P\not\!\!A-\frac{1}{{\cal D}^6}PB_5P\not\!\!AP\not\!\!A-\frac{1}{{\cal D}^8}PB_5P\not\!\!AP\tilde{B}_5P\not\!\!A\,.\nonumber\\
&&
\end{eqnarray}

Os termos que darão origem à indução de CS são aqueles que dependem somente de uma derivada do campo $A_\mu$\footnote{Termos do tipo Maxwell envolvem derivadas.}: $\not\!\!b\not\!\!\!A\not\!\!\partial\not\!\!A\gamma_5$ e que darão a estrutura do termo de CS. Essa contribuição tem origem no último termo da expressão acima. Assim, utilizando as formas explícitas (6.15-18), segue que
\vspace{-1cm}

\begin{eqnarray*}
&& (\not\!p+m)\not\!b\gamma_5(\not\!p+m)\not\!\!A(\not\!p+m)(\not\!b\gamma_5+i\!{\not\!\partial})(\not\!p+m)\not\!\!A=\\
&&\\
&&(\not\!p\not\!b\gamma_5\not\!p\not\!\!A+m\not\!p\not\!b\gamma_5\not\!\!A+m\not\!b\gamma_5\not\!p\not\!\!A+m^2\not\!b\gamma_5\not\!\!A)(\not\!p\not\!b\gamma_5+i\not\!p\not\!\partial+m\not\!b\gamma_5+im\not\!\partial)\\
&&\\
&&\hspace{11cm}\times(\not\!p\not\!\!A+m\not\!\!A  )\\
&&\\
&&=i(\not\!p\not\!b\gamma_5\not\!p\not\!\!A+m\not\!p\not\!b\gamma_5\not\!\!A+m\not\!b\gamma_5\not\!p\not\!\!A+m^2\not\!b\gamma_5\not\!\!A)(\not\!p\not\!\partial\not\!p\not\!\!A+m\not\!p\not\!\partial\not\!\!A\\
&&\\
&&\hspace{8.5cm}+m\not\!\partial\not\!p\not\!\!A+m^2\not\!\partial\not\!\!A  )+\cdots\\
&&\\
&&=+i\not\!p\not\!b\not\!p\not\!\!A\not\!p\not\!\partial\not\!p\not\!\!A\gamma_5+im^2\not\!p\not\!b\not\!p\not\!\!A\not\!\partial\not\!\!A\gamma_5+im^2\not\!p\not\!b\not\!\!A\not\!p\not\!\partial\not\!\!A\gamma_5\\
&&\\
&&im^2\not\!p\not\!b\not\!\!A\not\!\partial\not\!p\not\!\!A\gamma_5-im^2\not\!b\not\!p\not\!\!A\not\!p\not\!\partial\not\!\!A\gamma_5-im^2\not\!b\not\!p\not\!\!A\not\!\partial\not\!p\not\!\!A\gamma_5\\
&&\\
&&-im^2\not\!b\not\!\!A\not\!p\not\!\partial\not\!p\not\!\!A\gamma_5-im^4\not\!b\not\!\!A\not\!\partial\not\!\!A\gamma_5+\cdots
\end{eqnarray*}

Nas passagens acima, omitimos os termos que não contêm $\not\!\!\partial$ e aqueles que são ímpares no número de matrizes de Dirac pois, nessas condições, o cálculo do traço de termos desse tipo é nulo. Agora, vamos reduzir os termos da expressão acima de oito e seis para quatro matrizes $\gamma$ utilizando as propriedades $\not\!c\not\!d=-\not\!d\not\!c+2(c\cdot d)$ e $\not\!c^2=c^2$:
\newpage

\begin{eqnarray}
&& i\not\!p[-\not\!p\not\!b+2(b\cdot p)]\not\!\!A\not\!p[-\not p\not\!\partial+2(p\cdot \partial)]\not\!\!A\gamma_5+im^2\not\!p[-\not\!p\not\!b+2(b\cdot p) ]\not\!\!A\not\!\partial\gamma_5\nonumber\\
&&\nonumber\\
&&+im^2\not\!p\not\!b\not\!\!A\not\!p\not\!\partial\not\!\!A\gamma_5+im^2\not\!p\not\!b\not\!\!A[-\not\!p\not\!\partial+2(p\cdot\partial)]\not\!\!A\gamma_5\nonumber\\
&&\nonumber\\
&&-im^2\not\!b\not\!p[-\not\!p\not\!\!A+2(p\cdot A)]\not\!\partial\not\!\!A\gamma_5-im^2\not\!b\not\!p\not\!\!A\not\!\partial\not\!p\not\!\!A\gamma_5\nonumber\\
&&\nonumber\\
&&-im^2\not\!b[-\not\!p\not\!\!A+2(p\cdot A)]\not\!\partial\not\!p\not\!\!A\gamma_5-im^4\not\!b\not\!\!A\not\!\partial\not\!\!A\gamma_5\nonumber\\
&&\nonumber\\
&& =ip^4\not\!b\not\!\!A\not\!\partial\not\!\!A\gamma_5  -2ip^2\not\!b\not\!\!A(p\cdot\partial)\not\!p\not\!\!A\gamma_5 -2ip^2(b\cdot p)\not\!p\not\!\!A\not\!\partial\not\!\!A\gamma_5 \nonumber\\
&&\nonumber\\
&&+4i(b\cdot p)\not\!p\not\!\!A(p\cdot\partial)\not\!p\not\!\!A\gamma_5+2im^2(b\cdot p)\not\!p\not\!\!A\not\!\partial\not\!\!A\gamma_5+2im^2\not\!p\not\!b\not\!\!A(p\cdot\partial)\not\!\!A\gamma_5\nonumber\\
&&\nonumber\\
&&-2im^2(p\cdot A)\not\!b\not\!\partial\not\!p\not\!\!A\gamma_5-im^4\not\!b\not\!\!A\not\!\partial\not\!\!A\gamma_5
\end{eqnarray}

O próximo passo é inserir o resultado acima na ação (6.14), além de (6.17), para calcularmos a integral nos momentos. É preciso notar que, por contagem de potências, as integrais proporcionais a $p^4$ têm divergência logarítmica, enquanto que as proporcionais a $p^2$ são finitas:  
\begin{eqnarray*}
 \int_{inf} \frac{d^4p}{(2\pi)^4}\frac{p^4}{(p^2-m^2)^4}\hspace{0.8cm}\mbox{,}\hspace{0.8cm} \int_{fin} \frac{d^4p}{(2\pi)^4}\frac{p^2}{(p^2-m^2)^4}\hspace{0.8cm}\mbox{e}\hspace{0.8cm}\int_{fin} \frac{d^4p}{(2\pi)^4}\frac{1}{(p^2-m^2)^4}\,.
\end{eqnarray*}

Calculemos explicitamente o quinto termo (finito) da expressão (6.22). Utilizando (6.13), (6.20) e as integrais de Feynman calculadas explicitamente no Apêndice G, além do cálculo do traço das matrizes de Dirac, temos que:

\begin{eqnarray}
 &&-\frac{ie^2}{2}tr_D\int d^4x\int\frac{d^4p}{(2\pi)^4}\frac{2im^2(b\cdot p)\not\!p\not\!\!A\not\!\partial\not\!\!A\gamma_5}{(p^2-m^2)^4}           \nonumber\\
&&\nonumber\\
&=&e^2m^2\int d^4x\,tr_D(\gamma^\mu\gamma^\nu\gamma^\rho\gamma^\sigma\gamma_5)b_\alpha A_\nu\partial_\rho A_\sigma\int\frac{d^4p}{(2\pi)^4}\frac{p^\alpha p_\mu}{(p^2-m^2)^4}\nonumber\\
&&\nonumber\\
&=&4ie^2m^2\int d^4x\,\varepsilon^{\mu\nu\rho\sigma}\delta^\alpha_\mu b_\alpha A_\nu\partial_\rho A_\sigma\frac{-i}{192\pi^2m^2}\nonumber\\
&&\nonumber\\
&=&\frac{e^2}{48\pi^2}\int d^4x\,\varepsilon^{\mu\nu\rho\sigma}b_\mu A_\nu\partial_\rho A_\sigma
\end{eqnarray}

Os quatro primeiros termos da expressão (6.22) geram termos infinitos nas integrais para a ação efetiva. Para realizar os cálculos, vamos fazer a regularização\footnote{Vide Apêndice G.} $D=4-2\epsilon$ para o cálculo das integrais divergentes. Enquanto mantivermos $\epsilon\neq0$, essas integrais originalmente divergentes são mantidas finitas e assim podemos somá-las e subtraí-las.
\smallskip

Assim, calculamos a ação efetiva somando inicialmente as partes infinita, mantidas finitsa pela regularização, com as partes  finitas:
\newpage

\begin{eqnarray}
 &&S_{CS}=S_{CS}^{inf}+S_{CS}^{fin}\nonumber\\
&&\nonumber\\
&&=  -\frac{ie^2}{2}tr_D\int d^4x\{\int\frac{d^Dp}{(2\pi)^D}\frac{1}{(p^2-m^2)^4}  [ip^4\not\!b\not\!\!A\not\!\partial\not\!\!A\gamma_5 -2ip^2\not\!b\not\!\!A(p\cdot\partial)\not\!p\not\!\!A\gamma_5\nonumber\\
&&\nonumber\\
&& -2ip^2(b\cdot p)\not\!p\not\!\!A\not\!\partial\not\!\!A\gamma_5   +4i(b\cdot p)\not\!p\not\!\!A(p\cdot\partial)\not\!p\not\!\!A\gamma_5\}\nonumber\\
&&\nonumber\\
&&-\frac{ie^2}{2}tr_D\int d^4x\int\frac{d^4p}{(2\pi)^4}\frac{1}{(p^2-m^2)^4}\{2im^2(b\cdot p)\not\!p\not\!\!A\not\!\partial\not\!\!A\gamma_5 \nonumber\\
&&\nonumber\\
&&+2im^2\not\!p\not\!b\not\!\!A(p\cdot\partial)\not\!\!A\gamma_5-2im^2(p\cdot A)\not\!b\not\!\partial\not\!p\not\!\!A\gamma_5-im^4\not\!b\not\!\!A\not\!\partial\not\!\!A\gamma_5]\}\nonumber\\
&&\nonumber\\
&&=2e^2\int d^4x\,\varepsilon^{\mu\nu\rho\sigma}\left\{i.\frac{i}{24\pi^2}\left[\frac{1}{\epsilon}+\log\frac{4\pi}{m^2}-\gamma+{\cal O}(\epsilon)  \right]b_\mu A_\nu\partial_\rho A_\sigma\right.\nonumber\\
&& \left. \right.\nonumber\\
&&\left. -2i.\frac{i}{96\pi^2}\left[\frac{1}{\epsilon}+\log\frac{4\pi}{m^2}-\gamma+{\cal O}(\epsilon)  \right]b_\mu A_\nu\partial_\rho A_\sigma -2i.\frac{i}{96\pi^2}\left[\frac{1}{\epsilon}+\log\frac{4\pi}{m^2}-\gamma+{\cal O}(\epsilon)  \right]b_\mu A_\nu\partial_\rho A_\sigma \right.\nonumber\\
&& \left. \right.\nonumber\\ 
&&\left.+4i.\frac{i}{384\pi^2}\left[\frac{1}{\epsilon}+\log\frac{4\pi}{m^2}-\gamma+{\cal O}(\epsilon)  \right](b_\mu A_\nu\partial_\rho A_\sigma -b_\mu A_\nu\partial_\rho A_\sigma)    \right\}\nonumber\\
&&\nonumber\\
&& + 2e^2\int d^4x\,\varepsilon^{\mu\nu\rho\sigma}\left[ 2im^2.\frac{-i}{192\pi^2m^2}b_\mu A_\nu\partial_\rho A_\sigma  +2im^2.\frac{-i}{192\pi^2m^2}b_\nu A_\rho\partial_\mu A_\sigma   \right. \nonumber\\
&& \left.  \right.  \nonumber\\
&&  \left. -2im^2.\frac{-i}{192\pi^2m^2}b_\mu A_\rho\partial_\nu A_\sigma  -im^4.\frac{i}{96\pi^2m^4} b_\mu A_\nu\partial_\rho A_\sigma  \right]
\end{eqnarray}

Como a soma das contribuições das partes infinitas devidamente regularizadas se anulam,  não há necessidade de tomarmos o limite $m\,\to\,0$. Assim, obtemos o seguinte termo proveniente da ação efetiva \cite{Gom}:

\begin{equation}
 S_{CS}^{(3+1)D}=\frac{e^2}{12\pi^2}\int d^4x\varepsilon^{\mu\nu\rho\sigma}b_\mu A_\nu\partial_\rho A_\sigma\,.
\end{equation}
\vspace{0.1cm}

Portanto, concluimos que a adição de um termo com um campo de fundo que quebra de simetria de Lorentz à lagrangiana da teoria de calibre usual induz uma ação tipo Chern-Simons no espaço-tempo quadridimensional. Como é bem observado na literatura, este termo é finito e obtido por diversos métodos de regularização. A única diferença é a constante de proporcionalidade, que depende exclusivamente do tipo de método de regularização utilizado.  

\section{Birefringência dos fótons clássicos de CS}

Uma das consequências da quebra da simetria de Lorentz na EDQ em (3+1)$D$ provocadas pelo termo induzido corresponde ao fenômeno da birefringência dos fótons clássicos de CS. A lagrangiana desta teoria escrita em termos da ação (6.25), é dada por 
\begin{equation}
{\cal L}_{CS}^{(3+1)D}=-\frac{1}{4}F^{\mu\nu}F_{\mu\nu}-\frac{1}{2}\varepsilon^{\mu\nu\rho\sigma}\eta_\mu A_\nu \partial_\rho A_\sigma-A_\mu J^\mu\,,
\end{equation}
onde, foi utilizada a relação
\begin{equation}
 \eta_\mu=\frac{e^2}{6\pi^2}b_\mu\,.
\end{equation}

Em relação a esta lagrangiana, calculemos as derivadas

\begin{eqnarray}
 \frac{\partial{\cal L}}{\partial A_\nu}&=&-\frac{1}{2}\varepsilon^{\mu\nu\rho\sigma}\eta_\mu\partial_\rho A_\sigma-J^\nu\\
&&\nonumber\\
\frac{\partial{\cal L}}{\partial(\partial_\mu A_\nu)}&=&-\frac{1}{4}\left(\frac{\partial F_{\alpha\beta}}{\partial(\partial_\mu A_\nu)}F^{\alpha\beta} + F_{\alpha\beta}\frac{\partial F^{\alpha\beta}}{\partial(\partial_\mu A_\nu)}\right)-\frac{1}{2}\varepsilon^{\mu\nu\rho\sigma}\eta_\mu\frac{\partial(\partial_\rho A_\sigma)}{\partial(\partial_\mu A_\nu)}\nonumber\\
&&\nonumber\\
&=&-\frac{1}{2}( \delta^\mu_\alpha\delta^\nu_\beta -\delta^\mu_\beta\delta^\nu_\alpha)F^{\alpha\beta}-\frac{1}{2}\varepsilon^{\mu\nu\rho\sigma}\eta_\mu A_\nu \delta^\mu_\rho\delta^\nu_\sigma\nonumber\\
&&\nonumber\\
\partial_\mu\frac{\partial{\cal L}}{\partial(\partial_\mu A_\nu)}&=&-\partial_\mu F^{\mu\nu}+\frac{1}{2}\varepsilon^{\mu\nu\rho\sigma}\eta_\mu\partial_\rho A_\sigma
\end{eqnarray}

Rearranjando os índices de Lorentz, utilizando a antisimetria de $\varepsilon$ e aplicando o m\'etodo variacional através da equação de Euler-Lagrange, temos que 

\begin{equation}
\partial_\mu\frac{\partial{\cal L}}{\partial(\partial_\mu A_\nu)}=\frac{\partial{\cal L}}{\partial A_\nu}\,,
\end{equation}
à lagrangiana (6.26), conhecendo o resultado (5.2), obtemos
\begin{equation}
 \partial_\mu F^{\mu\nu}-\varepsilon^{\mu\nu\rho\sigma}\eta_\mu\partial_\rho A_\sigma=J^\nu\,.
\end{equation}

Nesta teoria, a corrente também é conservada, devido a antisimetria do tensor do campo eletromagnético e a identidade de Bianchi:
\begin{eqnarray}
 \varepsilon^{\mu\nu\rho\sigma}\partial_\mu\partial_\nu F_{\rho\sigma}=0\,.
\end{eqnarray}

A identidade (6.32) representada acima assegura que as equações de Maxwell homogêneas\footnote{Estamos utilizando $F^{\mu\nu\,\ast}=\dfrac{1}{2}\varepsilon^{\mu\nu\rho\sigma}F_{\rho\sigma}$, $F_{0i}=E^i$ e $F_{ij}=\dfrac{1}{2}\varepsilon_{jk\ell}B^\ell$\,.} $\partial_\mu F^{\mu\nu\,\ast}=0$  permaneçam inalteradas: 
\begin{eqnarray}
&& {\mbox{\boldmath{$\nabla$}}}\cdot{\bf B}=0\\
&&\nonumber\\
&&{\mbox{\boldmath{$\nabla$}}}\times{\bf E}+\frac{\partial {\bf B}}{\partial t}=0\,.
\end{eqnarray}

As modifica\c{c}\~oes geradas pelo quadrivetor acoplado $\eta^\mu=(\eta_0,{\mbox{\boldmath{$\eta$}}})$  podem ser interpretadas como uma adi\c{c}\~ao de um campo dependente do tempo na fonte de corrente usual:
\begin{eqnarray}
&&{\mbox{\boldmath{$\nabla$}}}\times{\bf B}-{\mbox{\boldmath{$\eta$}}}\times{\bf E}+\eta^0{\bf B}={\bf J}+\frac{\partial{\bf E}}{\partial t}\\
&&{\mbox{\boldmath{$\nabla$}}}\cdot{\bf E}+{\mbox{\boldmath{$\eta$}}}\cdot{\bf B}=\rho\,.
\end{eqnarray}

Estas equações não-homogêneas também preservam a invariância de calibre. Entretanto, elas demonstram um caráter que não é encontrado na teoria padrão: a introdução da quebra de simetria de Lorentz permite que os campos elétricos sirvam como fonte para correntes e que os campos magnéticos sirvam como fonte para cargas elétricas.
\smallskip

Analisemos agora as consequ\^encias da quebra da simetria de Lorentz sobre a equa\c{c}\~ao de movimento para o campo $A_\mu$ de CS. Na aus\^encia de fontes externas ($J^\mu=0$), utilizando $F_{\mu\nu}=\partial_\mu A_\nu-\partial_\nu A_\mu$, a equação de movimento (6.31) é dada por:
\begin{equation}
\square A^\nu-\partial^\nu(\partial_\mu A^\mu)-\varepsilon^{\mu\nu\rho\sigma}\eta_\mu\partial_\rho A_\sigma=0.
\end{equation}

Assim, com $k^\mu=(\omega,|{\bf k}|)$, escrevendo o campo de calibre no espaço dos momentos
\begin{equation}
A_\mu(x)=\int\frac{d^4k}{(2\pi)^4}A_\mu(k)\,e^{ik\cdot x}\,,
\end{equation} 
utilizando o calibre de Lorentz $\partial_\mu A^\mu=0$, obtemos a seguinte express\~ao
\begin{equation}
k^2+i\varepsilon^{\mu\nu\rho\sigma}\eta_\mu k_\sigma=0\,.
\end{equation}

\'E conveniente multiplicar a equa\c{c}\~ao acima por seu complexo conjugado:
\begin{eqnarray*}
(k^2-i\varepsilon^{\mu\nu\rho\sigma}\eta_\rho k_\sigma)(k^2+i\varepsilon_{\mu\nu\alpha\beta}\eta^\alpha k^\beta)&=&0\\
k^4+\varepsilon^{\mu\nu\rho\sigma}\varepsilon_{\mu\nu\alpha\beta}(\eta_\rho k_\sigma)(\eta^\alpha k^\beta)&=&0\,.
\end{eqnarray*}
a expressão acima é dada por
\begin{equation}
k^4+k^2\eta^2-(k\cdot\eta)^2=0\,,
\end{equation}
e a express\~ao (6.39) representa a rela\c{c}\~ao de dispers\~ao para os fótons de CS em $(3+1)$ dimens\~oes. 
\smallskip

Considere o caso particular $\eta^\mu=(\eta^0,{\bf 0})$ aplicado à equação (6.39). Resolvendo uma equação biquadrada para $\omega$, obtemos duas soluções:
\begin{equation}
\omega_\pm=\sqrt{|{\bf k}|(|{\bf k}|\pm\eta^0)}\,,
\end{equation}

Esta equa\c{c}\~ao indica que o o termo induzido de CS gera o efeito de birefringência, ou seja, a separação dos f\'otons em dois modos diferentes de polariza\c{c}\~oes para o vetor $|{\bf k}|$ com diferentes velocidades de grupo:
\begin{equation}
v_{g\pm}=\frac{\partial \omega_\pm}{\partial |{\bf k}|}= \frac{|{\bf k}|}{\omega_\pm}\left( 1\pm\frac{\eta_0}{2|{\bf k}|}  \right)    \,.
\end{equation}

O fato de existir f\'otons viajando com diferentes velocidades de grupo, além de ser uma forte indicação da violação de Lorentz, indica também evid\^encias para uma aparente instabilidade na teoria. Do resultado (6.42), vemos que $\omega$ se torna imagin\'ario para os casos em que $\eta_0>k$. Isso significa que h\'a solu\c{c}\~oes imaginárias e inst\'aveis, que não existem na eletrodinâmica convencional.
\smallskip

Na tentativa de detectar o campo de fundo $\eta_\mu$, Carrol, Field e Jackiw sugeriram \cite{Cae} a comparação dos resultados previstos pela teoria com dados experimentais. Assumindo o caso em que não há quebra de simetria rotacional, ou seja, $\eta^\mu=(\eta^0,{\bf 0})$, eles confrontaram dados geomagnéticos com o campo magnético da Terra obtido pelas soluçoes das equações modificadas na presença de fontes. Na mesma referência, Jackiw e colaboradores propuseram resolver a equação (6.35) para o caso em que ${\bf E}=0$. A solução obtida desta equação para o campo magnétrico é expandida em termos da posição para se obter um termo conhecido de dipolo acrescido de uma perturbação dependente de $\eta_0$. Conhecendo-se o valor do campo magnético na direção azimutal, é possível obter o resultado $\eta_0\lesssim6\times10^{-20}\,keV$. Analisando também a polarização da luz emitida por galáxias distantes, removendo o efeito Faraday devido suas rotações, o resultado obtido foi $\eta_0\lesssim6\times10^{-20}\,keV$. Maiores detalhes destes e de outros testes podem ser encontrados na referência \cite{Car}, além de resultados mais recentes apresentados por Kostelecký em \cite{Dat}. Apesar de todo o esforço e comparação de dados geomagnéticos e astrofísicos com resultados teóricos, nenhuma evidência de campos de quebra de simetria de Lorentz foram detectadas.

\chapter{Conclusão} 

Estudamos os poss\'iveis efeitos que a teoria do Modelo Padrão Extendido pode provocar em certos fen\^omenos quando novos elementos de quebra de simetria de Lorentz são introduzidos nas lagrangianas da teoria convencional. Estes efeitos foram analisados nos contextos da mecânica quântica e nas correções radiativas em um sistema de férmions interagindo com um campo de calibre. Notamos que tais efeitos são gerados pelo acoplamento axial $\not\!b\gamma_5$.
\smallskip

Constru\'imos um ferramental te\'orico, no setor da mat\'eria, que foi utilizado nos cálculos nos cap\'itulos posteriores. Analisamos algumas implicações da quebra da simetria de Lorentz na mecânica quântica, via níveis de Landau, através das mudanças de energia com e sem inversão do spin do el\'etron, e notamos que tais mudanças dependem diretamente de $|{\bf b}|$. O hamiltoniano de Dirac modificado pela teoria foi expandido no limite n\~ao-relativ\'istico, via m\'etodo FW, e foi dividido em um hamiltoniano de Pauli, n\~ao perturbado, e um hamiltoniano de intera\c{c}\~ao, dependendo da viola\c{c}\~ao de Lorentz, escrito em termos de $a_\mu$ e $b_\mu$. O efeito Zeeman foi estudado analisando-se os poss\'iveis efeitos produzidos por esses dois campos separadamente. Foi encontrado que o campo $a_\mu$ n\~ao provoca nenhuma mudan\c{c}a, uma vez que o mesmo redefine os zeros da energia e do momento. No caso do acoplamento axial, foi encontrada uma mudan\c{c}a na energia que depende do n\'umero qu\^antico $m_j$, na aus\^encia do campo magn\'etico. Nosso resultado é o dobro daquele encontrado na referência \cite{Man}. Na presen\c{c}a do campo magn\'etico, mesmo intenso, nenhuma corre\c{c}\~ao para a energia foi encontrada.
\smallskip

Calculamos também a ação induzida de Chern-Sompns no espaço quadrimensional. Verificamos que isso só é possível se introduzirmos o termo de quebra de simetria de Lorentz, pois o termo $\not\!b\gamma_5$ induz o cálculo de $tr_D[\gamma^\mu\gamma^\nu\gamma^\rho\gamma^\sigma\gamma_5]$, que gera a estrutura necessária $\varepsilon^{\mu\nu\rho\sigma}$ pertinente a este termo induzido. Nosso resultado está de acordo com outros obtidos na literatura, que podem diferenciar entre si, em relação a uma constante, se diferentes métodos de regularização forem utilizados. Em nosso cálculo, utilizando o método de regularização dimensional, notamos que os termos divergentes da ação induzida se cancelam mutuamente, e o limite $m\,\to\,0$ não se faz necessário e a ação induzida é finita.
\smallskip

As consequências geradas pela ação de Chern-Simons foram analisadas através do cálculo da velocidade de propagação dos fótons de Chern-Simons. Como resultado, notamos que tais velocidades são separadas em dois modos distintos de polarização. Também verificamos que as energias podem se tornar imaginárias para alguns valores específicos do campo de fundo.
\smallskip

É importante ressaltar que, apesar de todos os esforços teóricos realizados até aqui, na comparação de dados geomagnéticos e astrofísicos com resultados previstos pelas equações modificadas pelos termos de quebra de simetria de Lorentz, nenhuma evid\^{e}ncia experimental em relação a tais campos de fundo foram detectadas.

\appendix

\chapter{Convenções adotadas}

\begin{center}
 {\bf Matrizes de Dirac em $(2+1)D$}
\end{center}

Nesta dimensão, a métrica utilizada tem a forma $g^{\mu\nu}=diag(1,-1,-1)$. As matrizes de Dirac escolhidas são
\begin{equation}
 \gamma^0=\left(\begin{array}{cc}0&-i\\i&0\end{array}
 \right)\quad,\quad\gamma^1=\left(\begin{array}{cc}0&i\\i&0\end{array} \right)\quad\mbox{}\quad
\gamma^2=\left(\begin{array}{cc}i&0\\0&-i \end{array} \right)\,.
\end{equation}

A notação $tr_D$ indica que os traços são calculados sobre as matrizes $\gamma$ de Dirac. As mesmas têm as seguintes propriedades: $tr_D[I]=2$ e $tr_D[\gamma^\mu]=0$. Então, são úteis as seguintes relações:
\begin{eqnarray}
&&\{\gamma^\mu,\gamma^\nu  \}=2g^{\mu\nu}\,\\
&&\gamma_\mu\gamma^\nu\gamma^\mu=-\gamma^\nu\\
&&tr_D[\gamma^\mu\gamma^\nu]=2g^{\mu\nu}\,\\
&&tr_D[\gamma^\mu\gamma^\rho\gamma^\nu]=2i\varepsilon^{\mu\rho\nu}\,\\
&&tr_D[\gamma^\mu\gamma^\rho\gamma^\nu\gamma^\sigma]=2(g^{\mu\rho}g^{\nu\sigma}+g^{\rho\nu}g^{\sigma\mu}-g^{\mu\nu}g^{\rho\sigma})\,
\end{eqnarray}

\begin{center}
 {\bf Matrizes de Dirac em $(3+1)D$}
\end{center}

As matrizes de Dirac na representa\c{c}\~ao padrão são dadas por:
\begin{eqnarray*}
&&\gamma^0=\left(\begin{array}{cc}1&0\\0&-1\end{array}\right)\hspace{2eM}\hspace{2eM}
{\mbox{\boldmath{$\gamma$}}}=\left(\begin{array}{cc}0&{\mbox{\boldmath{$ \sigma$}}}\\-{\mbox{\boldmath{$ \sigma$}}}&0\end{array}\right)\,,\\
&&\\
&&\hspace{1cm}\gamma^5=i\gamma^0\gamma^1\gamma^2\gamma^3=\left(\begin{array}{cc}0&1\\1&0\end{array}\right)\,,
\end{eqnarray*}
onde $\gamma_5=i\gamma_0\gamma_1\gamma_2\gamma_3=-\gamma^5$ e tamb\'em satisfaz:
\begin{eqnarray*}
(\gamma^5)^2=1\quad\,\mbox{e}\quad\{\gamma^\mu,\gamma^5\}=0\,.
\end{eqnarray*}
\smallskip

As matrizes $\gamma$ de Dirac também obedecem as seguintes regras de anticomutação:
\begin{eqnarray}
\{\gamma^\mu,\gamma^\nu\}=2g^{\mu\nu}\,.
\end{eqnarray}

Duas matrizes $\gamma$ formam o seguinte objeto antissim\'etrico:
\begin{eqnarray*}
\sigma^{\mu\nu}=\frac{i}{2}[\gamma^\mu,\gamma^\nu]\,.
\end{eqnarray*}

Novamente, a notação $tr_D$ indica que os traços são calculados sobre as matrizes $\gamma$ de Dirac. Em $(3+1)D$, as mesmas têm as seguintes propriedades: $tr_D[I]=4$ e $tr_D[\gamma^\mu]=0$. Então, são úteis as seguintes relações:
\begin{eqnarray}
&&tr_D[\gamma^\mu\gamma^\nu]=4g^{\mu\nu}\\
&&tr_D[\gamma^\mu\gamma^\nu\gamma_5]=0\\
&&tr_D[\gamma^\mu\gamma^\nu\gamma^\sigma\gamma^\rho]=4\left[g^{\mu\nu}g^{\sigma\rho}-g^{\mu\sigma}g^{\nu\rho}+g^{\mu\rho}g^{\nu\sigma} \right]\\
&&tr_D[\gamma^\mu\gamma^\nu\gamma^\sigma\gamma^\rho\gamma_5]=4i\varepsilon^{\mu\nu\sigma\rho}
\end{eqnarray}

\chapter{Autofun\c{c}\~oes exatas do problema de Landau relativístico}

A equa\c{c}\~ao de movimento \cite{Ber} de um elétron de massa $m$ sujeito a um campo eletromagnético é dada por:
\begin{equation}
[\gamma^\mu(p_\mu-eA_\mu)-m]\psi=0\,.
\end{equation}

Aplicando o operador $\gamma^\nu(p_\nu-eA_\nu)+m$ pela esquerda, temos que
\begin{eqnarray}
[\gamma^\mu\gamma^\nu(p_\mu-eA_\mu)(p_\nu-eA_\nu)-m^2]\psi=0\,.
\end{eqnarray}

Utilizando a identidade $\gamma^\mu\gamma^\nu=g^{\mu\nu}+\sigma^{\mu\nu}$, segue que
\begin{eqnarray}
&&[(p-eA)^2+\sigma^{\mu\nu}(p_\mu-eA_\mu)(p_\nu-eA_\nu)-m^2]\psi=0\nonumber\\
&&\nonumber\\
&&\left[(p-eA)^2+\frac{e}{2}\sigma^{\mu\nu}(-A_\mu p_\nu+p_\nu
A_\mu+A_\nu p_\mu-p_\mu A_\nu)-m^2\right]\psi=0\nonumber\,,
\end{eqnarray}
onde o produto do segundo termo na primeira expressão foi anti-simetrizado. Sendo o tensor do campo eletromagnético dado por $F_{\mu\nu}=\partial_\mu A_\nu-\partial_\nu A_\mu$, temos a seguinte equação de segunda ordem:
\begin{equation}
\left[(p-eA)^2-\frac{ie}{2}\sigma^{\mu\nu}F_{\mu\nu}-m^2\right]\psi=0\,.
\end{equation}

Sendo $\sigma^{\mu\nu}=(\mbox{\boldmath{$\alpha$}},i{\bf \Sigma})$ e $F_{\mu\nu}=(-{\bf E}, {\bf B})$ as partes espaciais do produto $\sigma^{\mu\nu}F_{\mu\nu}$ e, no problema de Landau, ${\bf E}=0$ e $A^\mu=(0,B_0\,x\,\textbf{ê}_y)$, então, temos que
\begin{equation}
[(p-eA)^2+e{\bf\Sigma}\cdot{\bf B}-m^2]\psi=0\,.
\end{equation}

Assim, a equação de autovalores é
\begin{equation}
[p_x^2+(eB_0x-p_y)^2-e\sigma\,B_0]\psi=[(E^\pm)^2-p_z^2-m^2]\,\psi\,.
\end{equation}

O espectro de energia é o resultado da interação da partícula com este campo magnético intenso. Como o lado esquerdo da equação (B.5) possui um termo idêntico \`a equação de um oscilador harmônico, temos que as autoenergias relativísticas para os elétrons e pósitrons do problema são os conhecidos {\it níveis de Landau}:
\begin{equation}
E^\mp_{n,s}=\pm\sqrt{p_z^2+m^2+|e|\,B_0(2n+1-s)}\,,
\end{equation}
onde $\sigma\,\,\to\,\,s=\pm1$ indica os spins dos elétrons e pósitrons.
\smallskip

As autofunções n\~ao-perturbadas dos elétrons e dos pósitrons são obtidas atrav\'es de solu\c{c}\~oes estacion\'arias com energia $E$, da forma
\begin{equation}
\psi^-_{n,s}(x)=u^{(s)}(p)e^{-iE_{n,s}t}\,,
\end{equation}
onde
\begin{equation}
u=\left(\begin{array}{c}\varphi\\\eta\end{array}\right)\,.
\end{equation}

Assim, com as mudan\c{c}as de vari\'aveis
\begin{eqnarray*}
p_x\,\to\,-i\frac{d}{dx}\,,\hspace{1.2eM}\xi=\sqrt{eB_0}\left(x-\frac{p_y}{eB_0}\right)\,,\hspace{1.2eM}dx=\frac{1}{\sqrt{eB_0}}d\xi\,,\hspace{1.2eM}\alpha=\frac{(E^\pm)^2-m^2-p_z^2}{eB_0}\,,
\end{eqnarray*}
a equa\c{c}\~ao (B.5) \'e reescrita no espa\c{c}o das posi\c{c}\~oes, com $\varphi(x)=e^{i(p_yy+p_zz)}\,f(x)$, como
\begin{equation}
\left[\frac{d^2}{d\xi^2}-\xi^2\right]f(x)=-(\alpha+s)\,f(x)\,,\hspace{2eM}s=\pm1\,.
\end{equation}

Ent\~ao, as solu\c{c}\~oes acima s\~ao as mesmas do oscilador harm\^onico:

\begin{equation}
\varphi(x)=N\,e^{i(p_yy+p_zz)}\,e^{-\xi^2/2}\,H_n(\xi)\,,
\end{equation}
onde os $H_n(\xi)$ s\~ao os conhecidos {\it polin\^omios de Hermite}, que satisfazem a seguinte rela\c{c}\~ao de ortogonalidade:
\begin{eqnarray*}
\int\limits_{-\infty}^{+\infty}d\xi\,e^{-\xi^2}[H_n(\xi)]^2=2^nn!\sqrt{\pi}\,.
\end{eqnarray*}
\smallskip

Utilizando o espinor (2.33) com a condi\c{c}\~ao de acoplamento minimal $p_\mu\,\to\,p_\mu-eA_\mu$, a auto-fun\c{c}\~ao n\~ao-perturbada do el\'etron, normalizada, \'e dada por (B.7) com

\begin{equation}
u^{(s)}(p)=\frac{1}{\sqrt{2^nn!}}\left(\frac{eB_0}{\pi}\right)^{1/4}\sqrt{\frac{E^-_{n,s}+m}{2E^-_{n,s}}}\left(\begin{array}{c}H_n(\xi)\,e^{-\xi^2/2}\chi_s\\\frac{{\mbox{\boldmath{$\sigma$}}}\cdot({\bf
p}-e{\bf
A})}{\displaystyle{E^-_{n,s}+m}}H_n(\xi)e^{-\xi^2/2}\chi_s\end{array}\right)\,e^{i(p_yy+p_zz)}\,,
\end{equation}
que satisfaz

\begin{equation}
\int\psi^{-\dagger}_{n,s}(x)\psi^-_{n,s}(x)dx=\delta(p'-p)\,.
\end{equation}

\chapter{O método Foldy-Wouthuysen}

No trabalho clássico realizado por L.L. Foldy e S. A. Wouthuysen \cite{Fol} são propostas duas aplicações para partículas relativísticas: uma para partículas livres e outras para partículas sujeitas a campos eletromagnéticos externos. Em geral, transformações de uma função de onda $\Psi$ são obtidas através de um determinado operador unitário $U$:
\begin{equation}
\Psi'=U\Psi=e^{iS}\Psi\,.
\end{equation}

A equação de Dirac convencional pode ser trabalhada da seguinte forma:
\begin{eqnarray*}
H\Psi=i\frac{\partial}{\partial
t}\Psi\hspace{1eM},\hspace{1eM}H\,U^{-1}\Psi'=i\frac{\partial}{\partial
t}(U^{-1}\Psi')
\end{eqnarray*}
ou
\begin{eqnarray*}
(U\,H\,U^{-1})\Psi'=i\left(U\frac{\partial U^{-1}}{\partial
t}+\frac{\partial}{\partial t}\right)\,\Psi'\,,
\end{eqnarray*}
de onde identificamos
\begin{equation}
H'=U\,H\,U^{-1}-iU\,\frac{\partial U^{-1}}{\partial t}\,.
\end{equation}

O Hamiltoniano de Dirac separado em termos pares e ímpares é dado por (4.9):
\begin{equation}
H=m\gamma^0+{\cal P}+{\cal I}\,,
\end{equation}
onde ${\cal P}$ e ${\cal I}$ representam os operadores par e ímpar que obedecem as relações $[\gamma^0,{\cal P}]=0$ e
$\{\gamma^0,{\cal I}\}=0$. A ideia é propor um operador da forma:
\begin{equation}
S=-i\frac{\gamma^0{\cal I}}{{\cal P}}\,\theta\,.
\end{equation}

Este operador pode ser substituído em (C.2) e expandido para se obter
\begin{equation}
U^{\pm1}=\cos\theta\pm\frac{\gamma^0{\cal I}}{{\cal
P}}\,\sin\theta\,.
\end{equation}

Como $S$ independe do tempo, o Hamiltoniano (C.2) pode ser escrito
\begin{eqnarray}
H'&=&\left(\cos\theta+\frac{\gamma^0{\cal I}}{{\cal
P}}\,\sin\theta\right)H\left(\cos\theta-\frac{\gamma^0{\cal
I}}{{\cal P}}\,\sin\theta\right)\nonumber\\
&=&(m\gamma^0+{\cal I})\left(\cos\theta-\frac{\gamma^0{\cal
I}}{{\cal P}}\,\sin\theta\right)^2+{\cal P}\nonumber\\
&=&\gamma^0(m\,\cos2\theta+{\cal P}\sin2\theta)+{\cal
I}\left(\cos2\theta-\frac{m}{{\cal P}}\,\sin2\theta\right)+{\cal
P}\,.
\end{eqnarray}

O Hamiltoniano então será essencialmente par se tivermos
\begin{equation}
\tan2\theta=\frac{{\cal P}}{m}\,.
\end{equation}

Esta equa\c{c}\~ao tem duas solu\c{c}\~oes, $\theta_1$ e $\theta_2$, que diferem pelo fator $\pi/2$:
\begin{equation}
\tan\theta_1=\frac{{\cal P}}{E_p+m}\hspace{2eM},\hspace{2eM}\tan\theta_2=-\frac{{\cal P}}{E_p-m}\hspace{2eM},\hspace{2eM}E_p=\sqrt{m^2+|{\bf p}|^2}\,.
\end{equation}

Assim, tomando o limite superior, os c\'alculos do seno e do co-seno fornecem
\begin{equation}
\sin\theta=\frac{{\cal P}}{\sqrt{2E_p(E_p+m)}}\hspace{3eM},\hspace{3eM}\cos\theta=\sqrt{\frac{E_p+m}{2E_p}}\,.
\end{equation}

Portanto, de (C.5), obtemos a seguinte express\~ao para o operador $U$:
\begin{equation}
U=\frac{E_p+m+\gamma^0{\cal I}}{\sqrt{2E_p(E_p+m)}}\,,
\end{equation}
onde, na teoria de Dirac livre, ${\cal I}={\mbox{\boldmath{$\alpha$}}}\cdot{\bf p}$.
\smallskip

No limite não relativístico, de (C.9) e (C.4), vemos que o operador $S$ torna-se
\begin{equation}
 S=-\frac{i}{2m}\gamma^0{\cal I}\,.
\end{equation}

\chapter{O Acoplamento Spin-\'Orbita}

O momento angular total correspondente ao acoplamento spin-\'orbita, para o el\'etron, corresponde \`a soma dos momentos angulares orbital e de spin:
\begin{equation}
{\bf J}={\bf L}+{\bf S}\,.
\end{equation}

No caso do el\'etron, o acoplamento admite os seguintes autovalores:
\begin{equation}
j=\ell\pm\frac{1}{2}\hspace{2eM},\hspace{2eM}m_j=m\pm\frac{1}{2}\,.
\end{equation}

Sendo $j=\ell+\frac{1}{2}$, com $m_j=m\pm\frac{1}{2}$ fixo, podemos construir um autoestado de $\mathbf{J}$ na base
$|jm_j\rangle=|\ell m\rangle\otimes|sm_s\rangle$ e outro autoestado ortogonal $|j'm_{j'}\rangle$ com $j'=\ell-\frac{1}{2}$:
\begin{eqnarray}
|jm_j\rangle&=&\alpha|\ell m\rangle\otimes\left|\frac{1}{2}\frac{1}{2}\right\rangle+\beta|\ell m+1\rangle\otimes\left|\frac{1}{2}-\frac{1}{2}\right\rangle\\
&&\nonumber\\
|j'm_{j'}\rangle&=&\beta|\ell
m\rangle\otimes\left|\frac{1}{2}\frac{1}{2}\right\rangle-\alpha|\ell
m+1\rangle\otimes\left|\frac{1}{2}-\frac{1}{2}\right\rangle\,.
\end{eqnarray}

As equa\c{c}\~oes de autovalores para ${\bf J}$, ${\bf L}$ e ${\bf S}$, bem como para os operadores escada $L_{\pm}=L_x\pm iL_y$ e $S_{\pm}=S_x\pm iS_y$, ser\~ao utilizadas mais adiante:

\begin{eqnarray}
\mathbf{J}^2|jm_j\rangle&=&\hbar^2j(j+1)|jm_j\rangle\,;\\
\mathbf{L}^2|\ell m\rangle&=&\hbar^2\ell(\ell+1)|\ell m\rangle\,;\\
\mathbf{S}^2|sm\rangle&=&\hbar^2s(s+1)|sm\rangle\,;\\
J_\pm|jm_j\rangle&=&\hbar\sqrt{\frac{1}{2}(j\pm m_j+1)(j\mp m_j)}\,|jm_j\pm1\rangle=\hbar\,C_{j\pm m_j}|jm_j\pm1\rangle\,;\\
L_\pm|\ell m\rangle&=&\hbar\sqrt{\frac{1}{2}(\ell\pm m+1)(\ell\mp m)}\,|\ell m\pm1\rangle=\hbar C_{\ell\pm m}|\ell m\pm1\rangle\,;\\
S_\pm|sm_s\rangle&=&\hbar\sqrt{\frac{1}{2}(s\pm m_s+1)(s\mp
m_s)}\,|sm_s\pm1\rangle=\hbar C_{s\pm m_s}|sm_s\pm1\rangle\,.
\end{eqnarray}

Para se obter os coeficientes do acoplamento, \'e necess\'aria a utiliza\c{c}\~ao da identidade:
\begin{eqnarray}
\mathbf{J}^2=\mathbf{L}^2+\mathbf{S}^2+2L_zS_z+2L_+S_-+2L_-S_+\,.
\end{eqnarray}

Aplicando $\mathbf{J}^2$ ao estado (D.3), lan\c{c}ando m\~ao da identidade (D.11) e das equações de autovalores (D.5-10), obtemos, com $S_+\left|\frac{1}{2}\frac{1}{2}\right\rangle=0$ e $S_-\left|\frac{1}{2}-\frac{1}{2}\right\rangle=0$:

\begin{eqnarray}
\mathbf{J}^2|jm_j\rangle&=&\hbar^2j(j+1)\left[\alpha|\ell m\rangle\otimes\left|\frac{1}{2}\frac{1}{2}\right\rangle+\beta|\ell m+1\rangle\otimes\left|\frac{1}{2}-\frac{1}{2}\right\rangle\right]\nonumber\\
&\equiv&\alpha(\mathbf{L}^2+\mathbf{S}^2+2L_zS_z+2L_+S_-+2L_-S_+)|\ell m\rangle\otimes\left|\frac{1}{2}\frac{1}{2}\right\rangle\nonumber\\
&+&\beta(\mathbf{L}^2+\mathbf{S}^2+2L_zS_z+2L_+S_-+2L_-S_+)|\ell m+1\rangle\otimes\left|\frac{1}{2}-\frac{1}{2}\right\rangle\nonumber\\
&=&\alpha\left\{\left[\ell(\ell+1)+\frac{3}{4}+2m.\frac{1}{2}\right]|\ell m\rangle\otimes\left|\frac{1}{2}\frac{1}{2}\right\rangle+2C_{\ell m}C_{\frac{1}{2}-\frac{1}{2}}|\ell m+1\rangle\otimes\left|\frac{1}{2}-\frac{1}{2}\right\rangle\right\}\nonumber\\
&+&\beta\left\{\left[\ell(\ell+1)+\frac{3}{4}+2(m+1).\left(-\frac{1}{2}\right)\right]|\ell m+1\rangle\otimes\left|\frac{1}{2}-\frac{1}{2}\right\rangle\right.\nonumber\\
&+&\left.2C_{\ell-m}C_{\frac{1}{2}-\frac{1}{2}}|\ell m\rangle\otimes\left|\frac{1}{2}\frac{1}{2}\right\rangle\right\}\,.\nonumber\\
\end{eqnarray}

Uma rela\c{c}\~ao entre $\alpha$ e $\beta$ \'e obtida se agrupamos os coeficientes de cada estado:
\begin{eqnarray}
\alpha\left[j(j+1)-\ell(\ell+1)-m-\frac{3}{4}\right]-2C_{\ell-m}C_{\frac{1}{2}-\frac{1}{2}}\,\beta&=&0\\
-2\alpha C_{\ell
m}C_{\frac{1}{2}-\frac{1}{2}}+\beta\left[j(j+1)-\ell(\ell+1)+m+1-\frac{3}{4}\right]&=&0\,.
\end{eqnarray}

O determinante acima se anula para os valores de $j$ dados por $j=\ell\pm1/2$. Tomando $j=\ell+1/2$ na equa\c{c}\~ao (D.14), obtemos, com $C_{\ell m}$ e $C_{\frac{1}{2}-\frac{1}{2}}$ dados por (D.9) e (D.10), uma rela\c{c}\~ao entre $\alpha$ e $\beta$:
\begin{equation}
\frac{\alpha}{\beta}=\sqrt{\frac{\ell+m+1}{\ell-m}}\,.
\end{equation}

No entanto, os coeficientes $\alpha$ e $\beta$ do auto-estado (D.3) devem estar normalizados, ou seja,
\begin{equation}
|\alpha|^2+|\beta|^2=1\,.
\end{equation}

Ent\~ao, de (D.15) e (D.16), vemos que
\begin{equation}
\alpha=\sqrt{\frac{\ell+m+1}{2\ell+1}}\hspace{2eM}\mbox{e}\hspace{2eM}\beta=\sqrt{\frac{\ell-m}{2\ell+1}}\,.
\end{equation}

Portanto,
\begin{equation}
|jm_j\rangle=\sqrt{\frac{\ell+m+1}{2\ell+1}}\,|\ell m\rangle\otimes\left|\frac{1}{2}\frac{1}{2}\right\rangle+\sqrt{\frac{\ell-m}{2\ell+1}}\,|\ell m+1\rangle\otimes\left|\frac{1}{2}-\frac{1}{2}\right\rangle\,.
\end{equation}

Procedendo de modo an\'alogo, obtemos, para $j=\ell-1/2$ e
$m_j=m+1/2$:
\begin{equation}
|jm_j\rangle=\sqrt{\frac{\ell-m}{2\ell+1}}\,|\ell m\rangle\otimes\left|\frac{1}{2}\frac{1}{2}\right\rangle-\sqrt{\frac{\ell+m+1}{2\ell+1}}\,|\ell m+1\rangle\otimes\left|\frac{1}{2}-\frac{1}{2}\right\rangle\,.
\end{equation}

As fun\c{c}\~oes de onda para os casos $j=\ell\pm1/2$ podem ser obtidas no espa\c{c}o de coordenadas, atrav\'es da conex\~ao $\langle\theta\phi|jm_j\rangle=\psi_{j\ell\frac{1}{2}m_j}$\,,$\langle\theta\phi|\ell
m\rangle=Y_\ell^m(\theta,\phi)$ e, definindo a base
\begin{eqnarray*}
\left|\frac{1}{2}\frac{1}{2}\right\rangle\equiv|+\rangle=\left(\begin{array}{c}1\\0\end{array}\right)\hspace{2eM}\mbox{e}\hspace{2eM}\left|\frac{1}{2}-\frac{1}{2}\right\rangle\equiv|-\rangle=\left(\begin{array}{c}0\\1\end{array}\right)\,,
\end{eqnarray*}
obtemos:
\begin{eqnarray}
\psi_{j\ell\frac{1}{2}m_j}&=&\sqrt{\frac{\ell+m+1}{2\ell+1}}\,Y_\ell^m|+\rangle+\sqrt{\frac{\ell-m}{2\ell+1}}\,Y_\ell^{m+1}|-\rangle\nonumber\\
&&\nonumber\\
&=&\left(\begin{array}{c}\sqrt{\frac{\ell+m+1}{2\ell+1}}\,Y_\ell^m\\
\sqrt{\frac{\ell-m}{2\ell+1}}\,Y_\ell^{m+1}\end{array}\right)\,,\hspace{2eM}j=\ell+\frac{1}{2}\,,
\end{eqnarray}
e
\begin{eqnarray}
\psi_{j\ell\frac{1}{2}m_j}&=&\sqrt{\frac{\ell-m}{2\ell+1}}\,Y_\ell^m|+\rangle-\sqrt{\frac{\ell+m+1}{2\ell+1}}\,Y_\ell^{m+1}|-\rangle\nonumber\\
&&\nonumber\\
&=&\left(\begin{array}{c}\sqrt{\frac{\ell-m}{2\ell+1}}\,Y_\ell^m\\
-\sqrt{\frac{\ell+m+1}{2\ell+1}}\,Y_\ell^{m+1}\end{array}\right)\,,\hspace{2eM}j=\ell-\frac{1}{2}\,.
\end{eqnarray}

\chapter{Integração gaussiana em variáveis de Grassmann}

Quando lidamos com férmions, a estatística exige que utilizemos variáveis anticomutantes. Variáveis que obedecem a essa propriedade são denominadas {\it variá\-veis de Grassmann}. No entanto, em contraste com integrais gaussianas em variáveis comutantes, nas integrais gaussianas que envolvem variáveis de Grassmann, o determinante da matriz no expoente aparece no numerador. Pretendemos mostrar essa relação neste Apêndice.
\smallskip

Sejam $\xi$ e $\eta$ duas variáveis de Grassmann e $c$ uma variável complexa qualquer. São válidas as relações 
\begin{equation}
 \{\xi, \eta  \}=0\quad\,,\,\quad [\xi, c]=0\,.
\end{equation}
de onde segue automaticamente que $\xi^2=0$\,. Utilzando as definições
\begin{equation}
 \int d\xi\equiv0\quad\quad\,\mbox{e}\quad\quad\,\int d\xi(\xi)\equiv1\,, 
\end{equation}
obtemos
\begin{equation}
 \int d\xi\int d\eta(\eta\xi)=1\,.
\end{equation}

Como o campo de Dirac é definido em um espaço complexo, devemos ter
\begin{equation}
 (\xi\eta)^\ast=\eta^\ast\xi^\ast=-\xi^\ast\eta^\ast\,.
\end{equation}

Vamos calcular agora uma integral gaussiana simples em variáveis de Grassmann
\begin{eqnarray}
 \int d \xi^\ast d\xi\,e^{-\xi^\ast c\xi}&=&\int d \xi^\ast d\xi(1-\xi^\ast c\xi)\nonumber\\
&=&\int d \xi^\ast d\xi(1+\xi\xi^\ast c)\nonumber\\
&=&\int d \xi^\ast d\xi(\xi\xi^\ast c)=c
\end{eqnarray}

Note que a expansão da exponencial acima em série de Taylor é finita porque $(\xi\xi^\ast)^2=0$\,.
\smallskip

Agora, calculemos a integral introduzindo um fator $(\xi\xi^\ast)$:
\begin{eqnarray}
 \int d \xi^\ast d\xi(\xi\xi^\ast)\,e^{-\xi^\ast c\xi}&=&\int d \xi^\ast d\xi(\xi\xi^\ast)(1-\xi^\ast\xi c)\nonumber\\
&=&\int d \xi^\ast d\xi(\xi\xi^\ast c)\left(\frac{1}{c} + \xi\xi^\ast \right)=1\,.
\end{eqnarray}

Precisamos integrar agora sobre $n$ variáveis de Grassmann. Considere então $n$ variáveis de Grassmann $\xi_i$ e $C$ uma matriz unitária tal que $\eta_i=C_{ij}\xi_j$. Assim,
\begin{eqnarray*}
 \prod\limits_{i=1}^n\eta_i&=&\frac{1}{n!}\varepsilon^{ijk...\ell}\xi_1\xi_2...\xi_\ell\\
&=&\frac{1}{n!}\varepsilon^{ijk...\ell}C_{ii'}\xi_{i'}C_{jj'}\xi_{j'}...C_{\ell\ell'}\xi_{\ell'}\\
&=&\left[\frac{1}{n!}\varepsilon^{ijk...\ell}C_{ii'}C_{jj'}...C_{\ell\ell'}  \right] \left[\xi_{i'}\xi_{j'}...\xi_{\ell'}  \right]\\
&=&(\mbox{det}\,C)\prod_i\xi_i\,.
\end{eqnarray*}

Sendo $D\xi=\prod\limits_i d\xi_i$ e $D\xi^\ast=\prod\limits_id\xi_i^\ast$ e $C_{ij}$ uma matriz diagonalizada, utilizando (E.6), segue então que
\begin{eqnarray}
 \int D\xi^\ast D\xi\,e^{-\xi_i^\ast C_{ii}\xi_i}&=&\int D\xi^\ast D\xi\,e^{-\sum\limits_i\xi_i^\ast c_{ii}\xi_i}\nonumber\\
&=&\prod\limits_ic_{ii}=\mbox{det}\, C\,.
\end{eqnarray}

Portanto, concluímos que o determinante da integração gaussiana em variáveis de Grassmann surge no numerador.

\chapter{Prova da relação $\ln \mbox{det}\, Q=Tr\ln Q$}

Seja a matriz diagonal $P_D=\left(\begin{array}{ccc}p_1&\cdots&0\\\vdots&\ddots&\vdots\\0&\cdots&p_n\end{array}\right)$ e $Q_D$ definida como

\begin{eqnarray}
 Q_D&=&I+P_D+\frac{1}{2}P_D^2+\frac{1}{3!}P_D^3\cdots\nonumber\\
&=&e^{P_D}\,,
\end{eqnarray}
então,
\begin{equation}
 P_D=\ln\,Q_D\,.
\end{equation}

Desta forma, $Q_D$ também será diagonal:

\begin{equation}
 Q_D=\left(\begin{array}{ccc}e^{p_1}&\cdots&0\\\vdots&\ddots&\vdots\\0&\cdots&e^{p_n}\end{array}\right)=\left(\begin{array}{ccc}q_1&\cdots&0\\\vdots&\ddots&\vdots\\0&\cdots&q_n\end{array}\right)\,.
\end{equation}

Sendo $U$ uma matriz arbitrária, multipliquemos (F.1) pela esquerda por $U$ e pela direita por $U^{-1}$:
\begin{equation}
 UQ_DU^{-1}=I+UP_DU^{-1}+\frac{1}{2}(UP_DU^{-1})(UP_DU^{-1})+\cdots
\end{equation}

Agora, seja $Q=UQ_DU^{-1}$ e $P=UP_DU^{-1}$ as matrizes {\it não diagonais} resultantes dessa operação. Assim, de (F.4), temos que

\begin{eqnarray*}
 Q=I+P+\frac{1}{2}P^2+\cdots=e^P\,,
\end{eqnarray*}
ou
\begin{equation}
 P=\ln\,Q\,.
\end{equation}

Agora, veja que:
\begin{eqnarray}
\ln\,\mbox{det}\,Q&=&\ln\,\mbox{det}\,(UQ_DU^{-1})=\ln(q_1q_2\cdots q_n)\nonumber\\
&=&\ln\,e^{p_1+p_2+\cdots+p_n}\nonumber\\
&=&p_1+p_2+\cdots+p_n=\sum\limits_ip_i=Tr\,P\,.
\end{eqnarray}

Portanto, de (F.5) e (F.6), temos que
\begin{equation}
 \ln\,\mbox{det}\,Q=Tr\,\ln\,Q\,.
\end{equation}

\chapter{Integrais de Feynman regularizadas dimensionalmente}

As integrais de Feynman utilizadas nesta dissertação, representadas em uma dimensão $D$ qualquer no espaço de Minkowski \cite{Tqc} , são dadas por

\begin{eqnarray}
&& \int \frac{d^Dp}{(2\pi)^D}\frac{1}{(p^2-m^2)^\alpha}=\frac{(-1)^\alpha i}{(4\pi)^{D/2}m^{2\alpha-D}}\frac{\Gamma(\alpha-D/2)}{\Gamma(\alpha)}    \\
&&\nonumber\\
&&\int \frac{d^Dp}{(2\pi)^D}\frac{p_\mu p_\nu}{(p^2-m^2)^\alpha} = \frac{g_{\mu\nu}}{2}\frac{(-1)^{\alpha-1}i}{(4\pi)^{D/2}m^{2\alpha-D-2}}\frac{\Gamma(\alpha-D/2-1)}{\Gamma(\alpha)}  \\
&&\nonumber\\
&&\int \frac{d^Dp}{(2\pi)^D}\frac{p_\mu p_\nu p_\rho p_\sigma}{(p^2-m^2)^\alpha}=\frac{(-1)^\alpha i}{(4\pi)^{D/2}m^{2\alpha-D-4}} \frac{\Gamma(\alpha-D/2-2)}{4\Gamma(\alpha)}(g_{\mu\nu}g_{\rho\sigma}+g_{\mu\rho}g_{\nu\sigma}+g_{\mu\sigma}g_{\nu\rho})\nonumber\\
\end{eqnarray}
 
Em (2+1)$D$, a integral utilizada na dissertação é dada por (G.1) com $D$ = 3 e $\alpha$ = 2:
\begin{equation}
 \int \frac{d^3p}{(2\pi)^3}\frac{1}{(p^2-m^2)^2}=\frac{i}{8\pi|m|}\,.
\end{equation}

Para o caso em que $D$ = 4 e $\alpha$ = 4, as integrais (G.1) e (G.2) são finitas e resultam em
\begin{eqnarray}
 && \int \frac{d^4p}{(2\pi)^4}\frac{1}{(p^2-m^2)^4}=\frac{i}{96\pi^2m^4}\\
&&\nonumber\\
&& \int \frac{d^4p}{(2\pi)^4}\frac{p_\mu p_\nu}{(p^2-m^2)^4}=-\frac{i}{192\pi^2m^2}g_{\mu\nu}
\end{eqnarray}

Contraindo a última integral nos índices de Lorentz, onde $g_{\mu\nu}g^{\mu\nu}=4$, temos que
\begin{equation}
 \int \frac{d^4p}{(2\pi)^4}\frac{p^2}{(p^2-m^2)^4}=-\frac{i}{48\pi^2m^2}\,.
\end{equation}

A integral (G.3) é logaritmicamente divergente para o caso em que $D$ = 4 e $\alpha$ = 4. Assim, vamos analisar o comportamento desta integral na vizinhança de $D=4-2\epsilon$, expandido a expressão \cite{Pes}
\begin{equation}
 \left( \frac{1}{m^2} \right)^{2-\frac{D}{2}}=1-\left(1-\frac{D}{2}\right)\log(m^2)+\cdots
\end{equation}

A expansão da função $\Gamma(\epsilon)$ é representada por
\begin{equation}
 \Gamma(\epsilon)=\frac{1}{\epsilon}-\gamma+{\cal O}(\epsilon)\,,
\end{equation}
onde $\gamma\approx0,5772$ é a constante de Euler-Mascheroni. Sendo $C(m^2)$ a constante de (G.3), para $\alpha=4$ temos que
\begin{eqnarray}
 C(m^2)&=&\frac{i}{(4\pi)^2}\frac{\Gamma(\epsilon)}{4\Gamma(4)}\left(\frac{4\pi}{m^2}\right)^\epsilon = \frac{i}{(4\pi)^2}\frac{\Gamma(\epsilon)}{4\Gamma(4)}e^{\epsilon\log\left(\frac{4\pi}{m^2}\right)}  \nonumber\\
&&\nonumber\\
&=&\frac{i}{384\pi^2}\left[ \frac{1}{\epsilon}-\gamma+{\cal O}(\epsilon)  \right]   \left[1+\epsilon\log\frac{4\pi}{m^2}+\frac{1}{2}\epsilon^2\left(\log\frac{4\pi}{m^2}  \right)^2   \right]\nonumber\\
&&\nonumber\\
&=&\frac{i}{384\pi^2}\left[\frac{1}{\epsilon}+\log\frac{4\pi}{m^2}-\gamma+{\cal O}(\epsilon)   \right]
\end{eqnarray}

Portanto,

\begin{eqnarray}
 \int \frac{d^Dp}{(2\pi)^D}\frac{p_\mu p_\nu p_\rho p_\sigma}{(p^2-m^2)^4}=\frac{i}{384\pi^2}\left[\frac{1}{\epsilon}+\log\frac{4\pi}{m^2}-\gamma+{\cal O}(\epsilon)   \right](g_{\mu\nu}g_{\rho\sigma}+g_{\mu\rho}g_{\nu\sigma}+g_{\mu\sigma}g_{\nu\rho})\nonumber\\
\end{eqnarray}

Contraindo dois índices de Lorentz, obtemos
\begin{equation}
  \int \frac{d^Dp}{(2\pi)^D}\frac{p^2p_\mu p_\nu }{(p^2-m^2)^4}=\frac{i}{64\pi^2}\left[\frac{1}{\epsilon}+\log\frac{4\pi}{m^2}-\gamma+{\cal O}(\epsilon)   \right]g_{\mu\nu}
\end{equation}

Novamente, contraindo os dois últimos, temos que
\begin{equation}
  \int \frac{d^Dp}{(2\pi)^D}\frac{p^4}{(p^2-m^2)^4}=\frac{i}{16\pi^2}\left[\frac{1}{\epsilon}+\log\frac{4\pi}{m^2}-\gamma+{\cal O}(\epsilon)   \right]\,.
\end{equation}


\newpage

\addcontentsline{toc}{chapter}{Referências}

\begin{thebibliography}{12cm}

\bibitem{Ein} A. Einstein, Ann. der Phyzik 17, 981 (1905).

\bibitem{Sch} J. Schwinger, Phys. Rev. {\bf 82},(1951) 914.

\bibitem{Sam} V. A. Kostelecký, S. Samuel, Phys. Rev. D {\bf 40}, 11886 (1989).

\bibitem{Gib} L. K. Gibbons {\it et al.} Phys. Rev. D {\bf 55}, 6625 (1997).

\bibitem{Gre} O. W. Greenberg, Phys. Rev. Lett. {\bf 89}, 23 1602 (2002).

\bibitem{Col} D. Colladay, V. A. Kosteleck\'y, Phys. Rev. D {\bf 55}, 6760 (1997).

\bibitem{Des} S. Deser, R. Jackiw and S. Templeton, Ann. Phys. (NY) {\bf 140} (1982) 372.

\bibitem{Dun} G. V. Dunne, \textsf{arXiv:hep-ph/99021151}.

\bibitem{Car} S. M. Carroll, G. B. Field and R. Jackiw, Phys. Rev. D {\bf 41}, 1231 (1990).

\bibitem{Jac} R. Jackiw and V. A. Kostelecký, Phys. Rev. Lett. \textbf{82}, 3572 (1999).

\bibitem{Alt} B. Altschul, {\it Phys. Rev D} {\bf 70}, 056005 (2004).

\bibitem{Lin} T. Mariz, J. R. Nascimento, A. Yu. Petrov, L. Y. Santos, A. J. da Silva  \textit{Phys.Lett.B} 661:312-318,2008.

\bibitem{Str} Streatwer, R. F. and Wightman, A. S., $PCT$, {\it Spin, Statistics, an All That}, Benjamin Cummings, London 1964.

\bibitem{Rus} R. Bluhm, V. A. Kostelek\'y and N. Russel, {\it Phys. Rev. D} {\bf 57}, 3932-3943 (1998).

\bibitem{Ahi} R. Bluhm, V. A. Kostelek\'y and N. Russel, {\it Phys. Rev. Lett.} {\bf 79}, 1432 (1997).

\bibitem{Fol} L. L. Foldy, S. A. Wouthuysen, Phys. Rev., \textbf{78}, 29 (1950).

\bibitem{Kha} O. G. Kharlanov, V. Ch. Zhukovsky, \textsf{arXiv:hep-th/0705.3306v1}.

\bibitem{Man} M. M. Ferreira Jr., F. M. O. Moucherek \textsf{arXiv:hep-th/0601018v3}.

\bibitem{Tqc} Gomes, M. O. C. {\it Teoria Qu\^antica dos Campos}. Edusp, São Paulo, 2002.

\bibitem{Gra} Graddshteyn, I. S, Ryzhik, I. M. {\it Table of Integrals, Series and Products: Corrected and Enlarged Edition}. Academic Press, Inc. San Diego, 1980.

\bibitem{Gom} M. Gomes, T. Mariz, J. R. Nascimento, and A. J. da Silva {\it Phys. Rev. D} {\bf 77}, 105002 (2008). 

\bibitem{Dat} V. A. Kostelek\'y and N. Russel, \textsf{arXiv:hep-th/08010287v3.3306v1}.

\bibitem{Ber} V. B. Berestetskii, E. M. Lifshitz and L. P. Pitaevslii, {\it Quantum Electrodynamics - Landau and Lifshitz Course of Theoretical Physics}, Oxford, Elsevier Butterworth-Heinemann, 2006 (Second Edition).

\bibitem{Pes} M. E. Peskin, D. V, Schroeder, {\it An Introduction to Quantum Field Theory}, The Advanced Book Program, 1995.

\bibitem{Mar} T. Mariz, J.R. Nascimento, E. Passos  Braz.J.Phys. 36 (2006) 1171-1177.

\bibitem{Edq} R. Bluhm \textsf{arXiv:hep-th/0011272v1}.
\end{thebibliography}
\end{document}